\documentclass[%
 amsmath,amssymb,
 aps, physrev,
]{revtex4-2}
\usepackage{cases}
\usepackage{pgffor}
\usepackage{graphicx}
\usepackage{dcolumn}
\usepackage{bm}
\usepackage{float}
\usepackage{booktabs}
\usepackage[english]{babel}
\usepackage{dblfloatfix}

\usepackage{pgffor}
\usepackage[colorlinks=true, allcolors=blue]{hyperref}
\usepackage[table,xcdraw]{xcolor}
\title{Integrating Adam Optimizer with analog Ising Machines to improve solution quality and speed}

\begin{document}

\preprint{APS/123-QED}

\title{\textbf{Beyond Gradient Descent: Adam for Analog Ising Machines} 
}%

\author{Stijn Van Vooren}
\author{Guy Van der Sande}
\author{Guy Verschaffelt}%
 \affiliation{Applied Physics research group, Vrije Universiteit Brussel, Pleinlaan 2, 1050 Brussels, Belgium}
 \email{Stijn.Van.Vooren@vub.be}

\date{\today}

\begin{abstract}
As Moore's law reaches its limits, Ising machines offer a promising alternative computing approach for difficult optimization problems. However, many analog, time-continuous Ising machines rely on gradient-descent-like dynamics to find solutions, which can limit speed and robustness. We investigate whether momentum and Adam optimization can improve these systems. Since these optimizers are traditionally formulated in discrete time, we derive continuous-time versions suitable for analog, time-continuous Ising-machine dynamics. On Max-Cut benchmarks, we find that Adam-based dynamics substantially reduce time-to-target and improve solution quality compared with gradient-descent- and momentum-based dynamics. We further introduce a first-order continuous-time approximation of Adam that is intended as a simpler starting point for future physical implementations and while performing better than the full Adam formulation in a continuous-time setting. We also study a purely algorithmic discrete-time setting, where the performance gap is reduced on easier problem instances, while the Adam-based update rule performs best on harder weighted problem instances. These results identify continuous-time Adam dynamics as a powerful design principle for analog Ising machines.
\end{abstract}

\maketitle


\section{Introduction}
The miniaturization of transistors is approaching physical limits, making it increasingly more difficult and costly to continue the exponential growth of computing power predicted by Moore's Law \cite{moore1,moore2}. As a result, the focus is shifting towards enhancing computational performance through software, algorithms, and innovative hardware architectures \cite{moore3}. In this evolving landscape, Ising machines have emerged as a promising non-von-Neumann computing approach to address NP-hard optimization problems, statistical sampling and machine learning more efficiently than traditional digital computers \cite{faster1,faster2,faster3,faster4}.

Ising machines are physical or algorithmic systems based on the Ising spin model from statistical physics. They can be used for optimization, sampling, and machine learning; here we focus on binary optimization, where the goal is to minimize a cost function. In this setting, the problem is mapped \cite{mapping} onto the Ising Hamiltonian by choosing the couplings $J_{ij}$ and bias terms $b_i$ such that low-energy spin configurations correspond to low-cost solutions:
\begin{equation}\label{HamiltonianIsing124}
H_{\mathrm{Ising}}(\sigma)=-\frac{1}{2}\sum_{i,j}J_{ij}\sigma_i\sigma_j-\sum_i b_i\sigma_i,
\end{equation}
where $\sigma=\{\sigma_1,\sigma_2,\ldots,\sigma_N\}$ denotes the state of the system, $\sigma_i\in\{-1,+1\}$ are binary spin variables, $J_{ij}$ are the couplings between spins, $b_i$ are the bias terms, and $N$ is the number of spins. 

To situate our work more precisely, it is useful to categorize Ising machines by how the spin state is represented. In analog Ising machines, each spin is stored internally as a continuous variable $x_i$, and the binary Ising spin $\sigma_i \in \{-1,+1\}$ is obtained only after thresholding, for example through $\sigma_i=\mathrm{sign}(x_i)$. In binary Ising machines, the hardware state itself is already binary. Both categories exist in e.g. electrical and optical hardware \cite{oscillator_isingmachine1,continuousCIM,momentum1,momentum2,faster3,photonic_isingmachine1}. 

Our work is aimed at analog machines, where it is also useful to distinguish time-continuous realizations, whose states evolve continuously in time, from time-discrete realizations, where continuous internal variables are updated in clocked steps. This distinction is important in practice: in many time-discrete physical implementations, the evaluation of the coupling becomes a genuine bottleneck. In particular, the analog state must first be measured and converted from analog to digital, then the coupling multiplication is carried out in digital hardware such as an FPGA, and the result must finally be converted back and reinjected into the physical system. This means that the machine is no longer operating at the native bandwidth of the analog substrate; instead, its update rate is constrained by the latency and throughput of the ADC/DAC and FPGA loop. By contrast, time-continuous analog Ising machines implement the spin coupling directly in the physical system itself, making them the natural setting when one wants to exploit the intrinsic parallelism and high bandwidth of analog hardware \cite{faster2,faster4}.

Representative examples help illustrate the different categories. Among analog, time-continuous Ising machines, electrical oscillator-network implementations relax toward low-energy Ising states through the continuous dynamics of nonlinear oscillators \cite{oscillator_isingmachine1}, while all-optical coherent Ising machines based on injection-locked multicore-fiber lasers realize the couplings with spatial light modulators, and relax through coupled laser dynamics \cite{continuousCIM}. Analog, time-discrete Ising machines also use continuous internal variables, but update them in synchronized discrete steps. Examples include simulated bifurcation, typically implemented in digital hardware \cite{momentum1,momentum2}, and optical measurement-feedback coherent Ising machines, which use time-multiplexed DOPO pulses and compute the couplings once per cavity round trip using a CPU or FPGA \cite{faster3}. Binary Ising machines, by contrast, store the spin directly in bistable electrical or optical states; one large-scale optical example encodes spins in binary optical phases using spatial light modulation \cite{photonic_isingmachine1}. With this classification we want to make clear that our results are not tied to a single device, but to the broader class of analog Ising-machine architectures.

Many of these Ising machines rely heavily on gradient descent for their optimization. While gradient descent is a cornerstone of many optimization methods, it comes with notable challenges. One key limitation is that gradient descent can only move downhill in the energy landscape. This means it is prone to getting stuck in local minima, which are suboptimal solutions that hinder the system from reaching the global minimum---the true optimal solution. To remedy this issue, Ising machines typically incorporate noise into the system. This noise helps the system escape local minima by introducing stochastic fluctuations, allowing the machine to ``jump'' out of shallow wells and explore other regions of the energy landscape. While this approach can improve the system's ability to find the global minimum, it also introduces challenges in terms of the required noise amplitude and the tradeoff between exploration and exploitation. Too much noise can cause the system to fluctuate too much, preventing convergence to an optimal solution, while too little noise might not help escape local minima effectively \cite{LeenNoise}. While gradient descent-based Ising machines have demonstrated success, they remain constrained by slow convergence and a tendency to get trapped in local minima. Momentum and Adam optimizers offer significant advantages over gradient descent. Momentum can be intuitively understood as the inertia of a moving object. When an optimizer uses momentum, it accumulates velocity from previous gradients, allowing it to continue moving in the same direction even when encountering uphill slopes (local minima). This inertia helps the optimizer to escape local minima and maintain its trajectory towards the global minimum. The Adam optimizer combines the benefits of momentum and adaptive learning rates, adjusting the step size based on the first and second moments of the gradients. This results in faster convergence and better handling of noisy gradients, making Adam-based Ising machines potentially more efficient and robust for combinatorial optimization tasks.

Momentum has been introduced in some Ising machine implementations \cite{MomentumOriginal}. Incorporating momentum dynamics into simulated bifurcation, the Toshiba Bifurcation Machine achieves improved optimization performance \cite{momentum1,momentum2}. This technique is state of the art for SK (Sherrington--Kirkpatrick) optimization problems, according to recent benchmarks \cite{state_of_the_art}, while for other problem types the data is incomplete. While momentum-based optimizers have been widely successful in combinatorial optimization problems, the Adam optimizer has dominated machine learning and other fields, particularly in deep learning, natural language processing, and computer vision, due to its superior convergence speed and robustness. Despite this widespread success, Adam optimizer had not been applied to Ising machines until Brown et al. \cite{AdamIsing} introduced them in the context of a continuous-variable Coherent Ising Machine (CIM). In their work, Brown et al. demonstrated that momentum and Adam can significantly accelerate the optimization process in CIMs by modifying the feedback with more sophisticated update rules. Their simulations showed that these methods improve convergence speed, sample diversity, and stability over standard gradient descent in CIMs, particularly for poorly conditioned problems. Notably, they found that Adam, with its adaptive learning rate adjustments, enhances robustness against variations in feedback strength, making it more reliable for practical implementations. However, their work is specifically done for the CIM system. Our work is inspired by this, but is substantially broader. Rather than adapting momentum and Adam to this specific CIM architecture, we reformulate them for the general class of analog, time-continuous Ising machines.

This distinction matters because Adam is normally used as a discrete-time optimization algorithm. In purely algorithmic implementations, and also in many time-discrete physical Ising machines where the state is measured, processed digitally, and fed back through electronics such as an FPGA, porting these optimizers is comparatively straightforward: one can directly use the standard discrete momentum or Adam update rules known from the literature. In such cases, the main computation, often dominated by evaluating the coupling, is carried out digitally, so the effective update rate is tied to the speed of the digital control loop rather than to the native bandwidth of the analog substrate. Our focus, instead, is on analog Ising machines whose states evolve directly in continuous time. For this class of systems, the standard discrete formulations of momentum and Adam optimizer are not directly applicable. In this work, we therefore derive continuous-time versions of both optimizers and use them to construct analog, time-continuous Ising-machine dynamics. While continuous-time momentum is well established, continuous-time formulations of Adam have also been developed in the optimization literature \cite{barakat}. Our contribution here is to adapt this continuous-time perspective to the present Ising-machine setting, and to introduce a simplified first-order variant, which in our benchmarks performs even better than the full Adam-based formulation.

In addition, our work investigates a key aspect that was not explored by Brown et al., namely the impact of nonlinearity on performance. Ising machines rely on nonlinearity to enforce binary behavior, and different choices of nonlinear functions can drastically influence convergence speed and solution quality \cite{Orders}. We systematically analyze how different nonlinearities affect the optimization dynamics. This provides new insights into the interplay between nonlinearity and optimizer choice, further enhancing the efficiency of Ising machines across different analog, time-continuous implementations.

The remainder of this paper is organized as follows. We first place gradient-descent-based Ising machines in a general optimization framework and use this formulation to introduce momentum- and Adam-based dynamics. Since these optimizers are conventionally defined in discrete time, we recast them in continuous time so that they can be applied naturally to analog Ising machines, whose physical dynamics evolve continuously. In doing so, we also introduce a simplified first-order approximation of the continuous-time Adam dynamics, designed to be more suitable for physical implementation. We then benchmark the resulting Ising-machine models across several nonlinearities on standard Max-Cut problem instances, using time-to-target, solution quality, and success rate as performance measures. The benchmarks show that the first-order Adam-based dynamics, especially with the sigmoid nonlinearity, give large and consistent improvements over gradient-descent- and momentum-based Ising machines, both in convergence speed and in the quality of the solutions obtained. Finally, we also analyze a purely algorithmic discrete-time setting, where the continuous-time dynamics are implemented through numerical discretization and the timestep is treated as an additional hyperparameter.

\section{Gradient descent-based Ising machines (GD-IM)}

We begin with gradient descent-based Ising machines (GD-IM), which provide the starting point for the momentum- and Adam-based extensions introduced below. As discussed above, the discrete optimization problem is first mapped onto the Ising Hamiltonian of Eq.~(\ref{HamiltonianIsing124}), so that the binary problem can be written as
\begin{equation}
\min_{\sigma\in\{-1,+1\}^N}\left\{f(\sigma)=H_{\mathrm{Ising}}(\sigma)\right\} \; . 
\end{equation}
Solving binary optimization problems directly is computationally prohibitive due to the vast combinatorial solution space, making brute-force methods impractical to impossible. To address this, we leverage the analog nature of Ising machines by replacing the binary spins $\sigma_i\in\{-1,+1\}$ with continuous variables $x_i\in\mathbb{R}$. This turns the binary problem into the continuous optimization problem
\begin{equation}\label{fproblem}
\min_{\mathbf{x}\in\mathbb{R}^N} f(\mathbf{x}),
\end{equation}
with
\begin{equation}\label{isingcont}
f(\mathbf{x}) = H_{\mathrm{Ising,cont}}(\mathbf{x})
= -\frac{1}{2}\sum_{i,j} J_{ij} x_i x_j - \sum_i b_i x_i .
\end{equation}
This continuous formulation allows us to use gradient-based dynamics, while the original binary configuration is recovered only at readout by taking
\(\sigma_i=\operatorname{sign}(x_i)\). Henceforth, we work with the continuous variables
\(\mathbf{x}\in\mathbb{R}^N\) and reserve \(\boldsymbol{\sigma}\) for the binary readout. For simplicity, we neglect external biases throughout this paper, setting \(b_i=0\). Continuous-time Gradient descent-based Ising machines evolve to their lower energy states according to 
\begin{equation}
    \frac{d x_i}{d t} = - \left(\boldsymbol{\nabla} f(\boldsymbol{x})\right)_i.
\end{equation}
From Eq.~(\ref{fproblem}), the gradient becomes
\begin{equation}\label{classicalHamiltonian}
-\left(\boldsymbol{\nabla} f(\boldsymbol{x})\right)_i= \sum_{ j } J_{ij} x_j \quad .
\end{equation}
This formulation ensures that the system follows the negative gradient, decreasing the function value over time in the direction of steepest descent. In numerical simulations of continuous-time Ising machines this continuous evolution is discretized using e.g. Euler's method. 

For analog Ising machines, the continuous dynamics must remain consistent with the original binary problem, in which each spin can only take the values $+1$ or $-1$. Since analog spin variables can take a continuous range of values, a nonlinearity is introduced to steer them toward binary states and thereby recover meaningful solutions to the original problem. Böhm et al. showed that the choice of nonlinearity can strongly influence the computational performance of analog Ising machines \cite{Orders}. 
To examine this dependence systematically, we consider four nonlinear choices, following Böhm et al. \cite{Orders}, each of which modifies Eq.~(\ref{classicalHamiltonian}) so that the dynamics favor binary states. In all cases, $\alpha$ controls the local restoring force, $\beta$ sets the coupling strength, and $\gamma \zeta_i(t)$ is a Gaussian white-noise term that can help the system escape shallow local minima.

With the \textit{polynomial} choice, corresponding to the nonlinear term arising in coherent Ising machine (CIM) dynamics~\cite{Orders}, the gradient becomes
\begin{equation}\label{poly}\hspace*{-0.6cm}
-\left(\boldsymbol{\nabla} f(\boldsymbol{x})\right)_i=(\alpha-1)x_i-x_i^3+\beta \sum_{ j } J_{ij} x_j + \gamma \zeta_i(t),
\end{equation}
which introduces a double-well structure in the energy landscape, which effectively deepens the energy landscape near binary-valued amplitudes and raises it for intermediate values. For the \textit{sigmoid} nonlinearity, representative of neuron-inspired dynamics~\cite{Orders}, one obtains
\begin{equation}\label{sigmoid}\hspace*{-0.6cm}
-\left(\boldsymbol{\nabla} f(\boldsymbol{x})\right)_i=-x_i+\tanh \left( \alpha x_i+\beta \sum_{ j } J_{ij} x_j  + \gamma \zeta_i(t) \right),
\end{equation}
where the hyperbolic tangent bends large inputs toward bounded values instead of letting them grow without limit. In this way, small inputs are treated almost linearly, while larger ones are gradually compressed. The \textit{periodic} nonlinearity, associated with optoelectronic oscillator (OEO) implementations~\cite{Orders}, leads to
\begin{equation}\label{periodic}\hspace*{-0.6cm}
-\left(\boldsymbol{\nabla} f(\boldsymbol{x})\right)_i=-x_i + \cos^2 \left( \alpha x_i - \frac{\pi}{4}  + \beta \sum_j J_{ij} x_j + \gamma \zeta_i(t)\right) - \frac{1}{2},
\end{equation}
so that the restoring term becomes periodic, creating a repeated pattern of preferred amplitude regions. Finally, for the \textit{clipped} nonlinearity, corresponding to clipped-OEO dynamics~\cite{Orders}, the gradient becomes
\begin{equation}\label{clipped}\hspace*{-0.6cm}
    -\left(\boldsymbol{\nabla} f(\boldsymbol{x})\right)_i= \begin{cases} 
(\alpha - 1)x_i + \beta \sum_j J_{ij}x_j + \gamma \zeta_i(t), & |x_i| \leq 0.4 \\
0, & |x_i| > 0.4,
\end{cases}
\end{equation}
which imposes a hard cutoff on the amplitude and thereby tends to homogenize the spin magnitudes.

Together with one of these choices, the gradient-flow dynamics
\(
\frac{d\mathbf{x}}{dt} = -\boldsymbol{\nabla} f(\mathbf{x})
\)
define the \textit{Gradient Descent Ising Machine} (GD-IM). In the Momentum Ising Machine (MOM-IM), Adam Ising Machine (ADAM-IM), and first-order Adam Ising Machine (1-ADAM-IM), introduced below, the nonlinearity is kept fixed and only the optimization dynamics are modified.

\section{Theoretical results}
\subsection{Momentum-based Ising Machines (MOM-IM)}
To enable the physical implementation of the momentum optimizer in Ising machines, we derive its continuous-time formulation starting from the well-known discrete momentum optimizer \cite{MomentumOriginal}. A continuous formulation is essential for continuous-time analog Ising machines, where state evolution occurs dynamically rather than through discrete updates. By formulating the optimizer in continuous time, we establish a natural extension of the standard discrete momentum method, ensuring that it recovers the well-known discrete version under Euler discretization. The discrete momentum optimizer for an objective function 
$f(\mathbf{x}): \mathbb{R}^N \to \mathbb{R}$ 
is defined by the update rules \cite{MomentumOriginal}
\begin{equation}
\label{mom1}
\left\{
\begin{aligned}
    v_i^{(j+1)} &= \beta_1 v_i^{(j)}
                  + (1-\beta_1)\left(\boldsymbol{\nabla} f(\mathbf{x}^{(j)})\right)_i, \\[1mm]
    x_i^{(j+1)}&= x_i^{(j)} - \eta\, v_i^{(j+1)} ,
\end{aligned}
\right.
\end{equation}
where $\beta_1\in[0,1)$ is a damping parameter, $\eta$ is the learning rate, 
and $j$ denotes the iteration index.
To obtain a continuous-time representation, we first rewrite \eqref{mom1} by expressing the updates relative to their previous values:
\begin{equation}
\label{mom2}
\left\{
\begin{aligned}
v_i^{(j+1)} &= v_i^{(j)} + (1-\beta_1)\!\left((\boldsymbol{\nabla} f(\mathbf{x}^{(j)}))_i - v_i^{(j)}\right), \\[1mm]
x_i^{(j+1)} &= x_i^{(j)} - \eta\, v_i^{(j+1)}. 
\end{aligned}
\right.
\end{equation}
We then introduce an explicit timestep \(\Delta t>0\) and interpret the iteration index as sampling a trajectory at times \(t_j=j\,\Delta t\). This makes it possible to view the scheme as a forward Euler discretization of an underlying continuous-time system. Recall that for an ODE \(\dot y(t)=g(y(t),t)\), the forward Euler time step is
\begin{equation}
y(t+\Delta t)=y(t)+\Delta t\,g(y(t),t).
\end{equation}
Accordingly, we write
\begin{equation}
\left\{
\begin{aligned}
v_i^{(j+1)} 
    &= v_i^{(j)} 
      + \Delta t\,(1-\beta_1)\!\left((\boldsymbol{\nabla} f(\mathbf{x}^{(j)}))_i - v_i^{(j)}\right), \\[2mm]
x_i^{(j+1)} 
    &= x_i^{(j)} - \Delta t\,\eta\, v_i^{(j+1)}
      \;\approx\; x_i^{(j)} - \Delta t\,\eta\, v_i^{(j)} , 
\end{aligned}
\right.
\end{equation}
which reduces to \eqref{mom2} when \(\Delta t=1\).
In \eqref{mom1:xi-dt}, we replace \(v_i^{(j+1)}\) by \(v_i^{(j)}\) to obtain an explicit Euler form. Since \(v_i^{(j+1)}-v_i^{(j)}=\mathcal O(\Delta t)\), this modifies the \(x_i\)-update only by \(\mathcal O(\Delta t^2)\).
Interpreting $v_i^{(j)} = v_i(t_j)$ and $x_i^{(j)} = x_i(t_j)$, 
the updates take the Euler form
\begin{equation}
\left\{
\begin{aligned}
v_i(t+\Delta t) 
    &= v_i(t) + \Delta t\,(1-\beta_1)\!\left((\boldsymbol{\nabla} f(\mathbf{x}(t)))_i - v_i(t)\right), \\[2mm]
x_i(t+\Delta t) 
    &= x_i(t) - \Delta t\, \eta\, v_i(t).
\end{aligned}
\right.
\end{equation}
Taking the limit $\Delta t \to 0$ yields the continuous-time system, in vector notation
\begin{equation}\label{momentum_cont}
\left\{
\begin{aligned}
        \dfrac{d\mathbf{v}}{dt} &= (1-\beta_1)\left(\boldsymbol{\nabla} f - \mathbf{v}\right),\\[2mm]
        \dfrac{d\mathbf{x}}{dt} &= -\,\eta\, \mathbf{v},
\end{aligned}
\right.
\end{equation}
which matches the standard continuous-time momentum formulation found in 
\cite{momentum1,momentum2,ODE1,barakat}. Together with a suitable choice of nonlinearity (i.e. using Eq.~(\ref{poly}), (\ref{sigmoid}), (\ref{periodic}) or (\ref{clipped}) to specify $\boldsymbol{\nabla} f$) to enforce binary spin states, 
this defines the \textit{Momentum Ising Machine} (MOM-IM).
This system can be rewritten as a second-order differential equation by eliminating the auxiliary velocity variable, yielding
\begin{equation}
    \frac{d^2 x_i}{dt^2} + (1-\beta_1)\frac{dx_i}{dt} + \eta (1-\beta_1) (\nabla f(\mathbf{x}))_i = 0 \quad ,
\end{equation}
which corresponds to the classical heavy ball with friction model introduced by Polyak \cite{MomentumOriginal}. In this analogy, the optimization trajectory resembles a massive particle moving through a potential energy landscape defined by $f(\mathbf{x})$, with friction proportional to velocity. This formulation captures the role of momentum as a means of accelerating convergence: inertia allows the system to maintain direction across shallow regions of the landscape, while the friction term ensures eventual stabilization near a minimum. Compared to gradient descent, which lacks such temporal smoothing, this dynamic can more effectively traverse narrow valleys and escape suboptimal minima, often leading to faster convergence and less entrapment in local minima.

In the adiabatic limit $(1-\beta_1) \rightarrow + \infty$, we get $v_i \rightarrow (\nabla f(\mathbf{x}))_i$, and the continuous momentum optimizer Eq.~(\ref{momentum_cont}) becomes the gradient descent optimizer, if and only if $\eta = 1$. For this reason, to match time-scales of the two optimizers, such that we are making a fair comparison in time based performance measures, we set $\eta = 1$ throughout this work. 

\subsection{Adam-based Ising Machines (ADAM-IM) and First-order Adam Ising Machines (1-ADAM-IM)}
Just as in the momentum case, we start from the discrete Adam update rules
\cite{AdamOriginal} and rewrite them into a form suitable for interpretation as
a forward Euler discretization. Adam augments momentum by tracking a second
moment of the gradient together with bias-corrected estimates. The standard
updates read \cite{AdamOriginal}
\begin{equation}\label{adam_discrete}
\left\{
\begin{aligned}
g_i^{(j+1)} &= \left(\boldsymbol{\nabla} f(\mathbf{x}^{(j)})\right)_i,\\[1mm]
v_i^{(j+1)} &= \beta_1 v_i^{(j)} + (1-\beta_1) g_i^{(j+1)},\\[1mm]
w_i^{(j+1)} &= \beta_2 w_i^{(j)} + (1-\beta_2)\left(g_i^{(j+1)}\right)^2,\\[1mm]
\hat v_i^{(j+1)} &= \dfrac{v_i^{(j+1)}}{1-\beta_1^{\,j}},
\quad
\hat w_i^{(j+1)} = \dfrac{w_i^{(j+1)}}{1-\beta_2^{\,j}},\\[1mm]
x_i^{(j+1)} &= x_i^{(j)} - \eta\, 
      \dfrac{\hat v_i^{(j+1)}}{\sqrt{\hat w_i^{(j+1)}}+\epsilon}.
\end{aligned}
\right.
\end{equation}
Eliminating the auxiliary variables by substitution and rewriting the updates relative to their previous values, as in the momentum case, gives
\begin{equation}
\left\{
\begin{aligned}
v_i^{(j+1)} &= v_i^{(j)} + (1-\beta_1)
                \left((\boldsymbol{\nabla} f(\mathbf{x}^{(j)}))_i - v_i^{(j)}\right),\\[1mm]
w_i^{(j+1)} &= w_i^{(j)} + (1-\beta_2)
                \left((\boldsymbol{\nabla} f(\mathbf{x}^{(j)}))_i^2 - w_i^{(j)}\right),\\[1mm]
x_i^{(j+1)} 
    &= x_i^{(j)}
      - \eta\,\dfrac{\sqrt{1-\beta_2^{\,j}}}{1-\beta_1^{\,j}}
        \dfrac{v_i^{(j+1)}}{\sqrt{w_i^{(j+1)}}+\epsilon}.
\end{aligned}
\right.
\end{equation}
As in the momentum derivation, we first replace $v_i^{(j+1)}$ and $w_i^{(j+1)}$ in the $x_i$-update by their values at step $j$, which gives the first-order approximation
\begin{equation}\label{adam_Euler_approx}
x_i^{(j+1)} \approx
x_i^{(j)} - \eta\,\dfrac{\sqrt{1-\beta_2^{\,j}}}{1-\beta_1^{\,j}}
             \dfrac{v_i^{(j)}}{\sqrt{w_i^{(j)}}+\epsilon}.
\end{equation}
Since both $v_i$ and $w_i$ change by $\mathcal{O}(\Delta t)$ over a single step, this modifies the $x_i$-update only at order $\mathcal{O}(\Delta t^2)$ and therefore does not affect the continuous-time limit. We now introduce an explicit timestep $\Delta t>0$ and interpret the iteration index as sampling a trajectory at times $t_j:=j\,\Delta t$. In this parametrization, the original discrete updates are recovered for $\Delta t=1$, and the scheme becomes
\begin{equation}
\left\{
\begin{aligned}
v_i^{(j+1)} 
    &= v_i^{(j)}
      + \Delta t\,(1-\beta_1)
        \left((\boldsymbol{\nabla} f(\mathbf{x}^{(j)}))_i - v_i^{(j)}\right),\\[2mm]
w_i^{(j+1)} 
    &= w_i^{(j)}
      + \Delta t\,(1-\beta_2)
        \left((\boldsymbol{\nabla} f(\mathbf{x}^{(j)}))_i^2 - w_i^{(j)}\right),\\[2mm]
x_i^{(j+1)} 
    &= x_i^{(j)}
      - \Delta t\,\eta\,
        \dfrac{\sqrt{1-\beta_2^{\,j}}}{1-\beta_1^{\,j}}
        \dfrac{v_i^{(j)}}{\sqrt{w_i^{(j)}}+\epsilon}.
\end{aligned}
\right.
\end{equation}
Identifying \(v_i^{(j)}\) with \(v_i(t_j)\), \(w_i^{(j)}\) with \(w_i(t_j)\), and \(x_i^{(j)}\) with \(x_i(t_j)\), and then taking the limit \(\Delta t\to 0\), yields the continuous-time Adam dynamics, where powers, square roots, and divisions are understood componentwise
\begin{equation}\label{adam_cont}
\left\{
\begin{aligned}
        \dfrac{d\mathbf{v}}{dt}
            &=(1-\beta_1)\left(\boldsymbol{\nabla} f - \mathbf{v}\right),\\[2mm]
        \dfrac{d\mathbf{w}}{dt}
            &=(1-\beta_2)\left((\boldsymbol{\nabla} f)^2 - \mathbf{w}\right),\\[2mm]
        \dfrac{d\mathbf{x}}{dt}
            &= -\,\eta\,
              \dfrac{\sqrt{1-\beta_2^{\,t}}}{1-\beta_1^{\,t}}
              \dfrac{\mathbf{v}}{\sqrt{\mathbf{w}}+\epsilon}.
\end{aligned}
\right.
\end{equation}
Together with a suitable nonlinearity to enforce binary spin states, this
continuous-time formulation defines the \emph{Adam Ising Machine} (ADAM-IM).

\begin{center} 
    \begin{figure}[h!] 
        \centering 
        \includegraphics[width=0.6\linewidth]{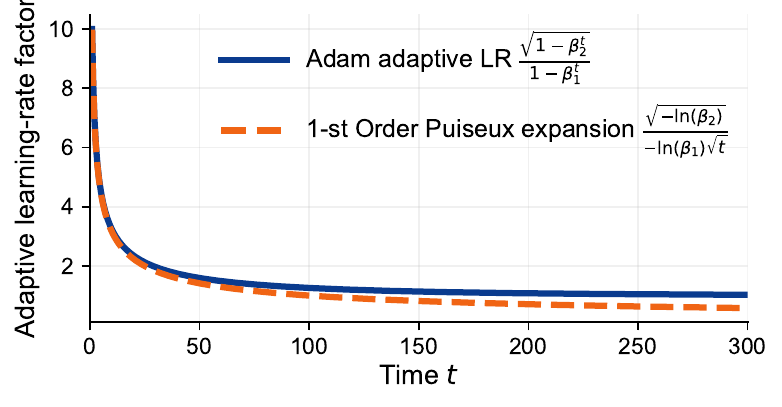} 
        \caption{Comparison of the Adam adaptive learning-rate factor and its first-order Puiseux approximation for $\beta_1=\beta_2=0.99$.} \label{Taylor} 
    \end{figure} 
\end{center}

The third line of Eq.~(\ref{adam_cont}) contains a time-dependent prefactor $\frac{\sqrt{1-\beta_2^t}}{1-\beta_1^t}$ that complicates both physical implementation and analysis, which we refer to as the \textit{adaptive learning rate}. To reduce this complexity, we examine its first-order Puiseux series expansion:
\begin{equation}\label{adam_prefactor_expansion}
    \frac{\sqrt{1-\beta_2^t}}{1-\beta_1^t}
    = \frac{\sqrt{-\ln(\beta_2)}}{-\ln(\beta_1)\sqrt{t}} 
    + \mathcal{O}(\sqrt{t}) .
\end{equation}
Motivated by this, we define a simplified variant
\begin{equation}
\label{one_adam_cont}
\left\{
\begin{aligned}
        \frac{d\mathbf{v}}{dt} &=  (1-\beta_1)\left(\boldsymbol{\nabla} f - \mathbf{v}\right) \\[6pt]
        \frac{d\mathbf{w}}{dt} &=  (1-\beta_2)\left(\left(\boldsymbol{\nabla} f\big(\mathbf{x}(t)\big)\right)^2 - \mathbf{w}\right) \\[6pt]
        \frac{d\mathbf{x}}{dt} &= -\eta\, 
        \frac{\mathbf{v}}{\sqrt{\mathbf{w}\,t}+\epsilon} \quad ,
\end{aligned}
\right.
\end{equation}
where the constant prefactor $\frac{\sqrt{-\ln(\beta_1)}}{-\ln(\beta_2)}$ is absorbed into $\eta$. Together with a suitable choice of nonlinearity to enforce binary spin states, 
this defines the \textit{First-order Adam Ising Machine} (1-ADAM-IM).
In Figure~\ref{Taylor}, we compare the Adam adaptive learning rate \(\frac{\sqrt{1-\beta_2^t}}{1-\beta_1^t}\) (shown in blue) with its first-order Puiseux expansion, \(\frac{\sqrt{-\ln(\beta_2)}}{-\ln(\beta_1)\sqrt{t}}\) (shown in orange), for \(\beta_1=\beta_2=0.99\). The two expressions show good agreement, indicating that the simplified form captures the behavior of the original function very well. Their main qualitative difference appears at large times: the exact Adam factor approaches $1$, whereas the first-order approximation decays to $0$. Consequently, the first-order formulation gradually reduces the effective step size of the dynamics. This may be advantageous after the initial descent, since the largest energy reductions typically occur early in the evolution, while further improvements near low-energy configurations require more local adjustments that are less likely to overshoot nearby lower-energy states.
The first-order formulation provides a simpler alternative to the full Adam-based Ising machine. We do not propose a concrete physical implementation of either continuous-time Adam formulation here. Nevertheless, comparing Eq.~(\ref{adam_cont}) with Eq.~(\ref{one_adam_cont}) shows a clear simplification: the full Adam state equation contains the prefactor depending on $\beta_1^t$ and $\beta_2^t$, whereas the first-order system uses the leading algebraic time dependence from Eq.~(\ref{adam_prefactor_expansion}), proportional to $t^{-1/2}$. It therefore provides a useful starting point for exploring how Adam-like dynamics could be incorporated into future analog Ising-machine hardware. In addition, $\beta_1$ and $\beta_2$ no longer appear in the state equation for $d\mathbf{x}/dt$, but only control the relaxation of the internal moment variables $\mathbf{v}$ and $\mathbf{w}$. This removes the direct connection to the discrete-time Adam coefficients and allows these parameters to be tuned more freely. As shown below, this simplified formulation improves both convergence speed and solution quality.

\subsection{Relations between Gradient descent, Momentum, Adam, and 1-Adam}
When comparing the time-to-target (TTT) of different continuous-time optimizers, one would ideally like the comparison to reflect differences in the optimizer dynamics themselves rather than a trivial rescaling of time. Exact matching of intrinsic timescales is generally not possible, because GD-IM, MOM-IM, ADAM-IM, and 1-ADAM-IM have different moment dynamics. For MOM-IM,
\begin{equation}
\dot{\mathbf{v}} = (1-\beta_1)\big(\nabla f(\mathbf{x}) - \mathbf{v}\big),
\qquad
\dot{\mathbf{x}} = -\,\eta\,\mathbf{v},
\end{equation}
the adiabatic limit \((1-\beta_1)\to\infty\) forces \(\mathbf{v}\to\nabla f(\mathbf{x})\), such that
\begin{equation}\label{gradient_descent}
\dot{\mathbf{x}} = -\,\eta\,\nabla f(\mathbf{x}),
\end{equation}
which coincides with GD-IM for \(\eta=1\). Thus MOM-IM reduces cleanly to GD-IM in the adiabatic limit. By contrast, ADAM-IM does not reduce cleanly to MOM-IM. Its \(x\)-equation contains the adaptive factor
\begin{equation}\label{adaptive_factor}
\frac{\sqrt{1-\beta_2^{\,t}}}{1-\beta_1^{\,t}}\,
\frac{1}{\sqrt{\mathbf{w}}+\epsilon}.
\end{equation}
For a reduction to MOM-IM, this factor would have to become a nonzero constant. This does not happen in general, since the adaptive factor (Eq.~(\ref{adaptive_factor})) depends explicitly on both time and the second-moment variable \(\mathbf{w}\). Even in the late-time regime \(t\to\infty\), one only obtains
\begin{equation}
\frac{\sqrt{1-\beta_2^{\,t}}}{1-\beta_1^{\,t}}\to 1,
\end{equation}
so that, if \(\mathbf{w}(t)\to \mathbf{w}_\infty\), the dynamics approach
\begin{equation}
\dot{\mathbf{x}}
\approx
-\,\eta\,\frac{\mathbf{v}}{\sqrt{\mathbf{w}_\infty}+\epsilon},
\end{equation}
that is, a rescaled momentum-like step. This is still not the same as MOM-IM. A similar obstruction appears for 1-ADAM-IM,
\begin{equation}
\dot{\mathbf{x}}
=
-\eta\,\frac{\mathbf{v}}{\sqrt{\mathbf{w}}\sqrt{t}+\epsilon}.
\end{equation}
For this equation to reduce to GD-IM, the right-hand side would have to be proportional to the gradient, as in Eq.~(\ref{gradient_descent}). Even when $\mathbf{v}$ follows $\nabla f$, this proportionality is prevented by the factor $(\sqrt{\mathbf{w}}\sqrt{t}+\epsilon)^{-1}$, which remains component- and time-dependent. Accordingly, we set \(\eta=1\) for GD-IM and MOM-IM, while for ADAM-IM and 1-ADAM-IM we leave \(\eta\) as a free hyperparameter.
Accordingly, absolute comparisons of \(\mathrm{TTT}\) should be interpreted with care. In particular, we place greater emphasis on relative trends in time-to-target across problem instances and on timescale-independent measures such as solution quality and success rate.
\section{Benchmark results}
All simulations in this work rely on a numerical Euler--Maruyama discretization of the continuous-time dynamical equations underlying each Ising-machine model. Every grid scan and Bayesian optimization evaluation averages \(400\) stochastic trajectories of \(10{,}000\)–\(15{,}000\) Euler time steps (\(10{,}000\) for g05 and \(15{,}000\) for \textsc{Gset}). The Euler timestep is chosen as $\Delta t=0.01$ for the continuous-time Ising machines and as a tunable parameter for the purely algorithmic comparisons. At the start of each run, all required initial state components are sampled independently from zero-mean Gaussian noise with standard deviation \(\sqrt{\Delta t}\). The sampled values are used directly for \(x_i(0)\) and, when present, \(v_i(0)\), while Adam-based \(w_i(0)\) is initialized using the absolute value of its sample. The system then evolves with fixed hyperparameters according to the corresponding dynamical equations; no annealing schedule is applied.

The success rate is defined as the fraction of runs that reach a prescribed target energy $E_{\mathrm{target}}$. We benchmark all Ising-machine variants on two well-established MAX-CUT problem families: the g05 problem instances from the BiqMac library~\cite{BiqMac}, which consist of unweighted problems of sizes $N=60, 80, 100$ with exactly known ground-state energies, and the \textsc{Gset} problem instances~\cite{Gset}, which include significantly larger sparse graphs of sizes $N=800$ and $N=2000$, both unweighted and weighted. For g05 problem instances we set $E_{\mathrm{target}}$ equal to the ground state, whereas for \textsc{Gset} we choose $E_{\mathrm{target}} = 0.995\,E_{\mathrm{best}}$, where $E_{\mathrm{best}}$ is the energy of the best cut value reported in the literature.

To quantify optimization speed, we first introduce the time-to-target \(\mathrm{TTT}\). For each individual run, we record the \emph{first-passage time}, i.e.\ the earliest time at which the trajectory first reaches a state whose energy satisfies the chosen target criterion. The quantity \(T_a\) denotes the average of these first-passage times over all runs that successfully reach the target, while the transient success rate \(SR_{\mathrm{tr}}\) is the fraction of runs that reach the target at least once during their evolution. Because independent repetitions can be used to increase the overall success probability, we define
\begin{equation}
\mathrm{TTT} =
\begin{cases}
T_a \dfrac{\log(0.01)}{\log(1 - SR_{\mathrm{tr}})}, & SR_{\mathrm{tr}} \le 0.99, \\
T_a, & SR_{\mathrm{tr}} > 0.99,
\end{cases}
\end{equation}
which represents the time required to reach the specified target with \(99\%\) probability. The time entering \(\mathrm{TTT}\) via $T_a$ depends on the setting. For the continuous-time Ising-machine models, time is measured as the simulated Euler time \(t=n\Delta t\) obtained from the Euler discretization. Throughout this work, we fix \(\Delta t=0.01\), so that all continuous-time methods are compared on the same numerical time scale and the discretization remains sufficiently fine to resolve the underlying dynamics. Thus, unless stated otherwise, \(\mathrm{TTT}\) refers to this Euler time. For the purely algorithmic discrete-time implementations discussed in Sec.~\ref{subsec:algorithmic_discrete}, by contrast, \(\Delta t\) is a tunable numerical parameter rather than a meaningful time variable, and we therefore report wall-clock CPU time instead, denoted by \(\mathrm{TTT}_{\mathrm{CPU}}\). For Bayesian optimization, we use \(-1/\mathrm{TTT}\) rather than the raw \(\mathrm{TTT}\) as the objective. This is because \(\mathrm{TTT}\) diverges when no target is reached, creating a discontinuity that is difficult for Bayesian optimization to model. By contrast, \(-1/\mathrm{TTT}\) maps such cases to a finite value \(0\), preserves the preference for low \(\mathrm{TTT}\), gives higher resolution in the low-\(\mathrm{TTT}\) region, and restores the minimization convention through the minus sign. The same construction is used for \(\mathrm{TTT}_{\mathrm{CPU}}\) in the algorithmic setting.

Each Ising-machine variant has a different number of free hyperparameters. To keep the comparison both fair and computationally tractable, we therefore use a two-stage procedure. First, for every nonlinearity and every Ising-machine variant, we perform a coarse \(30\times30\) grid scan over \((\alpha,\beta)\), with all remaining parameters fixed to baseline values. This stage is restricted to the g05 problem instances and provides an initial comparison of the nonlinearities at the coarse resolution of the grid scan. Second, after fixing the nonlinearity, we perform a full Bayesian optimization over all remaining free hyperparameters. Each Bayesian-optimization run uses \(1000\) parameter evaluations, of which the first \(300\) are random and the remaining \(700\) are chosen adaptively. For the final optimizer comparisons, the Bayesian optimization was repeated over seven distinct parameter ranges per optimizer. This was done to avoid relying on a single broad search domain: in practice, we found that very large parameter ranges made it difficult for Bayesian optimization to refine the optimum, since the relevant high-performing region could occupy only a small part of the full domain. Splitting the search into several smaller, range-focused runs improved the robustness of the tuning while keeping the total computational cost manageable. The choice of seven ranges was therefore a compromise between search coverage and the available computational budget. All optimizers are compared under the same computational budget, using the same number of Bayesian-optimization evaluations, stochastic trajectories per evaluation, and Euler steps per trajectory. Additional Bayesian-optimization details are provided in the Supplemental Material at [URL will be inserted by publisher]. Moreover, in our implementation the additional arithmetic required by MOM-IM, ADAM-IM, and 1-ADAM-IM is negligible compared to the cost of evaluating the coupling term \(\sum_j J_{ij}x_j\), which accounts for more than \(99\%\) of the runtime in our profiling. The optimizer-specific overhead scales only as \(\mathcal{O}(N)\), whereas for dense problems the coupling evaluation scales as \(\mathcal{O}(N^2)\), which explains why the coupling dominates the computational cost.
\subsection{Selection of nonlinearity}
To begin our analysis, we compare the performance of all nonlinearities for 
Adam‑based Ising machines using a systematic grid scan over the nonlinearity parameters.
Grid scans are computationally feasible when restricted to a small 
number of parameters. For this reason, we restrict our first evaluation to a 
$30\times30$ grid scan over $(\alpha,\beta)$ while keeping all other hyperparameters fixed to 
baseline values. This coarse but exhaustive scan provides a comparison of the 
four nonlinearities for each Ising‑machine variant, without relying on full Bayesian 
optimization. By varying only \((\alpha,\beta)\), we largely suppress the effect of optimizer-specific hyperparameters and isolate the role of the nonlinearity itself. Under Bayesian optimization, by contrast, the internal optimizer parameters are tuned as well, so the resulting performance reflects the combined effect of optimizer and nonlinearity.

\begin{figure}[!t]
    \centering 
    \hspace*{-0.05\linewidth}%
    \includegraphics[width=0.7\linewidth]{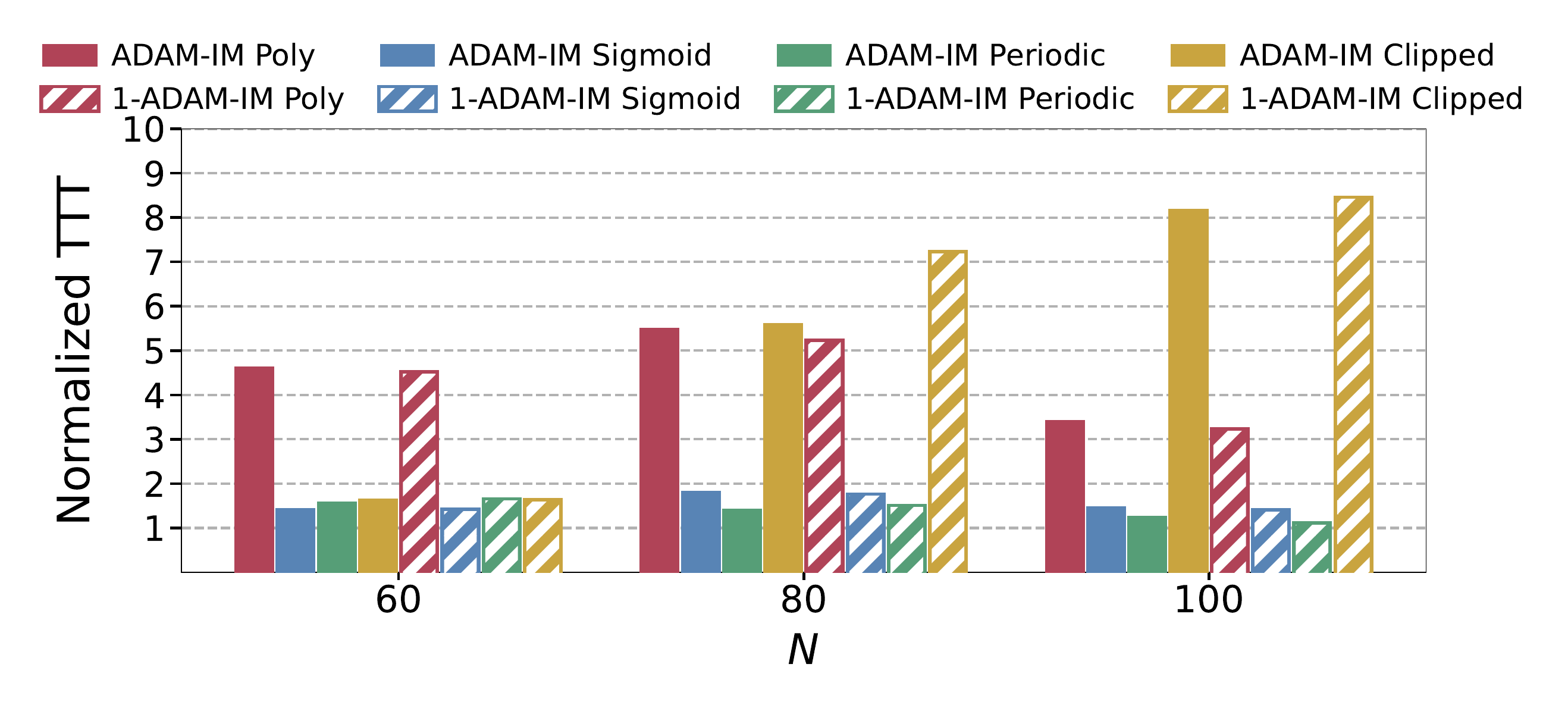}
    \caption{Grid-scan omparison of ADAM-IM and 1-ADAM-IM across the four nonlinearities on the g05 benchmark set. Bars show the normalized TTT, averaged over 10 problems. The scan is performed over $(\alpha,\beta)$ with $\gamma=0.005$, $\beta_1=\beta_2=0.99$, $\eta=1$ for ADAM-IM and $\eta=-\sqrt{-\ln(\beta_1)}/\ln(\beta_2)=9.97$ for 1-ADAM-IM.}
    \label{fig:g05_adam_gridscan_avg}
\end{figure}

\begin{figure*}[!t]
    \centering
    \begin{minipage}[t]{0.49\textwidth}
        \centering
        \includegraphics[width=\linewidth]{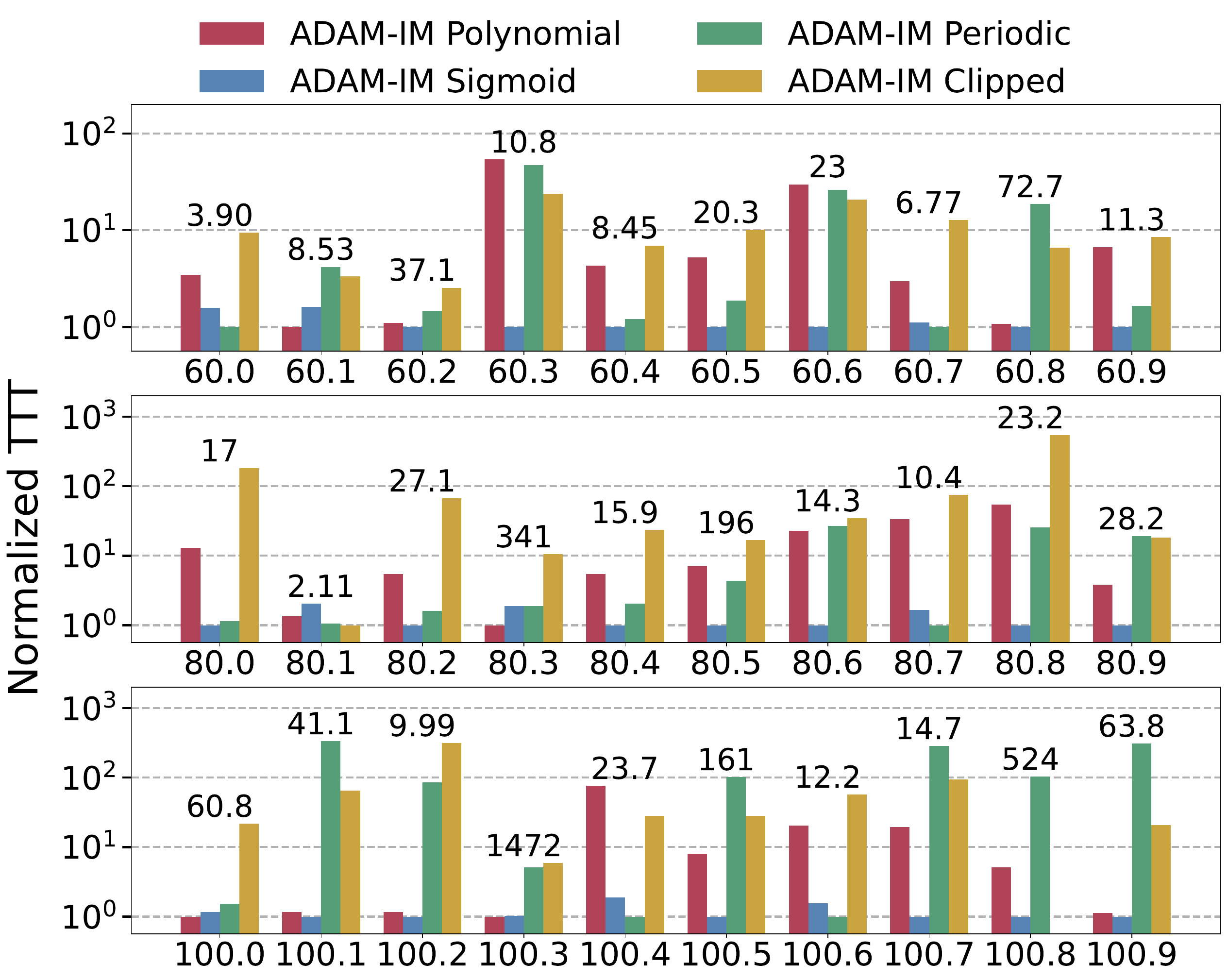}
        
        \vspace{1mm}
        \small (a) ADAM-IM
    \end{minipage}\hfill
    \begin{minipage}[t]{0.49\textwidth}
        \centering
        \includegraphics[width=\linewidth]{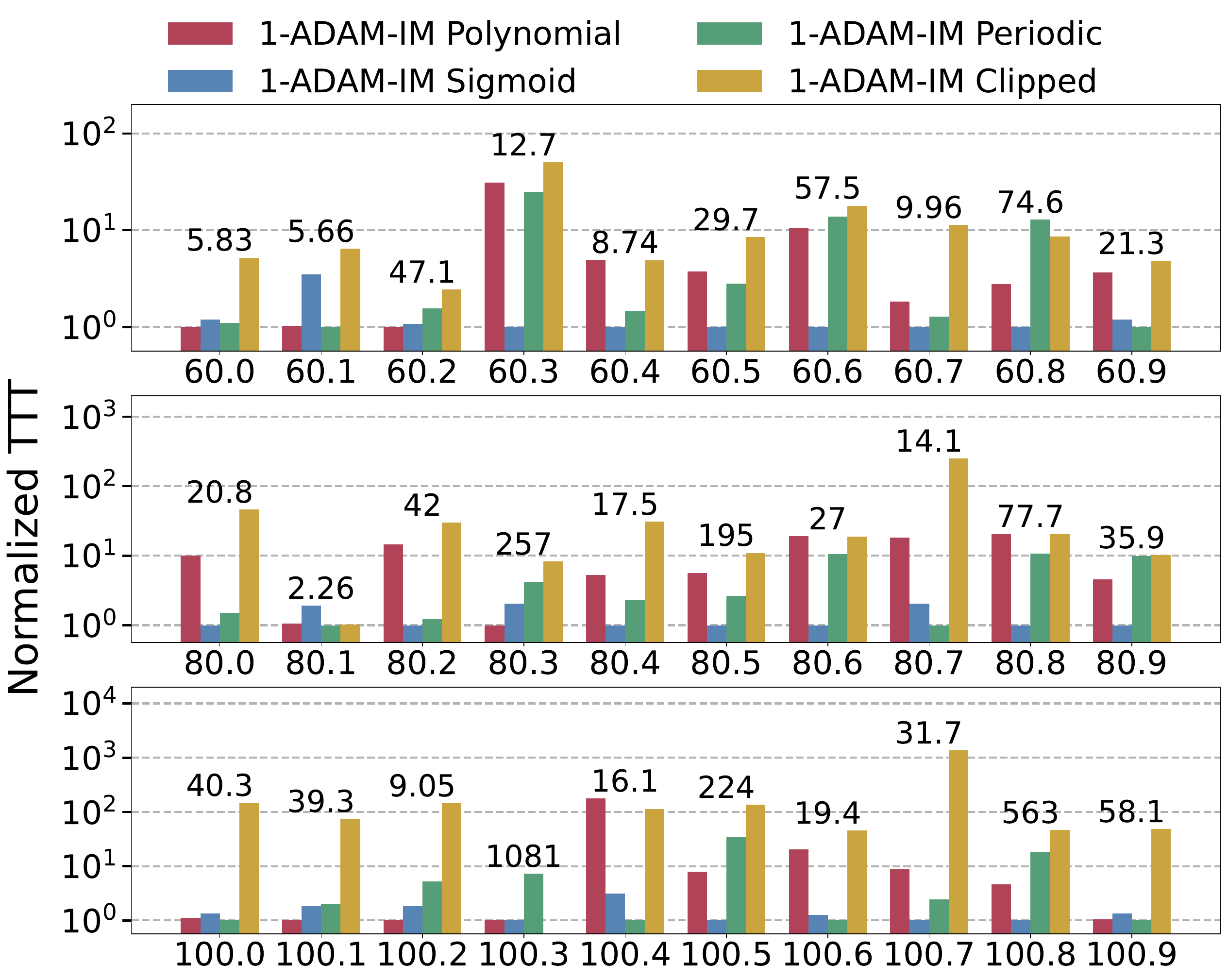}
        
        \vspace{1mm}
        \small (b) 1-ADAM-IM
    \end{minipage}
    \caption{Bayesian-optimized normalized time-to-target (TTT) on a logarithmic scale for all nonlinearities on the g05 benchmark set. Results are shown for (a) ADAM-IM and (b) 1-ADAM-IM. For each problem instance, TTT is normalized to the lowest value obtained on that problem instance, so that the fastest method is assigned the value of \(1\). The numerical labels above the bars indicate the corresponding unnormalized minimum TTT values used for this normalization.}
    \label{fig:g05_adam_nonlins_combined}
\end{figure*}

Figure~\ref{fig:g05_adam_gridscan_avg} summarizes the coarse grid-scan comparison across the four nonlinearities for ADAM-IM and 1-ADAM-IM on the g05 benchmark set. For each g05 problem instance and each optimizer--nonlinearity pair, we first determine the minimum \(\mathrm{TTT}\) obtained anywhere on the \(30\times30\) grid. This gives one best grid-scan value for every pair on every problem instance. For each problem instance, these best values are then normalized by the lowest value obtained on that same problem instance across all optimizer--nonlinearity pairs, so that the best pair has normalized \(\mathrm{TTT}=1\). We finally average the normalized values separately over the ten g05 problem instances with \(N=60\), \(N=80\), and \(N=100\). The full per-problem instance grid-scan data are provided in the Supplemental Material at [URL will be inserted by publisher].
Two points are immediately clear from this aggregated view. First, under this coarse grid scan, ADAM-IM and 1-ADAM-IM perform essentially identically: when both are evaluated at the same hyperparameter values and the 1-ADAM-IM learning rate is fixed from the first-order expansion, the simplified 1-ADAM-IM formulation reproduces the behavior of the full Adam dynamics. Second, the sigmoid and periodic nonlinearities emerge as the strongest overall choices and perform very similarly in this coarse comparison. By contrast, the polynomial and clipped nonlinearities perform weaker on average. The weaker performance of the clipped nonlinearity is notable, since clipped transfer functions have been shown to perform well in gradient-descent-based analog Ising machines by suppressing amplitude inhomogeneity~\cite{Orders}. This shows that the nonlinearity ranking depends not only on the nonlinearity itself, but also on its interaction with the optimizer dynamics.
\\
To complement the coarse grid-scan, we next perform a full Bayesian optimization over all hyperparameters, exluding the time step $\Delta t$, for each nonlinearity and compare the resulting time-to-target, \(\mathrm{TTT}\), on the g05 benchmark set. In Figure~\ref{fig:g05_adam_nonlins_combined}, \(\mathrm{TTT}\) is normalized separately for each benchmark problem instance by dividing by the lowest \(\mathrm{TTT}\) obtained on that problem instance, so that the fastest method is assigned the value \(1\). The figures therefore show relative performance on each problem instance rather than absolute \(\mathrm{TTT}\). They display the Bayesian-optimized results for (a) ADAM-IM and (b) 1-ADAM-IM, and thus reveal how the optimizer and the choice of nonlinearity jointly determine performance once all tunable parameters are refined beyond the \((\alpha,\beta)\) grid.

The Bayesian-optimized results refine the nonlinearity ranking suggested by the coarse scan. In both Adam-based optimizers, the performance of the sigmoid nonlinearity emerges as the clear overall winner: it achieves the lowest \(\mathrm{TTT}\) across the g05 problem instances and consistently outperforms the other nonlinearities. The polynomial nonlinearity, although clearly weaker in the coarse grid scan, can be tuned by Bayesian optimization to approach the sigmoid nonlinearity on some easier problem instances, but it does not match its overall robustness. By contrast, the periodic nonlinearity---which was nearly on par with the sigmoid nonlinearity in the grid scan---is now systematically outperformed once all hyperparameters are co-optimized. The clipped nonlinearity performs worst: in some cases it fails to reach the target at all, indicating limited tunability in the enlarged hyperparameter space. This shows that the nonlinearity ranking depends not only on the optimizer--nonlinearity combination, but also on how the corresponding hyperparameters are tuned.

\begin{figure}[!t]
    \centering
    \hspace*{-0.05\linewidth}%
    \includegraphics[width=0.7\linewidth]{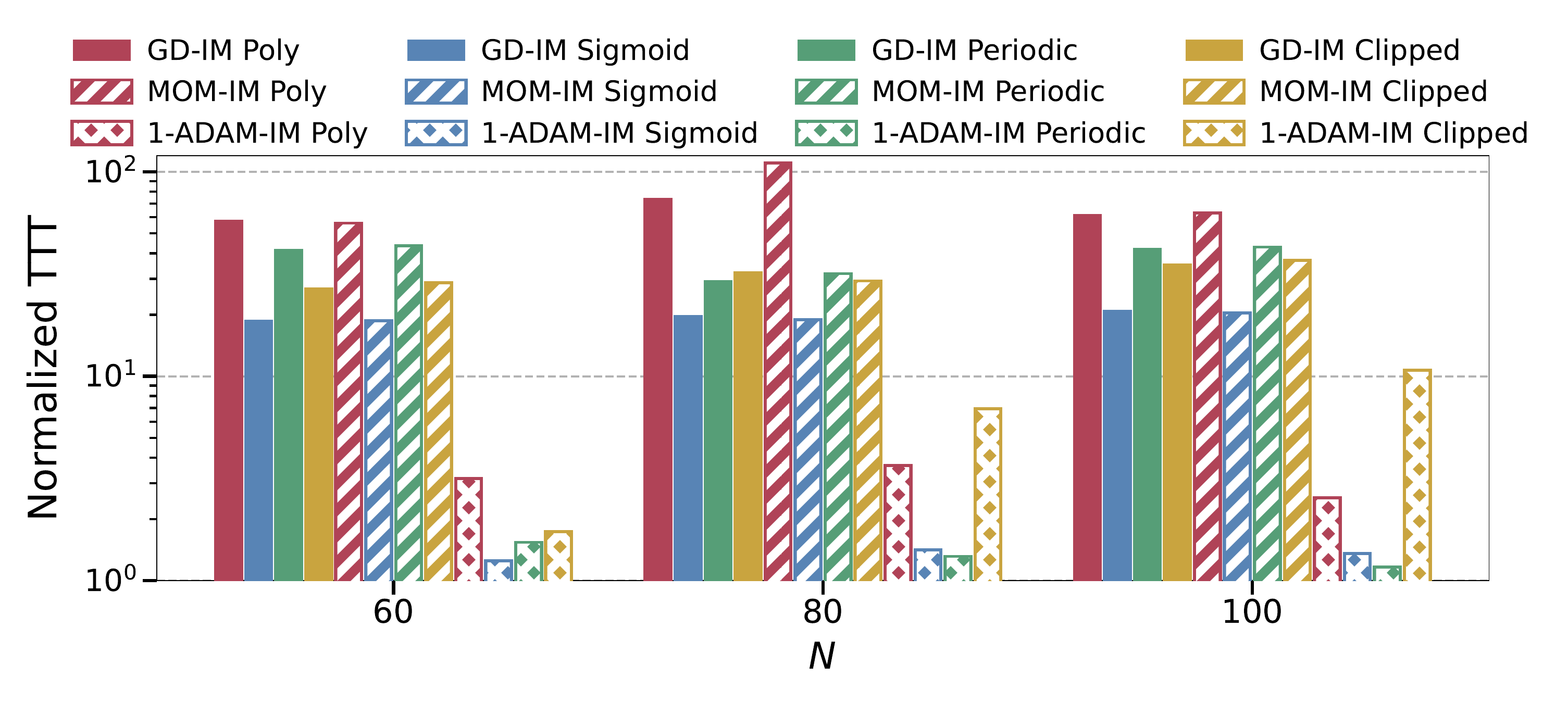}
    \caption{Grid-scan comparison of GD-IM, MOM-IM and 1-ADAM-IM across the four nonlinearities on the g05 benchmark set. Bars show the normalized TTT, averaged over 10 problems on a logarithmic scale. The scan is performed over $(\alpha,\beta)$ with $\gamma=0.005$, $\beta_1=\beta_2=0.99$, and $\eta=-\sqrt{-\ln(\beta_1)}/\ln(\beta_2)=9.97$ for 1-ADAM-IM.}
    \label{fig:g05_allopt_gridscan_avg}
\end{figure}

This naturally raises the question of whether the same nonlinearity preference also persists for the simpler optimizers GD-IM and MOM-IM. To address this, we compare all four nonlinearities across the three optimizers Gradient Descent (GD-IM), Momentum (MOM-IM), and First-Order Adam (1-ADAM-IM). As before, we perform a \(30\times30\) scan over \((\alpha,\beta)\) with all remaining parameters fixed. Figure~\ref{fig:g05_allopt_gridscan_avg} summarizes this grid scan comparison in a compact form. For each g05 problem instance and each method, we first determine the minimum time-to-target (TTT) over the grid scan. These per-problem instance minima are then normalized by the lowest TTT obtained on that same problem instance across all methods, and the resulting normalized values are finally averaged separately over the ten problem instances with \(N=60\), \(N=80\), and \(N=100\). The full per-problem instance grid-scan data are provided in the Supplemental Material at [URL will be inserted by publisher].
Two conclusions stand out clearly from this averaged normalized comparison. First, 1-ADAM-IM is by far the fastest optimizer throughout the coarse grid scan, with a speed advantage of roughly one to two orders of magnitude over both GD-IM and MOM-IM across all three problem sizes. Second, the nonlinearity ranking depends on the optimizer, but the sigmoid nonlinearity emerges as the clearest overall winner in this broader comparison. For GD-IM and MOM-IM, sigmoid is unambiguously the best-performing choice. Under 1-ADAM-IM, the periodic nonlinearity remains the closest competitor and is sometimes nearly as fast, but the sigmoid nonlinearity still yields the lowest TTT overall. Thus, while the periodic nonlinearity was also competitive in the Adam-based coarse scan, the present three-optimizer comparison more clearly singles out sigmoid as the most robust nonlinearity at coarse parameter grid-scan resolution. 

To test whether the same picture persists beyond the coarse grid scan, we also carried out the full Bayesian-optimization analysis for GD-IM and MOM-IM on the g05 benchmark set, optimizing all free hyperparameters jointly for each of the four nonlinearities. The detailed results are reported in the Supplemental Material at [URL will be inserted by publisher]. Qualitatively, however, they lead to the same conclusion as for the Adam-based variants: the sigmoid nonlinearity remains the strongest overall choice after full tuning. The polynomial nonlinearity can partly catch up on some easier problem instances, but is less consistently effective across the full problem set, while the periodic and clipped nonlinearities fall behind once all hyperparameters are co-optimized.

This outcome is consistent with earlier work by Böhm \emph{et\,al.}~\cite{Orders}, 
who compared the same four nonlinear transfer functions for GD-IM and argued in favor of 
the sigmoid nonlinearity.  
Their reasoning is that sigmoid, being a smooth saturating function, effectively suppresses 
amplitude inhomogeneity—a dominant source of performance degradation in analog Ising 
machines—and does so without introducing the attractor distortions or multistability 
issues sometimes seen in clipped and periodic systems.
\subsection{Optimizer comparison with the sigmoid nonlinearity}
Having established that the sigmoid nonlinearity is the most robust 
and consistently best-performing choice across all optimizers, we now restrict our attention 
to this nonlinearity and compare the optimizers themselves.  
The goal of this step is to assess how Gradient Descent, Momentum, Adam, and First-Order Adam 
behave when all their hyperparameters are fully tuned through Bayesian optimization, and 
when tested on benchmark sets that are more challenging.
To obtain a clearer picture, we therefore also consider the \textsc{Gset} benchmarks, which are 
significantly harder and come in two variants: unweighted and weighted MaxCut problem instances.  
These graphs are well known to produce more rugged energy landscapes and larger performance 
spreads, making them ideal for highlighting optimizer-induced differences.  

\begin{figure*}[!t]
    \centering
    
    \begin{minipage}[t]{0.49\textwidth}
        \centering
        \includegraphics[width=\linewidth]{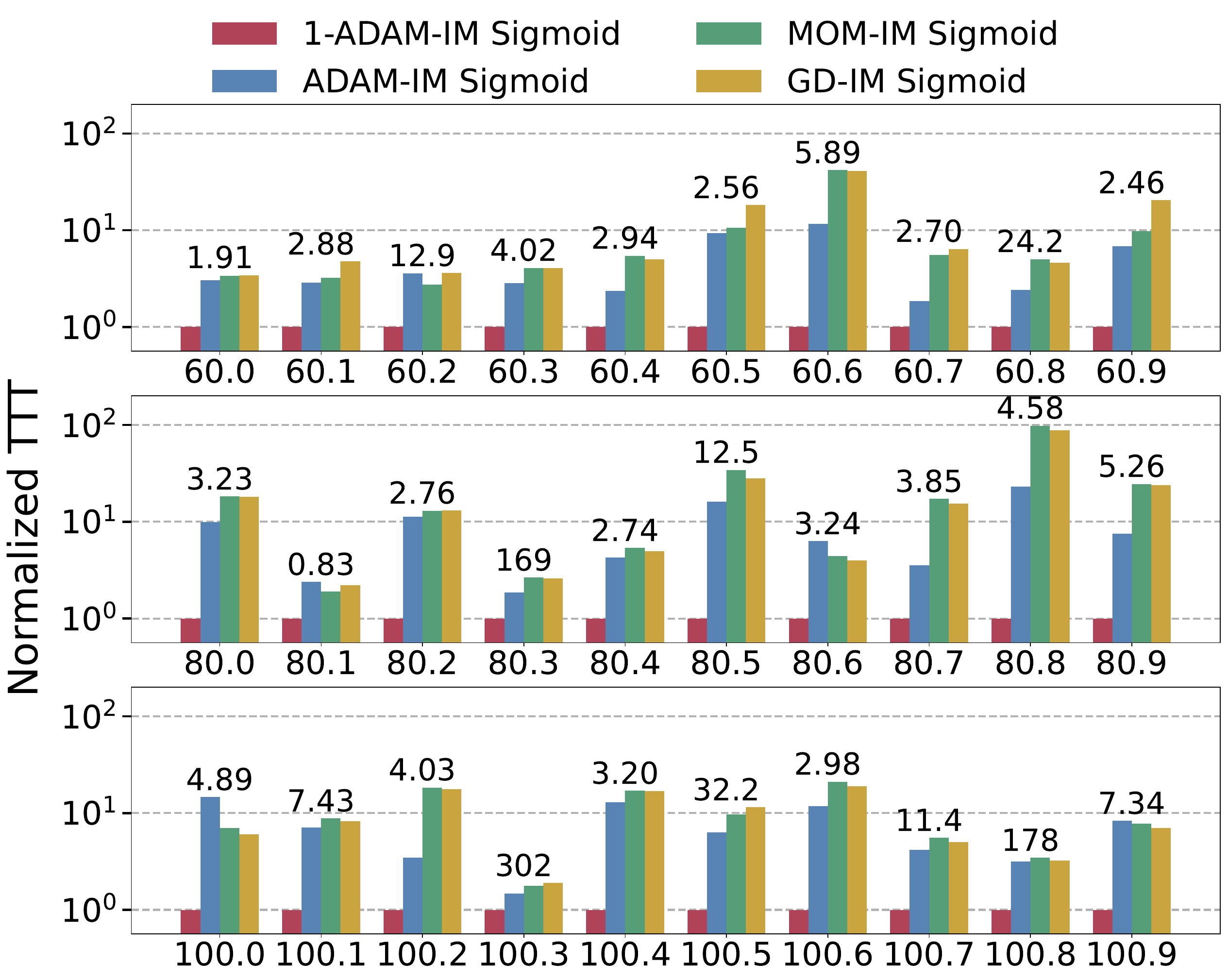}
        
        \vspace{1mm}
        \small (a) g05
    \end{minipage}\hfill
    \begin{minipage}[t]{0.49\textwidth}
        \centering
        \includegraphics[width=\linewidth]{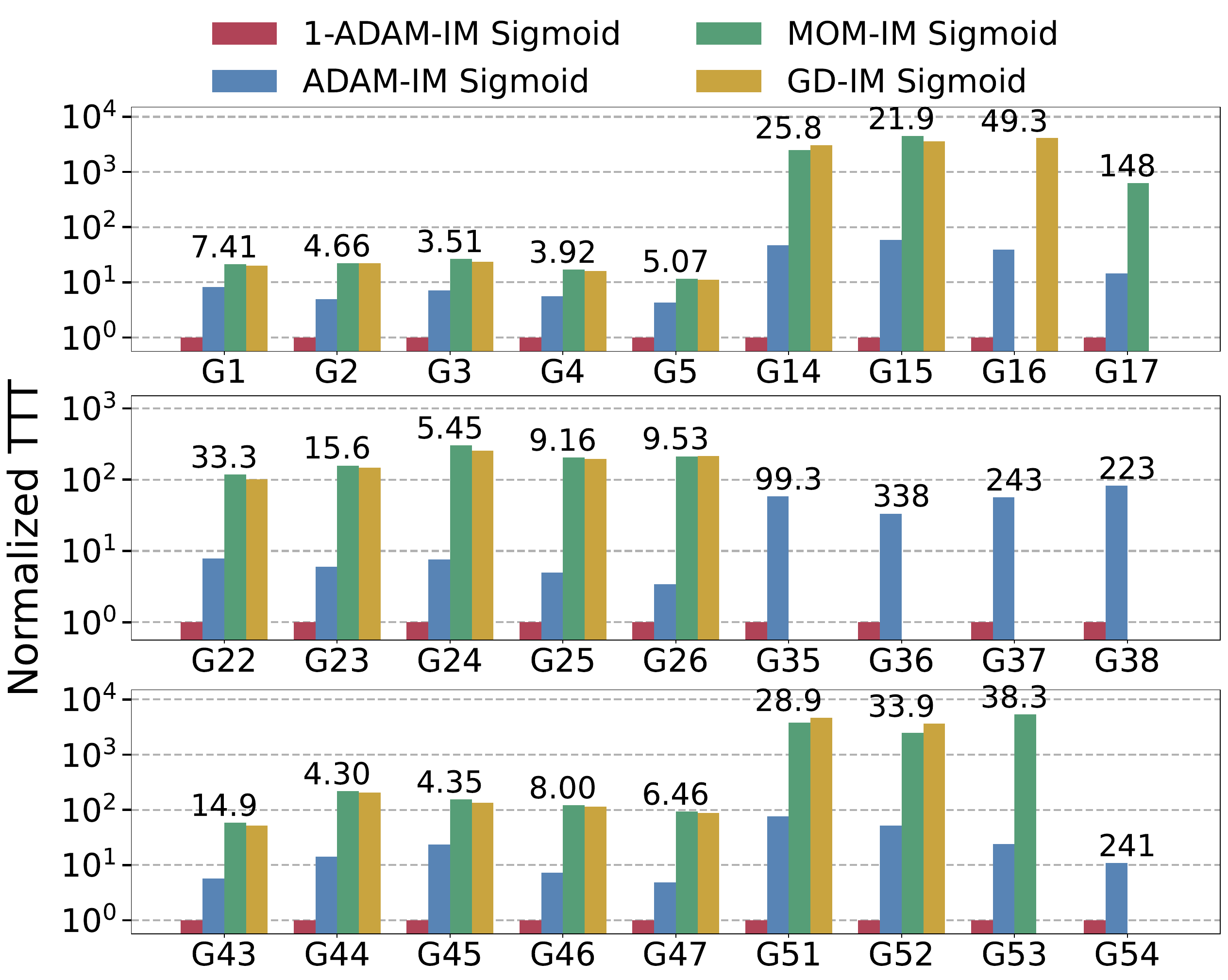}
        
        \vspace{1mm}
        \small (b) Unweighted \textsc{Gset}
    \end{minipage}
    
    \vspace{2mm}
    
    \begin{minipage}[t]{0.49\textwidth}
        \centering
        \includegraphics[width=\linewidth]{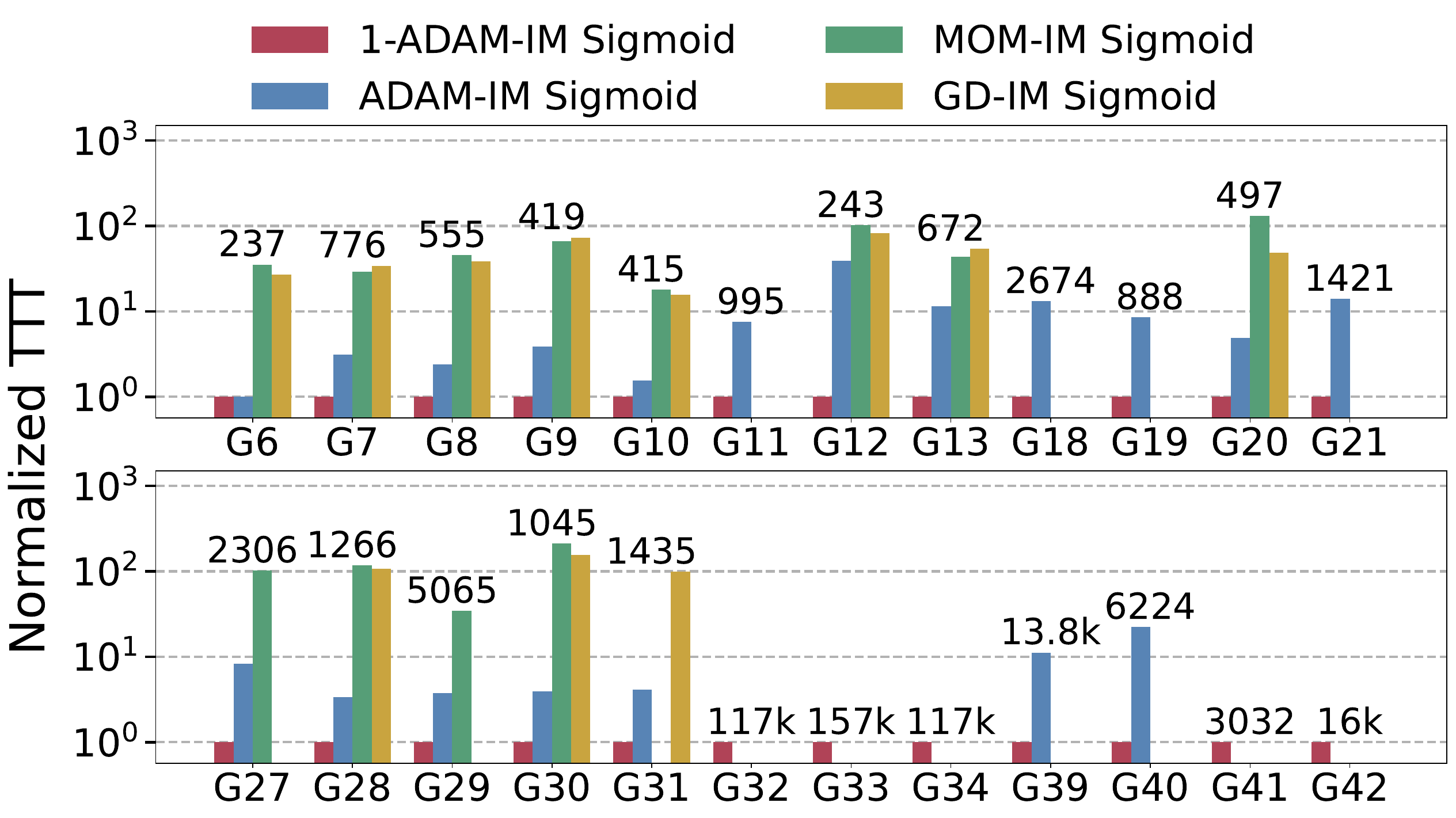}
        
        \vspace{1mm}
        \small (c) Weighted \textsc{Gset}
    \end{minipage}
    
\caption{Bayesian-optimized normalized \(\mathrm{TTT}\) for the continuous-time implementations with the sigmoid nonlinearity on (a) g05, (b) unweighted \textsc{Gset}, and (c) weighted \textsc{Gset}. Values are normalized per problem instance to the fastest method. In panel (a), all optimizers reach the ground-state target; in panels (b) and (c), bars are omitted when the optimizer does not reach \(99.5\%\) of the best-known cut value. Numerical labels indicate the unnormalized minimum \(\mathrm{TTT}\) used for normalization.}
    \label{fig:tts_combined_sigmoid}
\end{figure*}

Figure~\ref{fig:tts_combined_sigmoid} show the 
Bayesian-optimized time-to-target for GD-IM, MOM-IM, ADAM-IM, and 1-ADAM-IM on 
(a) g05, (b) unweighted and (c) weighted \textsc{Gset} problems. For every optimizer and every problem instance, the reported value corresponds to the best result
obtained across seven independent Bayesian optimizations performed over seven distinct 
hyperparameter ranges. 
A striking and universal observation across all problem instances is that 
1-ADAM-IM is the fastest optimizer for every single benchmark problem.  
The magnitude of this advantage depends strongly on problem difficulty, becoming more 
pronounced as we move from g05 to unweighted and finally weighted \textsc{Gset} graphs.

For the easier subset (G1–G5), 1-ADAM-IM is roughly one 
order of magnitude faster than the other methods, which again perform comparably among 
themselves.  
For medium-difficulty problem instances (G22–G26 and G43–G47), a stronger performance difference emerges.  
Here, 1-ADAM-IM is typically two orders of magnitude faster than GD-IM and MOM-IM, and 
about one order of magnitude faster than ADAM-IM; ADAM-IM itself tends to be roughly 
one order of magnitude faster than GD-IM and MOM-IM.  
On the hardest unweighted problems (G14–G17, G35–G38, G51–G54), the contrast becomes 
most pronounced: GD-IM and MOM-IM frequently fail to reach 99.5\% of the best-known 
cut value (in which case no bar is shown), and even when they do succeed, their
normalized times-to-target are three to four orders of magnitude larger than those of 
1-ADAM-IM.  
Both ADAM-IM and 1-ADAM-IM reliably reach the 99.5\% threshold on all these problem instances, 
with 1-ADAM-IM maintaining a consistent one- to two-order-of-magnitude advantage.

The same trend is amplified in the weighted \textsc{Gset} problems 
(Figure~\ref{fig:tts_combined_sigmoid} (c)), which are among the most challenging MaxCut 
benchmarks in our tests.  
For medium-hard problem instances (G6–G13, G18–G21, G27–G31), GD-IM and MOM-IM occasionally 
fail to meet the 99.5\% success criterion. When they do succeed, 1-ADAM-IM is again 
around two orders of magnitude faster, and typically maintains about a one-order of 
magnitude lead over ADAM-IM.  
For the hardest weighted problem instances (G32–G42), the gap becomes decisive: 
1-ADAM-IM is the only optimizer that consistently reaches 99.5\% of the best-known 
solution.  
ADAM-IM succeeds only sporadically, and GD-IM and MOM-IM almost never reach the target 
accuracy.  
This indicates that 1-ADAM-IM not only converges faster but also attains 
higher-quality solutions on the most difficult graphs. This motivates a more detailed investigation of how these optimizers differ in success 
rates, convergence profiles, and solution quality in the next section.

\begin{figure*}[!t]
    \centering
    \begin{minipage}[t]{0.49\textwidth}
        \centering
        \includegraphics[width=\linewidth]{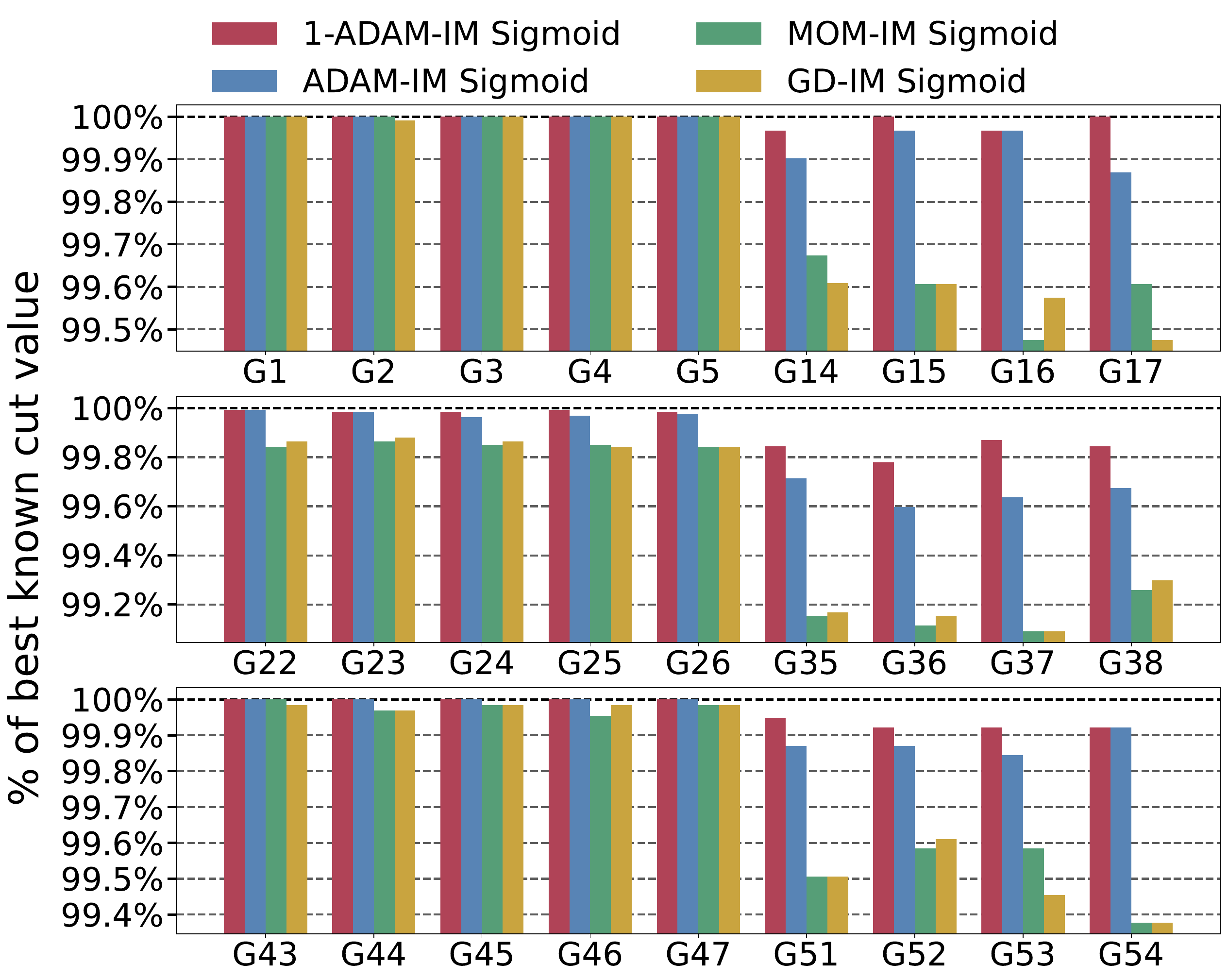}
        
        \vspace{1mm}
        \small (a) Unweighted \textsc{Gset}
    \end{minipage}\hfill
    \begin{minipage}[t]{0.49\textwidth}
        \centering
        \includegraphics[width=\linewidth]{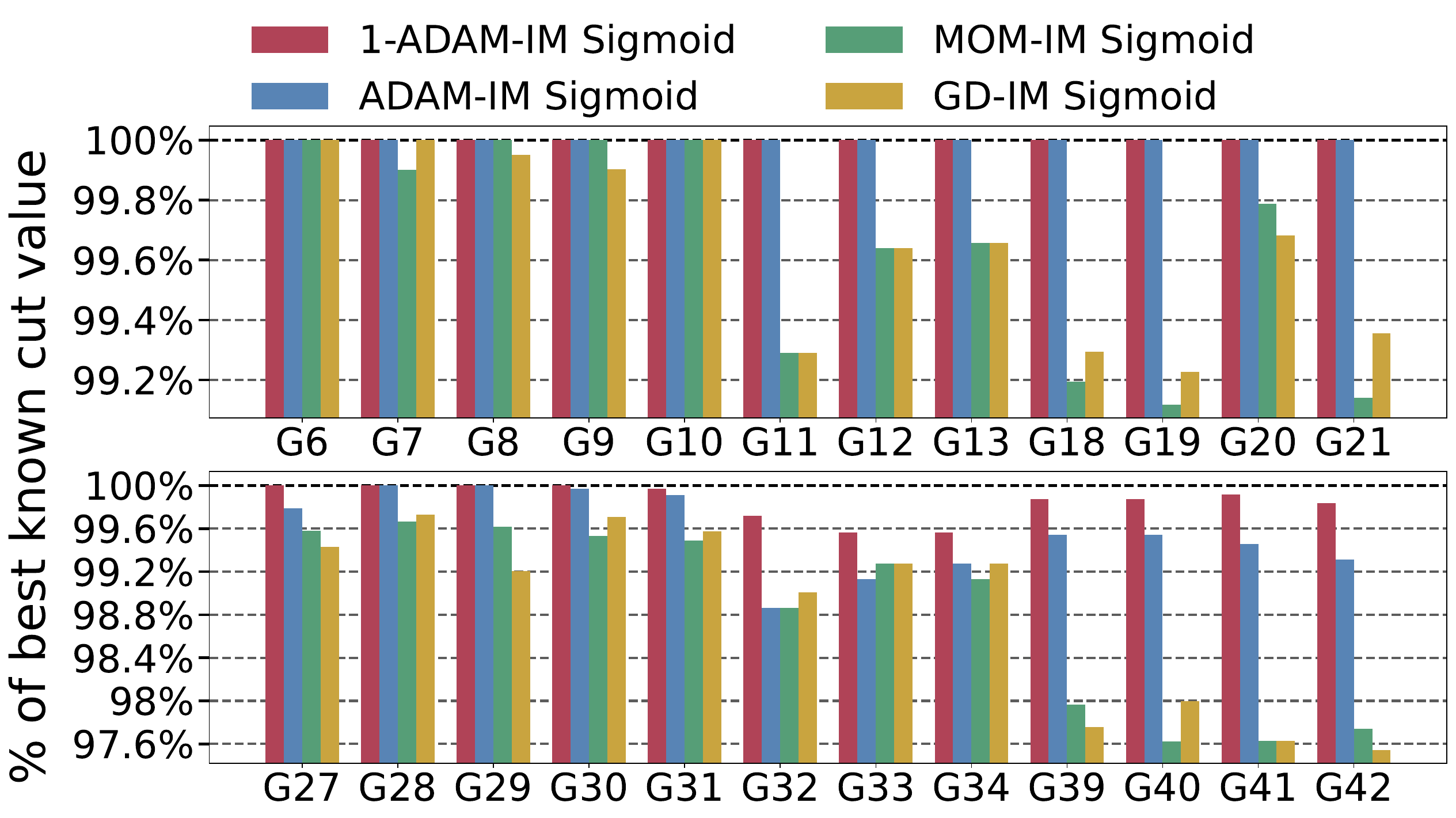}
        
        \vspace{1mm}
        \small (b) Weighted \textsc{Gset}
    \end{minipage}
\caption{Best solution quality obtained with the sigmoid nonlinearity on (a) unweighted and (b) weighted \textsc{Gset}. Values show the percentage of the best-known MaxCut value reached by each optimizer, using the best result over seven Bayesian-optimization runs.}
    \label{fig:Emin_Gset_combined}
\end{figure*}

To complement the time-to-target results, we now compare the achieved solution quality
for each optimizer on the unweighted and weighted \textsc{Gset} MaxCut benchmarks.  
Figure~\ref{fig:Emin_Gset_combined} displays, for each
problem instance, the percentage of the best-known cut value reached by GD-IM, MOM-IM, ADAM-IM,
and 1-ADAM-IM for the (a) unweighted and (b) weighted \textsc{Gset} problems. To ensure a fair comparison, each optimizer is given the same computational budget. 
For every optimizer and every problem instance, the reported value corresponds to the best result
obtained across seven independent Bayesian optimizations performed over seven distinct 
hyperparameter ranges, identical to those used in the previous TTT analysis.  
The corresponding absolute cut values, allowing direct comparison with the best-known values from the literature, are reported in the Supplemental Material at [URL will be inserted by publisher].
The solution-quality results in Figure~\ref{fig:Emin_Gset_combined} show that the speed advantage of 1-ADAM-IM is not obtained at the expense of final solution quality. On the easier \textsc{Gset} problem instances, all four optimizers often reach or nearly reach the best-known cuts, and the differences between methods are therefore small. On the harder problem instances, however, 1-ADAM-IM clearly emerges as the strongest method. ADAM-IM remains the closest competitor, but its \(\mathrm{TTT}\) is consistently higher than that of 1-ADAM-IM, while MOM-IM and GD-IM show substantially weaker performance. This hierarchy is already apparent on the harder unweighted graphs and becomes most striking on the weighted \textsc{Gset} benchmarks, where 1-ADAM-IM most often reaches the best-known cuts and separates itself decisively from the other optimizers. Overall, these results show that 1-ADAM-IM is not only the fastest optimizer considered here, but also gives the strongest final solution quality, especially on the most difficult benchmark problem instances.

To complement the time-to-target and solution-quality results, we also examine the
transient success rates of all four optimizers under the \textit{sigmoid} nonlinearity.
Throughout, we reuse exactly the same simulation data as in the previous Bayesian-optimization
study. We use $99.5\%$ of the best-known cut value for
\textsc{Gset} (unweighted and weighted) as target.

\begin{figure*}[!t]
    \centering
    
    \begin{minipage}[t]{0.49\textwidth}
        \centering
        \includegraphics[width=\linewidth]{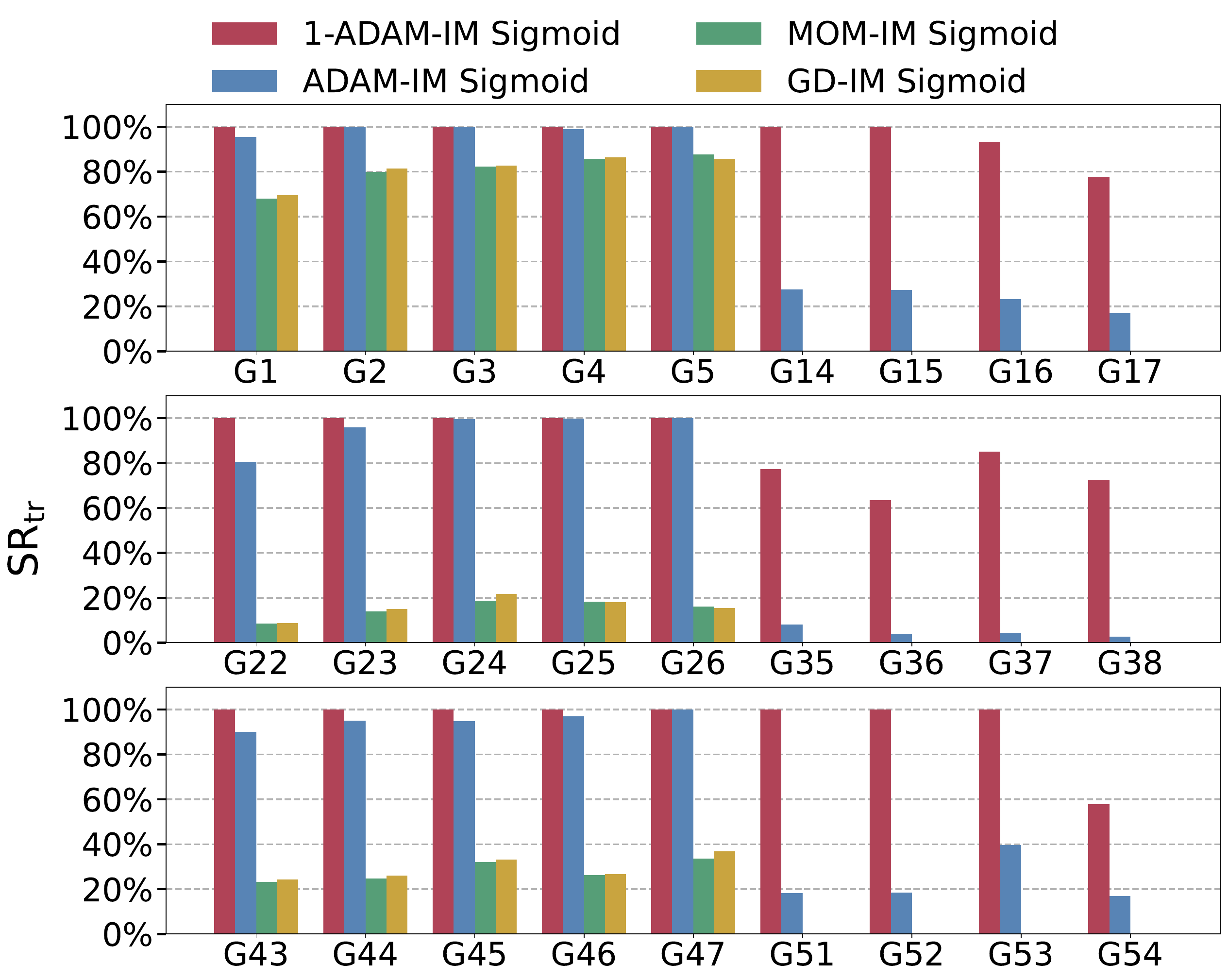}
        
        \vspace{1mm}
        \small (a) Unweighted \textsc{Gset}
    \end{minipage}\hfill
    \begin{minipage}[t]{0.49\textwidth}
        \centering
        \includegraphics[width=\linewidth]{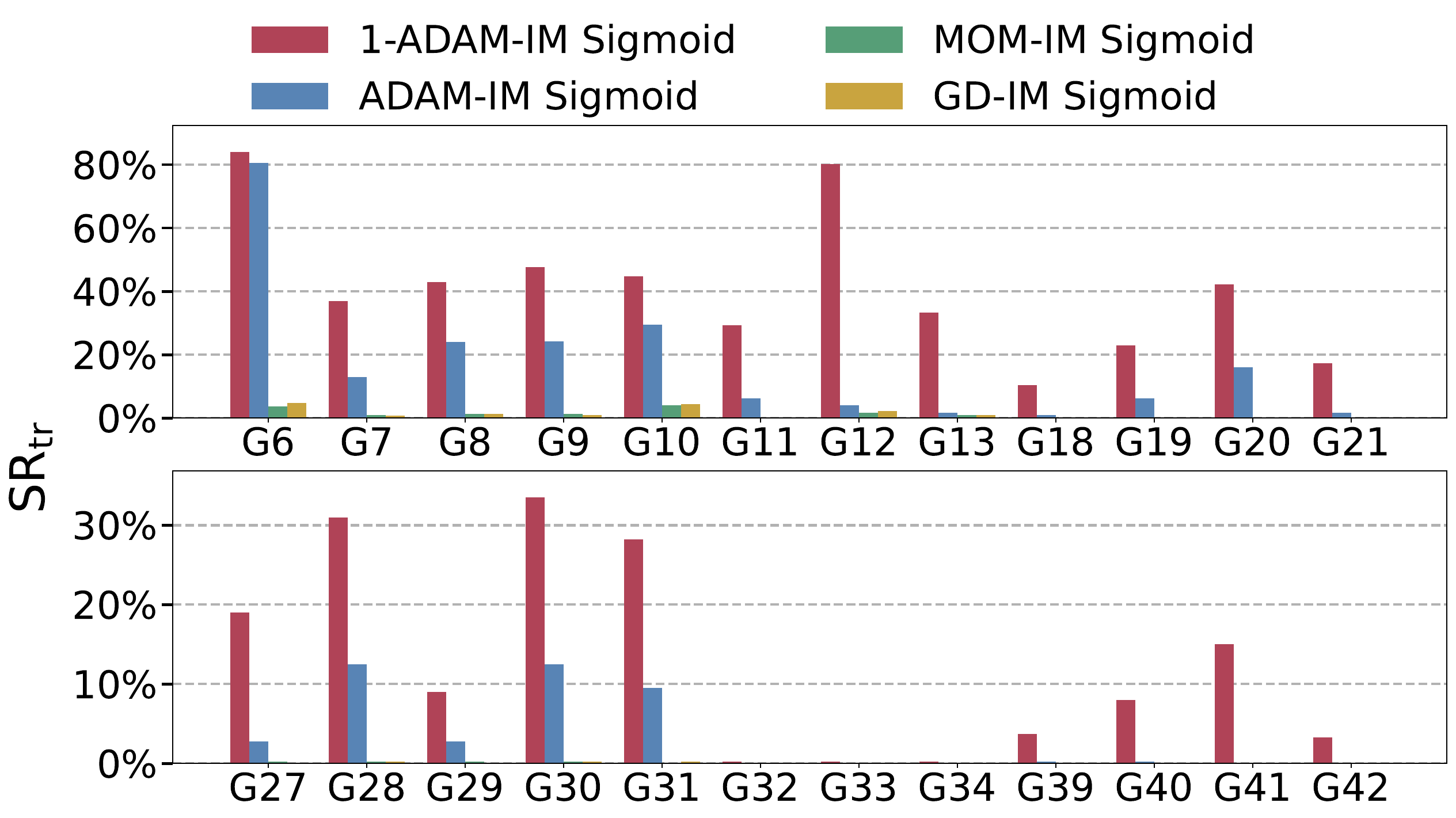}
        
        \vspace{1mm}
        \small (b) Weighted \textsc{Gset}
    \end{minipage}
        
\caption{Transient success rate with the sigmoid nonlinearity on (a) unweighted \textsc{Gset} and (b) weighted \textsc{Gset}. For each optimizer and problem instance, values show the best result over seven Bayesian-optimization runs.}
    \label{fig:srtr_combined}
\end{figure*}

Figure~\ref{fig:srtr_combined} presents the transient success rates
for (a) unweighted \textsc{Gset} and (b) weighted \textsc{Gset}. 
As seen in Figure~\ref{fig:srtr_combined}, 1-ADAM-IM achieves the highest transient success rate throughout all benchmark problem instances. On the unweighted \textsc{Gset} problem instances the difference becomes striking: 1-ADAM-IM remains at or near $100\%$ transient success on all problem instances, while GD-IM, MOM-IM, and ADAM-IM drop substantially. In particular, GD-IM and MOM-IM fall to $0\%$ on roughly half of these problem instances, whereas 1-ADAM-IM remains at $100\%$ or at least very high on those same problems. On the weighted \textsc{Gset} problem instances, which are the most challenging, the contrast becomes even stronger. Although 1-ADAM-IM no longer stays near $100\%$, it still maintains clearly elevated transient success rates, whereas the transient success rate for GD-IM and MOM-IM often fall to only a few percent or $0\%$. Note that some problem instances reached the target of $99.5\%$ of the best-known solution, as shown in Figure~\ref{fig:algorithmic_E_min}, yet appear to have a transient success rate close to zero in Figure~\ref{fig:algorithmic_srtr}. In fact, these values are nonzero but barely visible, since they are equal to $1/400$. For example on the harder G32--G34 problems, the transient success rate of 1-ADAM is nonzero but very small, making it barely visible in Fig.~\ref{fig:srtr_combined}(c). Overall, this consistently superior transient success rate is a major reason why 1-ADAM-IM so clearly outperforms the other optimizers in \(\mathrm{TTT}\). Since transient success is not itself a time-based measure, it also provides a timescale-independent confirmation that 1-ADAM-IM is the strongest method.

These findings reinforce the hierarchy already apparent in the time-to-target and solution quality results and show that 1-ADAM-IM combines both superior speed, superior solution quality and higher success rates, making it the best-performing optimizer across all benchmarks considered in this work.
\subsection{Discrete-time algorithmic implementations}
\label{subsec:algorithmic_discrete}
In the previous sections, we investigated the use of ADAM optimizer to improve analog IMs, that evolve toward low energy states in a continuous manner. However, many IMs evolve in discrete-time, in an iterative way, e.g. the measurement-feedback coherent Ising machines~\cite{faster3}. A natural extension is therefore to ask whether the same optimizer ideas are also useful in such discrete-time, iterative settings.

In the continuous-time simulations discussed above, the Euler--Maruyama timestep \(\Delta t\) is only a numerical discretization parameter. It must therefore remain sufficiently small: if \(\Delta t\) is too large, the simulated trajectory no longer approximates the underlying continuous-time analog dynamics. In a purely algorithmic implementation, this restriction disappears. The discrete update rule itself defines the algorithm, so \(\Delta t\) can be treated as an additional tunable parameter rather than as a resolution parameter. Since the resulting timestep no longer corresponds to a time increment, we compare algorithmic performance using wall-clock time-to-target, denoted by \(\mathrm{TTT}_{\mathrm{CPU}}\).

There are two possible ways to construct the discrete-time algorithms used in this comparison. One option is to start from the continuous-time equations derived above and discretize them with Euler--Maruyama, now allowing the timestep to be large and fully tuned. For GD-IM, we also test the corresponding fixed-timestep version with \(\Delta t=1\), so that only the learning-rate scale is tuned. The other option, available for MOM-IM, ADAM-IM and 1-ADAM-IM, is to use the standard discrete optimizer update rules from the literature \cite{MomentumOriginal,AdamOriginal}, given in Eqs.~(\ref{mom1}) and~(\ref{adam_discrete}). There are two possible ways to construct the discrete-time algorithms used in this comparison. One option is to start from the continuous-time equations derived above and discretize them with Euler--Maruyama, now allowing the timestep to be large and fully tuned. For GD-IM, we also test the corresponding fixed-timestep version with \(\Delta t=1\), so that only the learning-rate scale is tuned. The other option, available for MOM-IM, ADAM-IM, and 1-ADAM-IM, is to use the standard discrete optimizer update rules from the literature~\cite{MomentumOriginal,AdamOriginal}, given in Eqs.~(\ref{mom1}) and~(\ref{adam_discrete}). For each optimizer, we tested both available discrete-time variants and report only the better-performing one in the algorithmic results below.

This comparison leads to different choices for different optimizers. For GD-IM, MOM-IM, and 1-ADAM-IM, the Euler--Maruyama versions of Eqs.~(\ref{gradient_descent}), (\ref{momentum_cont}), and~(\ref{one_adam_cont}) give the best results and are therefore used below. In these cases, Bayesian optimization searches over \(\log_{10}\Delta t\), rather than over \(\Delta t\) directly, because useful timestep values can span several orders of magnitude. By contrast, \(\eta\) is optimized directly on a linear scale over the chosen learning-rate interval. For ADAM-IM, the standard discrete Adam update rule in Eq.~(\ref{adam_discrete}) performs best and is therefore used below. The full comparison between Euler--Maruyama and standard discrete implementations is given in the Supplemental Material at [URL will be inserted by publisher].

\begin{figure*}[!t]
    \centering
    \begin{minipage}[t]{0.49\textwidth}
        \centering
        \includegraphics[width=\linewidth]{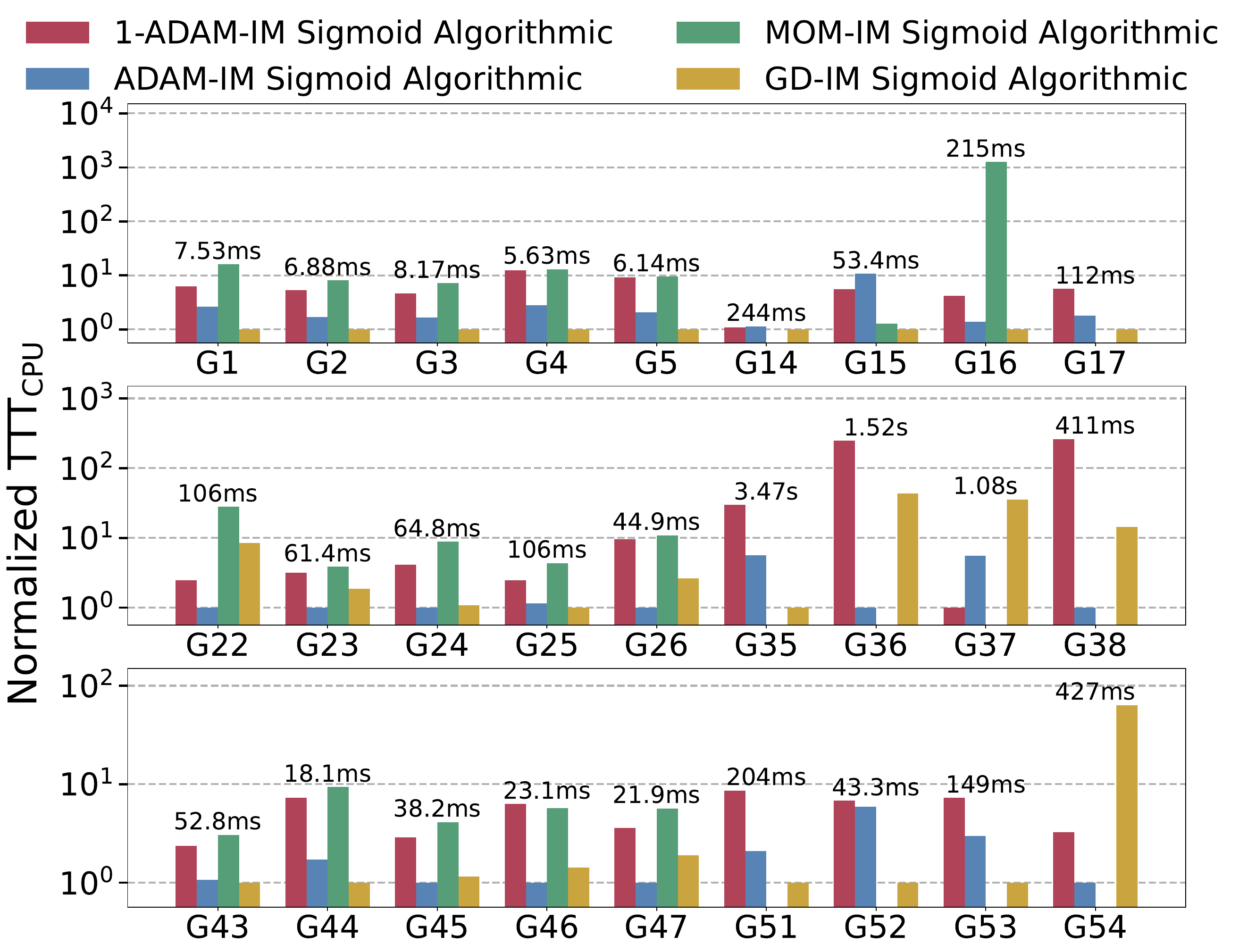}
        
        \vspace{1mm}
        \small (a) Unweighted \textsc{Gset}
    \end{minipage}\hfill
    \begin{minipage}[t]{0.49\textwidth}
        \centering
        \includegraphics[width=\linewidth]{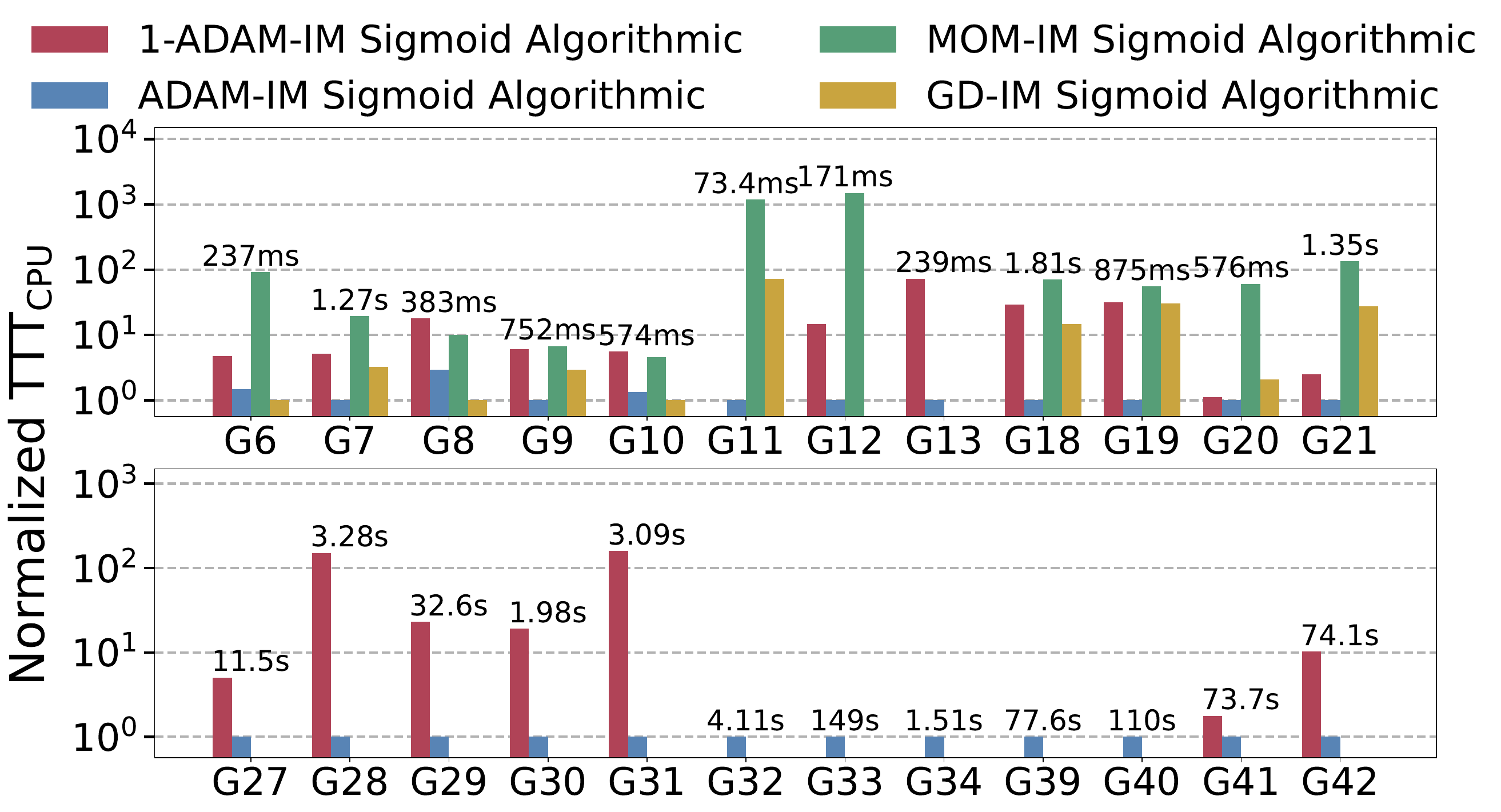}
        
        \vspace{1mm}
        \small (b) Weighted \textsc{Gset}
    \end{minipage}
\caption{Algorithmic discrete-time normalized \(\mathrm{TTT}_{\mathrm{CPU}}\) comparison after Bayesian optimization with the sigmoid nonlinearity on (a) unweighted \textsc{Gset} and (b) weighted \textsc{Gset}. Values are normalized per problem instance to the fastest method. Bars are omitted when the optimizer does not reach \(99.5\%\) of the best-known cut value. Numerical labels indicate the unnormalized minimum \(\mathrm{TTT}_{\mathrm{CPU}}\) used for normalization.}
\label{fig:algorithmic_TTS}
\end{figure*}

As in the continuous-time comparison, we restrict the algorithmic analysis to the sigmoid nonlinearity and optimize all free hyperparameters by Bayesian optimization. The detailed Bayesian-optimization settings are given in the Supplemental Material at [URL will be inserted by publisher]. Figure~\ref{fig:algorithmic_TTS} shows the resulting Bayesian-optimized normalized \(\mathrm{TTT}_{\mathrm{CPU}}\) for GD-IM, MOM-IM, ADAM-IM, and 1-ADAM-IM on the (a) unweighted \textsc{Gset} and (b) weighted \textsc{Gset} benchmarks.
The algorithmic \(\mathrm{TTT}_{\mathrm{CPU}}\) results show substantially smaller differences between optimizers than in the continuous-time setting. On the unweighted \textsc{Gset} problem instances, the performance differences generally remain within one order of magnitude, and all four optimizers perform comparably. On the weighted \textsc{Gset} problem instances, however, a clearer hierarchy emerges. ADAM-IM gives the lowest \(\mathrm{TTT}_{\mathrm{CPU}}\) overall, followed by 1-ADAM-IM, while GD-IM and especially MOM-IM are less competitive. This difference is most pronounced on the hardest weighted problem instances, shown in the second row of Fig.~\ref{fig:algorithmic_TTS}, where only ADAM-IM and 1-ADAM-IM consistently reach \(99.5\%\) of the best-known cut value. Among these two, ADAM-IM is the faster algorithmic method.

\begin{figure*}[!t]
    \centering
    \begin{minipage}[t]{0.49\textwidth}
        \centering
        \includegraphics[width=\linewidth]{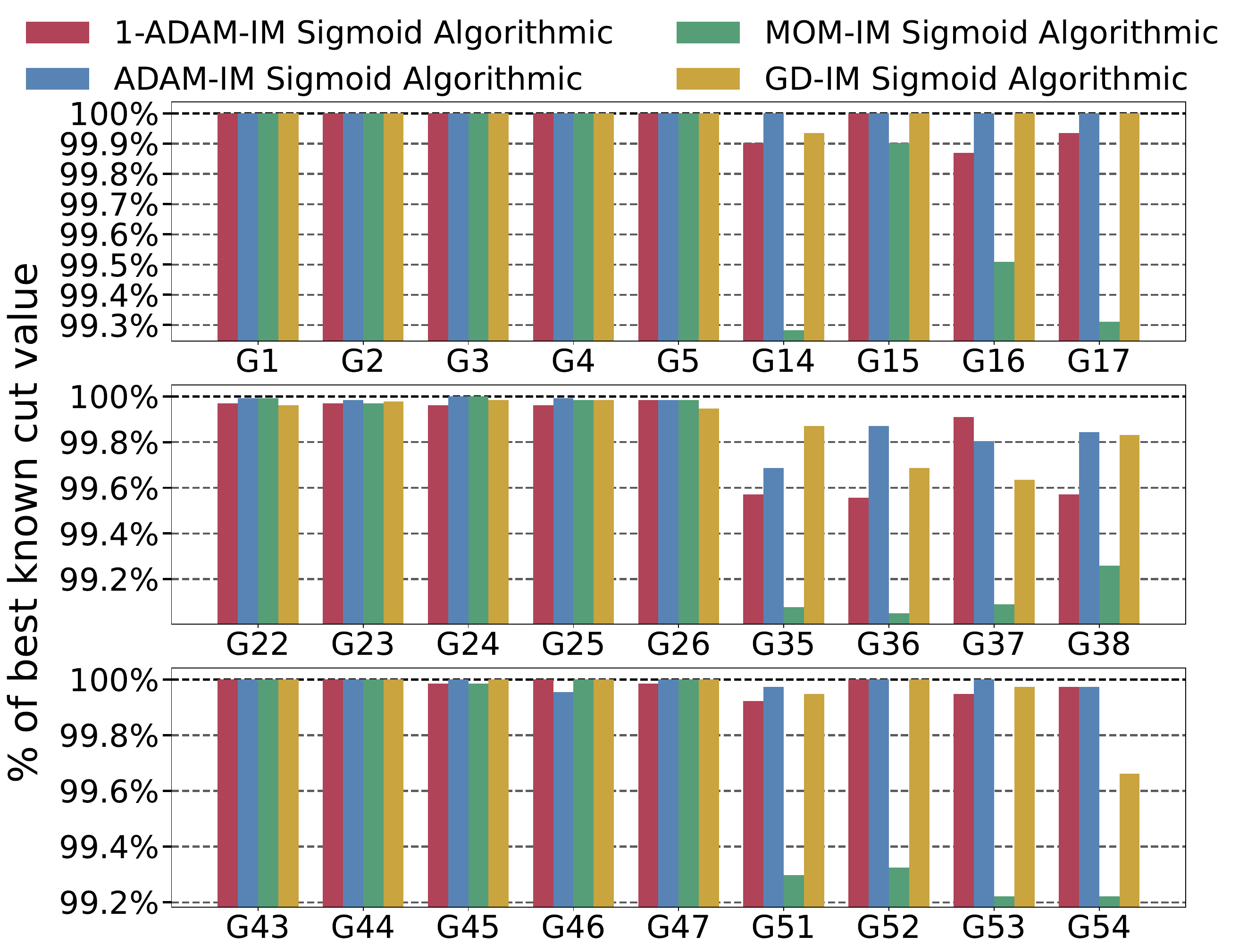}
        
        \vspace{1mm}
        \small (a) Unweighted \textsc{Gset}
    \end{minipage}\hfill
    \begin{minipage}[t]{0.49\textwidth}
        \centering
        \includegraphics[width=\linewidth]{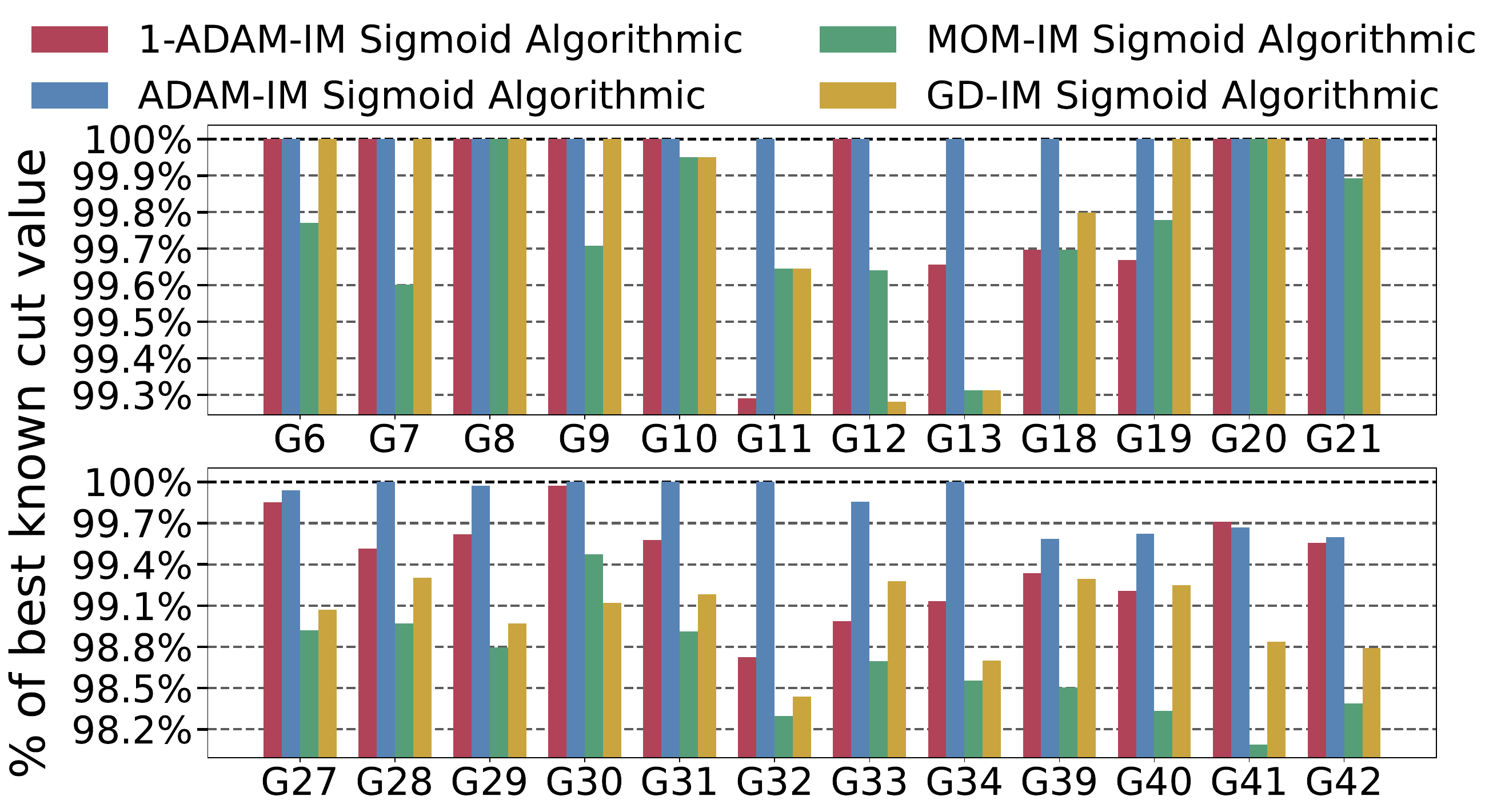}
        
        \vspace{1mm}
        \small (b) Weighted \textsc{Gset}
    \end{minipage}
\caption{Algorithmic discrete-time solution quality after Bayesian optimization obtained with the sigmoid nonlinearity on (a) unweighted and (b) weighted \textsc{Gset}. Values show the percentage of the best-known MaxCut value reached by each optimizer, using the best result over seven Bayesian-optimization runs.}
    \label{fig:algorithmic_E_min}
\end{figure*}

To determine whether this speed advantage is accompanied by better final solutions, we next compare the achieved solution quality. Figure~\ref{fig:algorithmic_E_min} shows the percentage of the best-known MaxCut value reached by each optimizer on the unweighted and weighted \textsc{Gset} problem instances.
The solution-quality results follow the same overall trend as those observed for the TTT. On the unweighted \textsc{Gset} problem instances, the optimizers perform similarly, although MOM-IM is somewhat weaker than the other methods. On the weighted problem instances, the differences become more pronounced: ADAM-IM reaches the best solutions most consistently, followed by 1-ADAM-IM, while GD-IM and MOM-IM obtain lower-quality solutions on the harder problems. Thus, the performance gap grows with problem difficulty.

\begin{figure*}[!t]
    \centering
    \begin{minipage}[t]{0.49\textwidth}
        \centering
        \includegraphics[width=\linewidth]{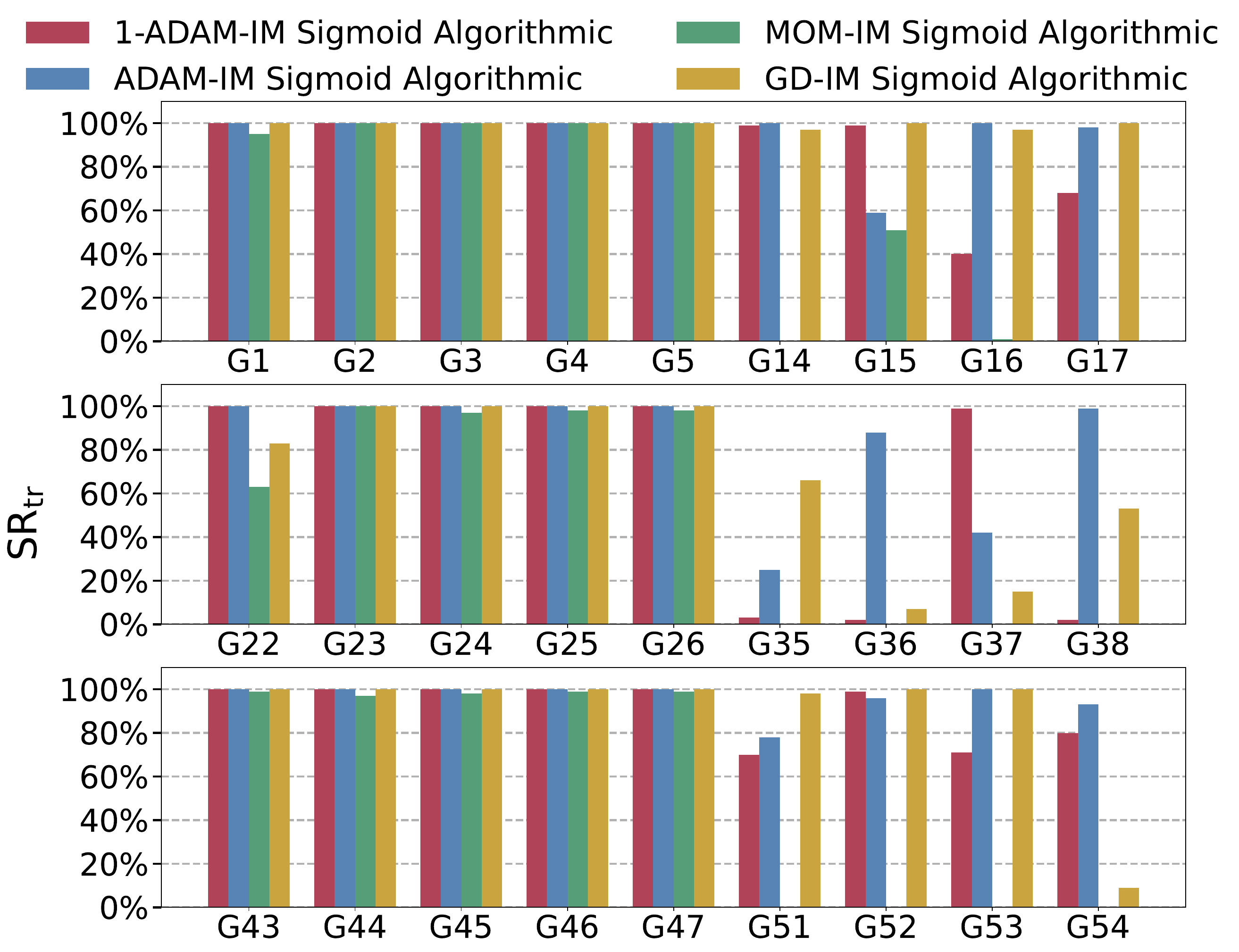}
        
        \vspace{1mm}
        \small (a) Unweighted \textsc{Gset}
    \end{minipage}\hfill
    \begin{minipage}[t]{0.49\textwidth}
        \centering
        \includegraphics[width=\linewidth]{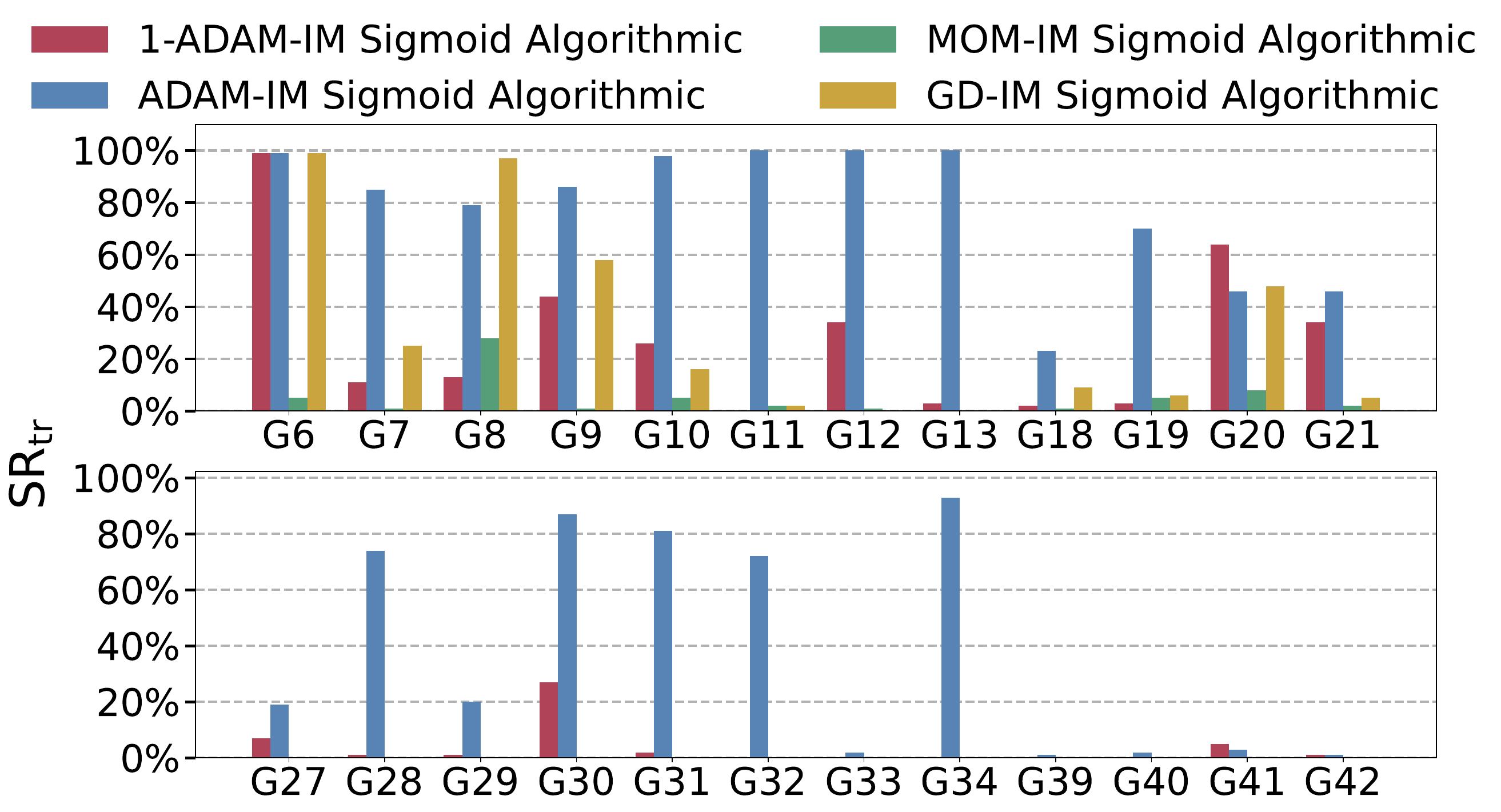}
        
        \vspace{1mm}
        \small (b) Weighted \textsc{Gset}
    \end{minipage}
\caption{Algorithmic discrete-time transient success rate after Bayesian optimization with the sigmoid nonlinearity on (a) unweighted \textsc{Gset} and (b) weighted \textsc{Gset}.}
    \label{fig:algorithmic_srtr}
\end{figure*}

Finally, we examine whether the same trend is visible in the transient success rate. Figure~\ref{fig:algorithmic_srtr} reports the best transient success rate obtained by each optimizer under the same Bayesian-optimization procedure. For the problem instances that reached the target of $99.5\%$ of the best known solution, which can be found in Figure~\ref{fig:algorithmic_E_min}, but the transient success rate appears to be zero on Figure~\ref{fig:algorithmic_srtr}, they are just hardly visible, and have a value of $1/400$.
The transient success-rate results reinforce the same picture. On the unweighted \textsc{Gset} problem instances, the performance of the different methods is quite similar, with MOM-IM again showing the weakest performance. On the easier weighted problem instances, ADAM-IM achieves the highest transient success rate, followed by 1-ADAM-IM, then GD-IM and MOM-IM. On the hardest weighted problem instances, the performance differences become larger: ADAM-IM is clearly the best-performing method, with 1-ADAM-IM as the closest competitor, while GD-IM and MOM-IM rarely reach the target. The behavior on G32 and G34 is particularly noteworthy: these problem instances are difficult for the other optimizers in our algorithmic comparison, yet ADAM-IM reaches the best-known cut value in Fig.~\ref{fig:algorithmic_E_min} and maintains substantial transient success rate in Fig.~\ref{fig:algorithmic_srtr} (around \(70\%\) for G32 and close to \(100\%\) for G34).

Overall, the algorithmic setting leads to a different conclusion from that of the continuous-time setting. When the target energy is reachable by all methods, as on the unweighted \textsc{Gset} problem instances, GD-IM can be tuned to achieve \(\mathrm{TTT}_{\mathrm{CPU}}\) comparable to the other methods, leaving less room for an adaptive optimizer to improve speed alone. The advantage of ADAM-IM becomes clear on the harder weighted problem instances, where GD-IM and MOM-IM increasingly fail to reach the target.

\newpage
\section{Discussion}

Our results show that optimizer performance depends strongly on whether the Ising machine is used as a continuous-time system or as a discrete iterative algorithm, and on how its hyperparameters are tuned. In the continuous-time setting, where the Euler timestep \(\Delta t\) is kept sufficiently small to track the underlying dynamics, 1-ADAM-IM consistently gives the lowest time-to-target, the highest solution quality, and the highest transient success rates. ADAM-IM is the closest competitor, while MOM-IM and GD-IM fall substantially behind on the harder problem instances.

The strong continuous-time performance of 1-ADAM-IM can be understood from the additional freedom introduced by the first-order formulation. In the full ADAM-IM dynamics, the parameters \((\beta_1,\beta_2)\) are restricted to the interval \([0,1)\), inherited from the discrete algorithm where they represent exponential moving-average decay factors. In contrast, the first-order formulation removes both the time-dependent debiasing term and the probabilistic interpretation of \((\beta_1,\beta_2)\), allowing them to take arbitrary real values. This enlarged parameter domain, with optimal values typically found in the range \(\beta_1,\beta_2 \in (-120,1)\), provides substantially more freedom to accelerate the internal dynamics. More negative values of these parameters, increase the relaxation rates \((1-\beta_1)\) and \((1-\beta_2)\). As a result, the moment variables \(v\) and \(w\) respond more quickly to changes in the gradient, producing faster dynamics and reducing the time needed to reach the target. A further possible reason is the different long-time behavior of the adaptive factor. In 1-ADAM-IM it decays to zero, whereas in ADAM-IM it approaches one. This slows the late-time 1-ADAM-IM dynamics, which may be beneficial near low-energy states where further improvements often require small local changes rather than large moves through the energy landscape.

The picture changes when the same optimizer ideas are used as purely discrete iterative algorithms. The algorithmic discrete-time setting leads to a different hierarchy. Once large discrete updates and standard optimizer rules are allowed, the performance gap between optimizers becomes smaller, especially on the unweighted \textsc{Gset} problem instances. On the harder weighted problem instances, however, ADAM-IM becomes the strongest method overall, followed by 1-ADAM-IM, while GD-IM and MOM-IM remain less reliable. Absolute time-to-target values should not be compared directly between the continuous-time and algorithmic regimes, since the former are measured in Euler time whereas the latter are measured in CPU time. Even for solution quality and transient success rate, one should be careful: algorithmic runs use a different tuning setup, different parameter ranges, only one Bayesian-optimization run rather than seven, fewer evaluations per run, and fewer stochastic trajectories per evaluation. The continuous-time figures should therefore be used to compare the continuous-time optimizers with each other, while the algorithmic figures should be used to compare the algorithmic optimizers with each other. They should not be read as a quantitative one-to-one comparison between continuous-time and algorithmic implementations.

With this caveat, the qualitative differences between the two regimes are still informative, especially when comparing the solution-quality and transient-success-rate trends in Figs.~\ref{fig:Emin_Gset_combined} and~\ref{fig:srtr_combined} with their algorithmic counterparts in Figs.~\ref{fig:algorithmic_E_min} and~\ref{fig:algorithmic_srtr}. Keeping these differences in tuning and measurement in mind, 1-ADAM-IM shows similar continuous-time and algorithmic results on the unweighted \textsc{Gset} problem instances, while its algorithmic version is somewhat weaker on the weighted problem instances. ADAM-IM shows the opposite trend: it improves in the algorithmic setting, especially on the weighted benchmarks. This is consistent with the implementation comparison in the Supplemental Material at [URL will be inserted by publisher], where the standard discrete Adam update, Eq.~(\ref{adam_discrete}), gives the best algorithmic ADAM-IM results compared to the Euler--Maruyama discretization of the continuous-time Adam dynamics. One possible reason is that the Euler--Maruyama discretization of the continuous-time Adam equations is not identical to the original discrete Adam rule, Eq.~(\ref{adam_discrete}), once large timesteps are allowed. In the derivation, the standard Adam update is first rewritten in terms of Euler updates, but before taking the continuous-time limit the newly updated moment variables \(v_i^{(j+1)}\) and \(w_i^{(j+1)}\) in the \(x_i\)-update are replaced by their previous-step values \(v_i^{(j)}\) and \(w_i^{(j)}\), as shown in Eq.~(\ref{adam_Euler_approx}). This changes the update only at order \(\mathcal{O}(\Delta t^2)\), so it is negligible in the continuous-time regime where \(\Delta t\) is kept small. In the algorithmic setting, however, \(\Delta t\) is allowed to be large, so this small-\(\Delta t\) equivalence no longer has to hold. The standard discrete Adam rule can therefore behave differently from the Euler--Maruyama discretization of the continuous-time Adam equations.

GD-IM also performs better in the algorithmic setting, especially on the easier problem instances. In the continuous-time setting, changing \(\eta\) would mainly rescale time: to keep the same numerical resolution of the underlying trajectory, \(\Delta t\) would have to be reduced so that \(\eta\Delta t\) remains sufficiently small. In the algorithmic setting, this restriction is absent. Here, \(\Delta t\) is itself an additional tunable step-size parameter, so the effective update scale can be adjusted through the product \(\eta\Delta t\). Because useful timestep values can span orders of magnitude, the Bayesian optimization searches over \(\log_{10}\Delta t\) rather than \(\Delta t\) directly, while \(\eta\) is tuned linearly over its chosen range. This gives GD-IM a genuinely larger search space than in the continuous-time setting, where \(\Delta t\) is fixed by numerical-resolution requirements. MOM-IM benefits from the same algorithmic freedom, since its Euler--Maruyama version can also use \(\Delta t\) as an additional tunable step-size parameter. Compared with the continuous-time setting, this improves MOM-IM's transient success rate on the unweighted problem instances and its solution quality on some harder weighted problem instances, but this extra tuning freedom is not sufficient to close the gap to the Adam-based methods. Overall, compared with the continuous-time setting, the algorithmic setting reduces the differences between optimizers on easier problem instances. On the harder weighted problem instances, however, the best algorithmic performance comes from the standard discrete Adam update, given by Eq.~(\ref{adam_discrete}).

A broader practical lesson from these results is that Ising-machine performance can depend strongly on how the hyperparameter search is organized. In our nonlinearity study, the coarse grid scans varied only the nonlinearity parameters \((\alpha,\beta)\), with all optimizer-specific parameters fixed. This showed that the periodic nonlinearity can appear competitive with the sigmoid nonlinearity for Adam-based IMs. After full Bayesian optimization, however, the nonlinearity parameters and optimizer parameters are tuned jointly, and the sigmoid nonlinearity consistently emerged as the strongest choice. Thus, the apparent ranking of nonlinearities can change once the optimizer dynamics and nonlinear transfer function are optimized together.

This observation also motivated the seven-range Bayesian-optimization strategy used for the final \textsc{Gset} comparison in the continuous-time setting. Instead of using one very broad search domain, we split the search into several smaller ranges, which made it easier for Bayesian optimization to refine high-performing regions while keeping the total computational cost manageable. More generally, these results show that the choice of optimizer, nonlinearity, and hyperparameter-search domain should be treated together: careful tuning is not just an implementation detail, but an important part of achieving strong Ising-machine performance. In practice, our results support a two-stage tuning protocol: first a light grid scan to set reasonable ranges, then range-split Bayesian optimization to refine the best-performing regions. Bayesian optimization improves upon the coarse grid-scan baseline for all optimizers, with the largest gain observed for 1-ADAM-IM. This is practically important because the refinement is not more sample-expensive than the grid scan itself: the coarse scan uses \(30\times30=900\) parameter evaluations, whereas each Bayesian-optimization run uses \(1000\) evaluations. Moreover, the overhead of the Bayesian-optimization procedure itself is negligible compared with the cost of generating these samples. In other words, Bayesian optimization may appear much more sophisticated, but in our setting it operates at essentially the same sampling budget while delivering systematically better performance. This makes the combination of a coarse scan for orientation followed by full Bayesian optimization a particularly effective protocol.

A related practical point is the choice of Bayesian-optimization objective. In this work, the Bayesian optimization is based on a target-reaching metric, namely \(\mathrm{TTT}\). This is appropriate for the benchmark setting used here, where the g05 ground states are known and the \textsc{Gset} targets can be defined relative to best-known cut values. In a practical deployment, however, such a reference value is usually not known in advance. In that case, the Bayesian optimization would need a different objective, for example one that balances the best energy reached against the runtime needed to reach it.

Finally, because the continuous-time results are expressed partly through time-to-target, it is important to separate optimizer effects from simple time-scale effects. As discussed earlier, the different optimizers do not evolve on exactly the same natural timescale, so absolute differences in \(\mathrm{TTT}\) measured in Euler time should be interpreted with some care. In particular, part of the observed difference between GD-IM, MOM-IM, ADAM-IM, and 1-ADAM-IM could in principle come from how the natural time scale of each optimizer is set, rather than from the optimizer dynamics alone. That caveat, however, cannot explain the main trend in our results. If the observed differences simply reflected different effective time scales, one would expect a simple global rescaling of time, \(t\mapsto c\,t\), for each optimizer. Such a rescaling would multiply all \(\mathrm{TTT}\) values of that optimizer by the same constant across benchmark problem instances. It would therefore not explain why the advantage of ADAM-IM and especially 1-ADAM-IM becomes progressively larger as the problems become harder, from g05 to \textsc{Gset} and then to the harder \textsc{Gset} problem instances. This conclusion is further supported by the solution-quality and transient-success-rate results, which are not themselves time-based measures: 1-ADAM-IM consistently reaches higher cut values and higher success rates, often by a substantial margin. An additional check in the Supplemental Material at [URL will be inserted by publisher] shows that this conclusion does not change when the \(\mathrm{TTT}\) values are rescaled to account for possible differences in effective time scale.

Taken together, these results show that Adam-type dynamics do more than rescale gradient descent: they change the optimization behavior in a way that becomes increasingly beneficial on harder problem instances. In the continuous-time setting, the 1-ADAM-IM formulation is the strongest method across all benchmarks, improving not only time-to-target but also solution quality and transient success rate. In the algorithmic setting, the advantage shifts to the standard discrete Adam update, which gives the best performance on the harder weighted instances. This makes Adam-type optimization a promising design principle both for improving analog, time-continuous Ising machines and for guiding future algorithmic Ising-machine variants.

\section*{Acknowledgements}
We thank Toon Sevenants for helpful discussions and for feedback on the figures and their presentation. This research was funded by the Research Foundation Flanders (FWO) under Grants No. G0A6L25N and No. G006020N. Additional funding was provided by the EOS project “Photonic Ising Machines.” This project (EOS No. 40007536) has received funding from the FWO and F.R.S.-FNRS under the Excellence of Science (EOS) programme. The computational resources and services used in this work were provided by the VSC (Flemish Supercomputer Center), funded by the Research Foundation Flanders (FWO) and the Flemish Government – department WEWIS. - De rekeninfrastructuur en dienstverlening gebruikt in dit werk, werd voorzien door het VSC (Vlaams Supercomputer Centrum), gefinancierd door het FWO en de Vlaamse regering – departement WEWIS.

\section*{Code and data availability}
The code and processed data used in this work are available at\\
\url{https://github.com/StijnVanVooren/Adam-analog-ising-machines}.
The repository contains the continuous-time and discrete-time Ising-machine simulation code, Bayesian-optimization scripts, processed benchmark results, and plotting code. The original benchmark instances are not redistributed with the repository; the g05 instances are available from the BiqMac library, and the \textsc{Gset} instances are available from the standard \textsc{Gset} MaxCut benchmark collection.

\clearpage

\setcounter{section}{0}
\setcounter{figure}{0}
\setcounter{table}{0}
\setcounter{equation}{0}

\renewcommand{\thesection}{S\arabic{section}}
\renewcommand{\thefigure}{S\arabic{figure}}
\renewcommand{\thetable}{S\arabic{table}}
\renewcommand{\theequation}{S\arabic{equation}}

\title{Supplemental Material for\\
``Beyond Gradient Descent: Adam for Analog Ising Machines''}

\author{Stijn Van Vooren}
\email{Stijn.Van.Vooren@vub.be}
\affiliation{Applied Physics research group, Vrije Universiteit Brussel, Pleinlaan 2, 1050 Brussels, Belgium}

\author{Guy Van der Sande}
\affiliation{Applied Physics research group, Vrije Universiteit Brussel, Pleinlaan 2, 1050 Brussels, Belgium}

\author{Guy Verschaffelt}
\affiliation{Applied Physics research group, Vrije Universiteit Brussel, Pleinlaan 2, 1050 Brussels, Belgium}

\date{\today}

\begin{center}
{\Large \textbf{Supplemental Material for\\
``Beyond Gradient Descent: Adam for Analog Ising Machines''}}
\end{center}
\tableofcontents
\section{Bayesian-optimization protocol and implementation details}
\label{subsec:supp_bo_details}

In the main text, benchmark comparisons are based on a two-stage procedure consisting of a coarse grid scan followed by a full Bayesian optimization. We collect here the implementation details of the Bayesian-optimization stage.

Each Ising-machine variant has a different number of free hyperparameters. The nonlinearities contribute the parameters \((\alpha,\beta,\gamma)\), momentum adds \(\beta_1\), and Adam-based machines add \((\beta_1,\beta_2,\eta)\). After the coarse nonlinearity-selection stage described in the main text, we therefore perform a full Bayesian optimization over all remaining free hyperparameters, allowing the optimizer dynamics and the selected nonlinearity to be tuned jointly in a single search procedure.

Bayesian optimization is implemented with the \textit{BayesOpt} C++ library~\cite{BayesOpt}. We use a Student's t-process surrogate with Normal--Inverse--Gamma hyperpriors, a constant mean, and a composite kernel consisting of a Matern--ARD(3) term plus a rational-quadratic term. The surrogate hyperparameters are learned during the optimization run. To ensure numerical stability in the presence of stochastic trajectory averages, the surrogate noise level is fixed to \(10^{-3}\).

Each Bayesian-optimization run begins with \(300\) randomly chosen parameter evaluations in order to explore the search space broadly. These are followed by \(700\) Bayesian-optimization iterations that refine the search around the most promising regions, for a total of \(1000\) sampled parameter settings per run. For the continuous-time physical benchmarks, each Bayesian-optimization run begins with \(300\) randomly chosen parameter evaluations, followed by \(700\) Bayesian-optimization iterations, for a total of \(1000\) sampled parameter settings per run. Each objective-function evaluation is an average over \(400\) independent stochastic trajectories. Depending on the benchmark instance, each trajectory is evolved for \(10{,}000\) to \(20{,}000\) Euler steps. We use \(-1/\mathrm{TTT}\) rather than the raw \(\mathrm{TTT}\) as the objective. This is because \(\mathrm{TTT}\) diverges when no target is reached, creating a discontinuity that is difficult for Bayesian optimization to model. By contrast, \(-1/\mathrm{TTT}\) maps such cases to a finite value \(0\), preserves the preference for low \(\mathrm{TTT}\), gives higher resolution in the low-\(\mathrm{TTT}\) region, and restores the minimization convention through the minus sign. The same construction is used for \(\mathrm{TTT}_{\mathrm{CPU}}\) in the algorithmic setting.

For the nonlinearity comparisons on the g05 benchmark set, we use a single Bayesian-optimization range per optimizer. For the final optimizer comparisons reported in the main text, however, we repeat the entire Bayesian-optimization procedure over seven distinct search domains for the sigmoid nonlinearity parameters \((\alpha,\beta)\), listed in Table~\ref{tab:bo_alpha_beta_ranges}. Each \((\alpha,\beta)\) range is used for all optimizers, while the remaining search dimensions are optimizer-specific. For each optimizer and benchmark instance, we retain the best result over these independent Bayesian-optimization runs. This reduces boundary lock-in and makes the optimization less sensitive to the precise choice of search domain. In the implementation, infinitesimal offsets from the endpoints are used to avoid evaluating exactly at the boundaries.

\begin{table}[h]
\centering
\caption{Bayesian-optimization search domains for the sigmoid nonlinearity parameters \((\alpha,\beta)\). Each range is used for all optimizers in the final optimizer comparison.}
\label{tab:bo_alpha_beta_ranges}
\begin{ruledtabular}
\begin{tabular}{ccc}
Range & \(\alpha\) & \(\beta\) \\
\hline
1 & \([-2,-1]\) & \([0,1]\) \\
2 & \([-1,0]\)  & \([0,1]\) \\
3 & \([0,1]\)   & \([0,1]\) \\
4 & \([1,2]\)   & \([0,1]\) \\
5 & \([-1,0]\)  & \([1,2]\) \\
6 & \([0,1]\)   & \([1,2]\) \\
7 & \([1,2]\)   & \([1,2]\) \\
\end{tabular}
\end{ruledtabular}
\end{table}
The remaining physical continuous-time search domains are listed in Table~\ref{tab:physical_bo_ranges}. The timestep is not optimized in the physical simulations; it is fixed to \(\Delta t=10^{-2}\) so that the Euler--Maruyama discretization remains a faithful approximation of the underlying continuous-time dynamics. For MOM-IM, the momentum parameter is denoted by \(\beta_1\) in the paper, corresponding to \(\theta\) in the implementation.

\begin{table}[h]
\centering
\caption{Optimizer-specific Bayesian-optimization search domains for the physical continuous-time comparisons. The \((\alpha,\beta)\) ranges are given separately in Table~\ref{tab:bo_alpha_beta_ranges}.}
\label{tab:physical_bo_ranges}
\begin{ruledtabular}
\begin{tabular}{cccccc}
Optimizer & \(\log_{10}\gamma\) & \(\beta_1\) & \(\beta_2\) & \(\eta\) & \(\Delta t\) \\
\hline
GD-IM 
& \([-10,2]\) & -- & -- & \(1\) & \(10^{-2}\) \\

MOM-IM 
& \([-10,2]\) & \([-200,1]\) & -- & \(1\) & \(10^{-2}\) \\

ADAM-IM 
& \([-10,2]\) & \([0,1]\) & \([0,1]\) & \([1,200]\) & \(10^{-2}\) \\

1-ADAM-IM 
& \([-10,2]\) & \([-200,1]\) & \([-200,1]\) & \([1,200]\) & \(10^{-2}\) \\
\end{tabular}
\end{ruledtabular}
\end{table}
For the algorithmic discrete-time comparisons, the Bayesian-optimization protocol is slightly reduced in order to keep the computational cost manageable. Each run uses \(200\) random initial evaluations followed by \(600\) Bayesian-optimization iterations, for a total of \(800\) sampled parameter settings. Each objective-function evaluation averages \(100\) stochastic trajectories. The optimized parameters depend on the optimizer and on whether the Euler--Maruyama or standard discrete implementation is used. When an Euler--Maruyama implementation is used, \(\Delta t\) is optimized on a logarithmic scale through \(\log_{10}\Delta t\). For standard discrete optimizer formulations, we fix \(\Delta t=1\). The learning-rate parameter \(\eta\), when varied, is optimized on a linear scale. The search ranges used for the algorithmic comparisons are listed in Table~\ref{tab:algorithmic_bo_ranges}.

\begin{table*}[h]
\centering
\caption{Bayesian-optimization search domains used for the algorithmic discrete-time comparisons. The timestep range applies only to Euler--Maruyama implementations; for standard discrete formulations, \(\Delta t=1\).}
\label{tab:algorithmic_bo_ranges}
\begin{ruledtabular}
\begin{tabular}{lccccccc}
Optimizer / setting & \(\alpha\) & \(\beta\) & \(\beta_1\) & \(\beta_2\) & \(\eta\) & \(\log_{10}\gamma\) & \(\log_{10}\Delta t\) \\
\hline
GD-IM, unweighted \textsc{Gset} 
& \([-2,2]\) & \([0,3]\) & -- & -- & \([1,200]\) & \([-10,2]\) & \([-3,2]\) \\

GD-IM, weighted \textsc{Gset} 
& \([-2,2]\) & \([0,3]\) & -- & -- & \([1,200]\) & \([-10,2]\) & \(\Delta t=1\) \\

MOM-IM 
& \([-2,2]\) & \([0,3]\) & \([-200,1]\) & -- & \(1\) & \([-10,2]\) & \([-3,2]\) \\

ADAM-IM 
& \([-2,2]\) & \([0,3]\) & \([0,1]\) & \([0,1]\) & \([1,200]\) & \([-10,2]\) & \(\Delta t=1\) \\

1-ADAM-IM 
& \([-2,2]\) & \([0,3]\) & \([-200,1]\) & \([-200,1]\) & \([1,200]\) & \([-10,2]\) & \([-3,2]\) \\
\end{tabular}
\end{ruledtabular}
\end{table*}

All optimizers are compared under the same computational budget: the same number of Bayesian-optimization runs, the same number of sampled parameter settings per run, and the same number of stochastic trajectories per objective-function evaluation. In our implementation, the additional arithmetic required by MOM-IM, ADAM-IM, and 1-ADAM-IM is negligible compared with the cost of evaluating the coupling term \(\sum_j J_{ij}x_j\), which accounts for more than \(99\%\) of the runtime in our profiling. The optimizer-specific overhead scales only as \(\mathcal{O}(N)\), whereas for dense problems the coupling evaluation scales as \(\mathcal{O}(N^2)\) and remains the dominant cost more generally.

\section{Additional coarse grid-scan comparison of nonlinearities for ADAM-IM and 1-ADAM-IM}
\label{sec:supp_gridscan_adam}

To complement the main-text Bayesian-optimization results, we provide here the full coarse grid-scan comparison across nonlinearities for the two Adam-based variants. This scan is restricted to the g05 benchmark set and is intended to probe the standalone effect of the nonlinearity before optimizer-specific hyperparameters are jointly refined. As in the main text, we vary only the nonlinearity parameters \((\alpha,\beta)\), while keeping all remaining hyperparameters fixed to their baseline values.

Table~\ref{tab:supp_adam_gridscan} reports, for each g05 instance and for each of the four nonlinearities, the minimum time-to-solution obtained on the \(30\times30\) \((\alpha,\beta)\) grid for ADAM-IM and 1-ADAM-IM. At this coarse level, the two Adam-based formulations behave very similarly overall, and the ranking of nonlinearities is likewise similar for both. In particular, the periodic and sigmoid nonlinearities perform best most often, whereas the clipped nonlinearity is less consistent and the polynomial nonlinearity is generally weaker in this limited-parameter scan. These coarse results provide the baseline against which the full Bayesian-optimization comparison in the main text should be interpreted.

\begin{table}[H]
\caption{\textbf{g05 benchmark set: Time-to-solution (TTS) across all four nonlinearities for ADAM-IM and 1-ADAM-IM.}
For each g05 instance, we report the minimum \(TTS\) found over a \(30\times30\) grid scan in \((\alpha,\beta)\), with \(\gamma=0.005\) fixed, for each of the four nonlinearities. Each grid point is averaged over 400 runs of \(10{,}000\) Euler steps with \(\Delta t=0.01\). Colored cells indicate the within-row ranking: green (best), yellow (second best), orange (second worst), and red (worst). For both ADAM-IM and 1-ADAM-IM, we set \(\beta_1=\beta_2=0.99\); for 1-ADAM-IM, we set \(\eta=\frac{\sqrt{-\ln(\beta_1)}}{-\ln(\beta_2)}=9.97\). All values are reported in milliseconds.}
\label{tab:supp_adam_gridscan}
\begin{center}
\begin{tabular}{lllllllll}
\cline{2-9}
\multicolumn{1}{l|}{}                       & \multicolumn{4}{c|}{\textbf{ADAM-IM}}                                                                                                                                         & \multicolumn{4}{c|}{\textbf{1-ADAM-IM}}                                                                                                                                \\ \cline{2-9} 
\multicolumn{1}{l|}{}                       & \multicolumn{1}{l|}{\textbf{Poly}}          & \multicolumn{1}{l|}{\textbf{Sig}} & \multicolumn{1}{l|}{\textbf{Per}} & \multicolumn{1}{l|}{\textbf{Clip}} & \multicolumn{1}{l|}{\textbf{Poly}} & \multicolumn{1}{l|}{\textbf{Sig}} & \multicolumn{1}{l|}{\textbf{Per}} & \multicolumn{1}{l|}{\textbf{Clip}} \\ \hline
\multicolumn{1}{|l|}{ \textbf{g\_05\_60.0} } & \multicolumn{1}{c}{\cellcolor[HTML]{FE0000}6.05} & \multicolumn{1}{c}{3.33} & \multicolumn{1}{c}{\cellcolor[HTML]{34FF34}1.91} & \multicolumn{1}{c}{3.62} & \multicolumn{1}{c}{\cellcolor[HTML]{FFCB2F}5.7} & \multicolumn{1}{c}{3.12} & \multicolumn{1}{c}{\cellcolor[HTML]{F8FF00}2.1} & \multicolumn{1}{c}{4.29} \\ \cline{1-1}
\multicolumn{1}{|l|}{ \textbf{g\_05\_60.1} } & \multicolumn{1}{c}{1.47} & \multicolumn{1}{c}{1.4} & \multicolumn{1}{c}{1.55} & \multicolumn{1}{c}{\cellcolor[HTML]{FFCB2F}1.74} & \multicolumn{1}{c}{1.54} & \multicolumn{1}{c}{\cellcolor[HTML]{F8FF00}1.39} & \multicolumn{1}{c}{\cellcolor[HTML]{34FF34}1.36} & \multicolumn{1}{c}{\cellcolor[HTML]{FE0000}1.91} \\ \cline{1-1}
\multicolumn{1}{|l|}{ \textbf{g\_05\_60.2} } & \multicolumn{1}{c}{\cellcolor[HTML]{FFCB2F}5.35} & \multicolumn{1}{c}{\cellcolor[HTML]{34FF34}3.56} & \multicolumn{1}{c}{4.17} & \multicolumn{1}{c}{\cellcolor[HTML]{FE0000}5.6} & \multicolumn{1}{c}{5.29} & \multicolumn{1}{c}{\cellcolor[HTML]{F8FF00}3.65} & \multicolumn{1}{c}{4.43} & \multicolumn{1}{c}{4.94} \\ \cline{1-1}
\multicolumn{1}{|l|}{ \textbf{g\_05\_60.3} } & \multicolumn{1}{c}{\cellcolor[HTML]{FFCB2F}86.11} & \multicolumn{1}{c}{\cellcolor[HTML]{34FF34}4.88} & \multicolumn{1}{c}{6.82} & \multicolumn{1}{c}{12.1} & \multicolumn{1}{c}{\cellcolor[HTML]{FE0000}95.99} & \multicolumn{1}{c}{\cellcolor[HTML]{F8FF00}4.96} & \multicolumn{1}{c}{6.95} & \multicolumn{1}{c}{9.22} \\ \cline{1-1}
\multicolumn{1}{|l|}{ \textbf{g\_05\_60.4} } & \multicolumn{1}{c}{\cellcolor[HTML]{FE0000}5.48} & \multicolumn{1}{c}{\cellcolor[HTML]{F8FF00}2.24} & \multicolumn{1}{c}{2.85} & \multicolumn{1}{c}{3.87} & \multicolumn{1}{c}{\cellcolor[HTML]{FFCB2F}5.2} & \multicolumn{1}{c}{\cellcolor[HTML]{34FF34}2.22} & \multicolumn{1}{c}{2.7} & \multicolumn{1}{c}{3.55} \\ \cline{1-1}
\multicolumn{1}{|l|}{ \textbf{g\_05\_60.5} } & \multicolumn{1}{c}{\cellcolor[HTML]{FFCB2F}11.47} & \multicolumn{1}{c}{7.15} & \multicolumn{1}{c}{7.86} & \multicolumn{1}{c}{\cellcolor[HTML]{34FF34}6.07} & \multicolumn{1}{c}{\cellcolor[HTML]{FE0000}11.63} & \multicolumn{1}{c}{6.72} & \multicolumn{1}{c}{8.09} & \multicolumn{1}{c}{\cellcolor[HTML]{F8FF00}6.29} \\ \cline{1-1}
\multicolumn{1}{|l|}{ \textbf{g\_05\_60.6} } & \multicolumn{1}{c}{\cellcolor[HTML]{FFCB2F}84.02} & \multicolumn{1}{c}{74.81} & \multicolumn{1}{c}{\cellcolor[HTML]{34FF34}40.05} & \multicolumn{1}{c}{\cellcolor[HTML]{F8FF00}42.04} & \multicolumn{1}{c}{52.28} & \multicolumn{1}{c}{\cellcolor[HTML]{FE0000}88.76} & \multicolumn{1}{c}{80.25} & \multicolumn{1}{c}{46.25} \\ \cline{1-1}
\multicolumn{1}{|l|}{ \textbf{g\_05\_60.7} } & \multicolumn{1}{c}{\cellcolor[HTML]{FE0000}13.3} & \multicolumn{1}{c}{2.67} & \multicolumn{1}{c}{\cellcolor[HTML]{34FF34}1.93} & \multicolumn{1}{c}{4.91} & \multicolumn{1}{c}{\cellcolor[HTML]{FFCB2F}12.09} & \multicolumn{1}{c}{2.68} & \multicolumn{1}{c}{\cellcolor[HTML]{F8FF00}2.06} & \multicolumn{1}{c}{5.37} \\ \cline{1-1}
\multicolumn{1}{|l|}{ \textbf{g\_05\_60.8} } & \multicolumn{1}{c}{\cellcolor[HTML]{FE0000}24.27} & \multicolumn{1}{c}{8.92} & \multicolumn{1}{c}{\cellcolor[HTML]{FFCB2F}21.66} & \multicolumn{1}{c}{\cellcolor[HTML]{34FF34}4.75} & \multicolumn{1}{c}{21.52} & \multicolumn{1}{c}{8.48} & \multicolumn{1}{c}{20.3} & \multicolumn{1}{c}{\cellcolor[HTML]{F8FF00}6.72} \\ \cline{1-1}
\multicolumn{1}{|l|}{ \textbf{g\_05\_60.9} } & \multicolumn{1}{c}{\cellcolor[HTML]{FE0000}20.37} & \multicolumn{1}{c}{8.63} & \multicolumn{1}{c}{7.29} & \multicolumn{1}{c}{\cellcolor[HTML]{F8FF00}6.79} & \multicolumn{1}{c}{\cellcolor[HTML]{FFCB2F}17.03} & \multicolumn{1}{c}{8.59} & \multicolumn{1}{c}{7.98} & \multicolumn{1}{c}{\cellcolor[HTML]{34FF34}5.49} \\ \cline{1-1}
\multicolumn{1}{|l|}{ \textbf{g\_05\_80.0} } & \multicolumn{1}{c}{\cellcolor[HTML]{FE0000}42.49} & \multicolumn{1}{c}{19.24} & \multicolumn{1}{c}{\cellcolor[HTML]{F8FF00}14.82} & \multicolumn{1}{c}{15.95} & \multicolumn{1}{c}{\cellcolor[HTML]{FFCB2F}35.0} & \multicolumn{1}{c}{17.96} & \multicolumn{1}{c}{\cellcolor[HTML]{34FF34}12.88} & \multicolumn{1}{c}{28.38} \\ \cline{1-1}
\multicolumn{1}{|l|}{ \textbf{g\_05\_80.1} } & \multicolumn{1}{c}{\cellcolor[HTML]{FFCB2F}0.57} & \multicolumn{1}{c}{0.36} & \multicolumn{1}{c}{0.44} & \multicolumn{1}{c}{0.39} & \multicolumn{1}{c}{\cellcolor[HTML]{FE0000}0.73} & \multicolumn{1}{c}{\cellcolor[HTML]{F8FF00}0.33} & \multicolumn{1}{c}{0.46} & \multicolumn{1}{c}{\cellcolor[HTML]{34FF34}0.27} \\ \cline{1-1}
\multicolumn{1}{|l|}{ \textbf{g\_05\_80.2} } & \multicolumn{1}{c}{\cellcolor[HTML]{FFCB2F}136.42} & \multicolumn{1}{c}{\cellcolor[HTML]{34FF34}20.45} & \multicolumn{1}{c}{\cellcolor[HTML]{F8FF00}20.62} & \multicolumn{1}{c}{81.04} & \multicolumn{1}{c}{\cellcolor[HTML]{FE0000}137.61} & \multicolumn{1}{c}{22.09} & \multicolumn{1}{c}{22.16} & \multicolumn{1}{c}{69.4} \\ \cline{1-1}
\multicolumn{1}{|l|}{ \textbf{g\_05\_80.3} } & \multicolumn{1}{c}{\cellcolor[HTML]{FFCB2F}115.39} & \multicolumn{1}{c}{87.91} & \multicolumn{1}{c}{62.76} & \multicolumn{1}{c}{\cellcolor[HTML]{34FF34}34.16} & \multicolumn{1}{c}{\cellcolor[HTML]{FE0000}118.11} & \multicolumn{1}{c}{81.93} & \multicolumn{1}{c}{100.22} & \multicolumn{1}{c}{\cellcolor[HTML]{F8FF00}48.59} \\ \cline{1-1}
\multicolumn{1}{|l|}{ \textbf{g\_05\_80.4} } & \multicolumn{1}{c}{\cellcolor[HTML]{FFCB2F}40.94} & \multicolumn{1}{c}{\cellcolor[HTML]{34FF34}12.49} & \multicolumn{1}{c}{14.05} & \multicolumn{1}{c}{40.08} & \multicolumn{1}{c}{39.53} & \multicolumn{1}{c}{\cellcolor[HTML]{F8FF00}12.54} & \multicolumn{1}{c}{13.98} & \multicolumn{1}{c}{\cellcolor[HTML]{FE0000}57.0} \\ \cline{1-1}
\multicolumn{1}{|l|}{ \textbf{g\_05\_80.5} } & \multicolumn{1}{c}{\cellcolor[HTML]{FFCB2F}135.32} & \multicolumn{1}{c}{125.37} & \multicolumn{1}{c}{\cellcolor[HTML]{F8FF00}75.98} & \multicolumn{1}{c}{\cellcolor[HTML]{34FF34}59.61} & \multicolumn{1}{c}{\cellcolor[HTML]{FE0000}142.83} & \multicolumn{1}{c}{126.82} & \multicolumn{1}{c}{101.48} & \multicolumn{1}{c}{76.63} \\ \cline{1-1}
\multicolumn{1}{|l|}{ \textbf{g\_05\_80.6} } & \multicolumn{1}{c}{123.98} & \multicolumn{1}{c}{36.73} & \multicolumn{1}{c}{\cellcolor[HTML]{F8FF00}25.93} & \multicolumn{1}{c}{\cellcolor[HTML]{FE0000}237.8} & \multicolumn{1}{c}{\cellcolor[HTML]{FFCB2F}149.08} & \multicolumn{1}{c}{36.1} & \multicolumn{1}{c}{\cellcolor[HTML]{34FF34}22.92} & \multicolumn{1}{c}{120.19} \\ \cline{1-1}
\multicolumn{1}{|l|}{ \textbf{g\_05\_80.7} } & \multicolumn{1}{c}{111.81} & \multicolumn{1}{c}{23.28} & \multicolumn{1}{c}{\cellcolor[HTML]{34FF34}5.45} & \multicolumn{1}{c}{\cellcolor[HTML]{FFCB2F}154.75} & \multicolumn{1}{c}{92.64} & \multicolumn{1}{c}{23.59} & \multicolumn{1}{c}{\cellcolor[HTML]{F8FF00}6.17} & \multicolumn{1}{c}{\cellcolor[HTML]{FE0000}255.62} \\ \cline{1-1}
\multicolumn{1}{|l|}{ \textbf{g\_05\_80.8} } & \multicolumn{1}{c}{\cellcolor[HTML]{FFCB2F}265.74} & \multicolumn{1}{c}{\cellcolor[HTML]{F8FF00}70.94} & \multicolumn{1}{c}{118.85} & \multicolumn{1}{c}{180.58} & \multicolumn{1}{c}{\cellcolor[HTML]{FE0000}283.59} & \multicolumn{1}{c}{\cellcolor[HTML]{34FF34}57.69} & \multicolumn{1}{c}{105.9} & \multicolumn{1}{c}{245.7} \\ \cline{1-1}
\multicolumn{1}{|l|}{ \textbf{g\_05\_80.9} } & \multicolumn{1}{c}{\cellcolor[HTML]{FE0000}63.56} & \multicolumn{1}{c}{\cellcolor[HTML]{34FF34}23.5} & \multicolumn{1}{c}{30.86} & \multicolumn{1}{c}{38.73} & \multicolumn{1}{c}{\cellcolor[HTML]{FFCB2F}54.58} & \multicolumn{1}{c}{\cellcolor[HTML]{F8FF00}24.71} & \multicolumn{1}{c}{26.9} & \multicolumn{1}{c}{36.57} \\ \cline{1-1}
\multicolumn{1}{|l|}{ \textbf{g\_05\_100.0} } & \multicolumn{1}{c}{26.99} & \multicolumn{1}{c}{22.72} & \multicolumn{1}{c}{\cellcolor[HTML]{F8FF00}22.61} & \multicolumn{1}{c}{\cellcolor[HTML]{FFCB2F}181.64} & \multicolumn{1}{c}{28.22} & \multicolumn{1}{c}{23.39} & \multicolumn{1}{c}{\cellcolor[HTML]{34FF34}21.36} & \multicolumn{1}{c}{\cellcolor[HTML]{FE0000}186.43} \\ \cline{1-1}
\multicolumn{1}{|l|}{ \textbf{g\_05\_100.1} } & \multicolumn{1}{c}{30.31} & \multicolumn{1}{c}{33.45} & \multicolumn{1}{c}{\cellcolor[HTML]{34FF34}25.92} & \multicolumn{1}{c}{\cellcolor[HTML]{FE0000}128.37} & \multicolumn{1}{c}{28.95} & \multicolumn{1}{c}{30.85} & \multicolumn{1}{c}{\cellcolor[HTML]{F8FF00}26.05} & \multicolumn{1}{c}{\cellcolor[HTML]{FFCB2F}71.44} \\ \cline{1-1}
\multicolumn{1}{|l|}{ \textbf{g\_05\_100.2} } & \multicolumn{1}{c}{\cellcolor[HTML]{F8FF00}6.74} & \multicolumn{1}{c}{8.8} & \multicolumn{1}{c}{8.28} & \multicolumn{1}{c}{\cellcolor[HTML]{FFCB2F}243.13} & \multicolumn{1}{c}{\cellcolor[HTML]{34FF34}6.5} & \multicolumn{1}{c}{8.46} & \multicolumn{1}{c}{8.01} & \multicolumn{1}{c}{\cellcolor[HTML]{FE0000}288.43} \\ \cline{1-1}
\multicolumn{1}{|l|}{ \textbf{g\_05\_100.3} } & \multicolumn{1}{c}{\cellcolor[HTML]{FFCB2F}535.04} & \multicolumn{1}{c}{450.77} & \multicolumn{1}{c}{416.69} & \multicolumn{1}{c}{\cellcolor[HTML]{F8FF00}400.19} & \multicolumn{1}{c}{\cellcolor[HTML]{FE0000}784.09} & \multicolumn{1}{c}{494.97} & \multicolumn{1}{c}{\cellcolor[HTML]{34FF34}346.06} & \multicolumn{1}{c}{438.42} \\ \cline{1-1}
\multicolumn{1}{|l|}{ \textbf{g\_05\_100.4} } & \multicolumn{1}{c}{\cellcolor[HTML]{FE0000}130.59} & \multicolumn{1}{c}{37.17} & \multicolumn{1}{c}{\cellcolor[HTML]{34FF34}26.02} & \multicolumn{1}{c}{\cellcolor[HTML]{FFCB2F}124.61} & \multicolumn{1}{c}{119.41} & \multicolumn{1}{c}{37.3} & \multicolumn{1}{c}{\cellcolor[HTML]{F8FF00}26.64} & \multicolumn{1}{c}{109.22} \\ \cline{1-1}
\multicolumn{1}{|l|}{ \textbf{g\_05\_100.5} } & \multicolumn{1}{c}{136.72} & \multicolumn{1}{c}{\cellcolor[HTML]{FE0000}145.45} & \multicolumn{1}{c}{99.62} & \multicolumn{1}{c}{\cellcolor[HTML]{34FF34}82.92} & \multicolumn{1}{c}{\cellcolor[HTML]{FFCB2F}140.46} & \multicolumn{1}{c}{135.34} & \multicolumn{1}{c}{\cellcolor[HTML]{F8FF00}82.96} & \multicolumn{1}{c}{116.76} \\ \cline{1-1}
\multicolumn{1}{|l|}{ \textbf{g\_05\_100.6} } & \multicolumn{1}{c}{\cellcolor[HTML]{FE0000}181.31} & \multicolumn{1}{c}{\cellcolor[HTML]{F8FF00}28.4} & \multicolumn{1}{c}{30.8} & \multicolumn{1}{c}{64.56} & \multicolumn{1}{c}{\cellcolor[HTML]{FFCB2F}161.03} & \multicolumn{1}{c}{\cellcolor[HTML]{34FF34}26.12} & \multicolumn{1}{c}{28.74} & \multicolumn{1}{c}{65.18} \\ \cline{1-1}
\multicolumn{1}{|l|}{ \textbf{g\_05\_100.7} } & \multicolumn{1}{c}{\cellcolor[HTML]{FE0000}257.54} & \multicolumn{1}{c}{50.39} & \multicolumn{1}{c}{\cellcolor[HTML]{F8FF00}22.97} & \multicolumn{1}{c}{222.63} & \multicolumn{1}{c}{\cellcolor[HTML]{FFCB2F}229.64} & \multicolumn{1}{c}{50.78} & \multicolumn{1}{c}{\cellcolor[HTML]{34FF34}21.36} & \multicolumn{1}{c}{176.88} \\ \cline{1-1}
\multicolumn{1}{|l|}{ \textbf{g\_05\_100.8} } & \multicolumn{1}{c}{641.96} & \multicolumn{1}{c}{536.13} & \multicolumn{1}{c}{\cellcolor[HTML]{FE0000}767.98} & \multicolumn{1}{c}{549.82} & \multicolumn{1}{c}{\cellcolor[HTML]{FFCB2F}749.28} & \multicolumn{1}{c}{\cellcolor[HTML]{34FF34}429.45} & \multicolumn{1}{c}{573.94} & \multicolumn{1}{c}{\cellcolor[HTML]{F8FF00}469.38} \\ \cline{1-1}
\multicolumn{1}{|l|}{ \textbf{g\_05\_100.9} } & \multicolumn{1}{c}{48.78} & \multicolumn{1}{c}{44.37} & \multicolumn{1}{c}{43.05} & \multicolumn{1}{c}{\cellcolor[HTML]{FFCB2F}335.68} & \multicolumn{1}{c}{46.58} & \multicolumn{1}{c}{\cellcolor[HTML]{F8FF00}42.51} & \multicolumn{1}{c}{\cellcolor[HTML]{34FF34}36.94} & \multicolumn{1}{c}{\cellcolor[HTML]{FE0000}351.9} \\ \cline{1-1}
\end{tabular}
\end{center}
\end{table}

\section{Additional coarse grid-scan comparison of nonlinearities for GD-IM, MOM-IM and 1-ADAM-IM}
\label{subsec:supp_gridscan_allopts}
Following the Adam-only coarse grid-scan comparison in Sec.~\ref{sec:supp_gridscan_adam}, we here provide the full per-instance results for the broader comparison between GD-IM, MOM-IM, and 1-ADAM-IM across all four nonlinearities on the g05 benchmark set. As in the main text, this coarse scan is intended to isolate the role of the nonlinearity before performing the full Bayesian optimization. We therefore vary only the nonlinearity parameters \((\alpha,\beta)\) on a \(30\times30\) grid, while keeping all remaining optimizer-specific hyperparameters fixed to their baseline values.

Table~\ref{tab:supp_allopt_gridscan} reports, for each g05 instance and for each optimizer--nonlinearity combination, the minimum time-to-solution obtained anywhere on this grid. These full per-instance data complement the averaged normalized comparison shown in the main text. They make clear that even at this coarse level, 1-ADAM-IM already substantially outperforms both GD-IM and MOM-IM, while among the nonlinearities the sigmoid is overall the strongest choice and the periodic nonlinearity remains the closest competitor, especially on the easier instances. By contrast, the polynomial and clipped nonlinearities are generally less competitive in this scan.

\begin{table}[H]
\centering
\caption{\textbf{g05 benchmark set: Time-to-solution (TTS) across all four nonlinearities for Gradient Descent, Momentum, and First-Order Adam.}
For each g05 instance, we report the minimum \(\mathrm{TTS}\) found over a \(30\times30\) grid scan in \((\alpha,\beta)\), with \(\gamma=0.005\) fixed, for each of the four nonlinearities. Here, \(T_a\) denotes the mean first-passage time to the ground-state energy and \(SR_{\mathrm{tr}}\) the transient success rate. Each grid point is averaged over 400 runs of \(10{,}000\) Euler steps with \(\Delta t=0.01\). Colored cells indicate the within-row ranking: green (best), yellow (second best), orange (second worst), and red (worst). For MOM-IM, we set \(\beta_1=0.99\); for 1-ADAM-IM, we set \(\beta_1=\beta_2=0.99\) and \(\eta=\frac{\sqrt{-\ln(\beta_1)}}{-\ln(\beta_2)}=9.97\). All values are reported in Euler time and rounded to one decimal place.}
\label{tab:supp_allopt_gridscan}
\begin{tabular}{lllllllllllll}
\cline{2-13}
\multicolumn{1}{l|}{}                       & \multicolumn{4}{c|}{\textbf{GD-IM}}                                                                                                                    & \multicolumn{4}{c|}{\textbf{MOM-IM}}                                                                                                                            & \multicolumn{4}{c|}{\textbf{1-ADAM-IM}}                                                                                                                                \\ \cline{2-13} 
\multicolumn{1}{l|}{}                       & \multicolumn{1}{l|}{\textbf{Poly}} & \multicolumn{1}{l|}{\textbf{Sig}} & \multicolumn{1}{l|}{\textbf{Per}} & \multicolumn{1}{l|}{\textbf{Clip}} & \multicolumn{1}{l|}{\textbf{Poly}} & \multicolumn{1}{l|}{\textbf{Sig}} & \multicolumn{1}{l|}{\textbf{Per}} & \multicolumn{1}{l|}{\textbf{Clip}} & \multicolumn{1}{l|}{\textbf{Poly}} & \multicolumn{1}{l|}{\textbf{Sig}} & \multicolumn{1}{l|}{\textbf{Per}} & \multicolumn{1}{l|}{\textbf{Clip}} \\ \hline
\multicolumn{1}{|l|}{ \textbf{g\_05\_60.0} } & \multicolumn{1}{c}{\cellcolor[HTML]{FFCB2F}189.6} & \multicolumn{1}{c}{31.5} & \multicolumn{1}{c}{44.0} & \multicolumn{1}{c}{49.8} & \multicolumn{1}{c}{\cellcolor[HTML]{FE0000}197.9} & \multicolumn{1}{c}{31.8} & \multicolumn{1}{c}{49.1} & \multicolumn{1}{c}{56.8} & \multicolumn{1}{c}{6.0} & \multicolumn{1}{c}{\cellcolor[HTML]{F8FF00}4.0} & \multicolumn{1}{c}{\cellcolor[HTML]{34FF34}3.1} & \multicolumn{1}{c}{6.9} \\ \cline{1-1}
\multicolumn{1}{|l|}{ \textbf{g\_05\_60.1} } & \multicolumn{1}{c}{30.9} & \multicolumn{1}{c}{48.4} & \multicolumn{1}{c}{\cellcolor[HTML]{FE0000}150.0} & \multicolumn{1}{c}{44.3} & \multicolumn{1}{c}{30.4} & \multicolumn{1}{c}{50.8} & \multicolumn{1}{c}{\cellcolor[HTML]{FFCB2F}134.6} & \multicolumn{1}{c}{40.0} & \multicolumn{1}{c}{\cellcolor[HTML]{34FF34}1.6} & \multicolumn{1}{c}{\cellcolor[HTML]{F8FF00}1.8} & \multicolumn{1}{c}{2.0} & \multicolumn{1}{c}{3.1} \\ \cline{1-1}
\multicolumn{1}{|l|}{ \textbf{g\_05\_60.2} } & \multicolumn{1}{c}{81.8} & \multicolumn{1}{c}{115.7} & \multicolumn{1}{c}{\cellcolor[HTML]{FFCB2F}237.1} & \multicolumn{1}{c}{103.7} & \multicolumn{1}{c}{78.1} & \multicolumn{1}{c}{114.6} & \multicolumn{1}{c}{\cellcolor[HTML]{FE0000}249.5} & \multicolumn{1}{c}{107.2} & \multicolumn{1}{c}{\cellcolor[HTML]{F8FF00}5.6} & \multicolumn{1}{c}{\cellcolor[HTML]{34FF34}4.8} & \multicolumn{1}{c}{6.6} & \multicolumn{1}{c}{7.2} \\ \cline{1-1}
\multicolumn{1}{|l|}{ \textbf{g\_05\_60.3} } & \multicolumn{1}{c}{\cellcolor[HTML]{FE0000}754.3} & \multicolumn{1}{c}{83.8} & \multicolumn{1}{c}{51.8} & \multicolumn{1}{c}{135.3} & \multicolumn{1}{c}{\cellcolor[HTML]{FFCB2F}607.1} & \multicolumn{1}{c}{89.3} & \multicolumn{1}{c}{53.6} & \multicolumn{1}{c}{138.3} & \multicolumn{1}{c}{91.9} & \multicolumn{1}{c}{\cellcolor[HTML]{34FF34}6.4} & \multicolumn{1}{c}{\cellcolor[HTML]{F8FF00}10.1} & \multicolumn{1}{c}{14.5} \\ \cline{1-1}
\multicolumn{1}{|l|}{ \textbf{g\_05\_60.4} } & \multicolumn{1}{c}{102.9} & \multicolumn{1}{c}{76.6} & \multicolumn{1}{c}{97.2} & \multicolumn{1}{c}{\cellcolor[HTML]{FFCB2F}125.0} & \multicolumn{1}{c}{100.7} & \multicolumn{1}{c}{68.1} & \multicolumn{1}{c}{90.0} & \multicolumn{1}{c}{\cellcolor[HTML]{FE0000}128.1} & \multicolumn{1}{c}{5.3} & \multicolumn{1}{c}{\cellcolor[HTML]{34FF34}2.8} & \multicolumn{1}{c}{\cellcolor[HTML]{F8FF00}3.9} & \multicolumn{1}{c}{5.6} \\ \cline{1-1}
\multicolumn{1}{|l|}{ \textbf{g\_05\_60.5} } & \multicolumn{1}{c}{160.5} & \multicolumn{1}{c}{124.0} & \multicolumn{1}{c}{\cellcolor[HTML]{FFCB2F}219.4} & \multicolumn{1}{c}{134.7} & \multicolumn{1}{c}{160.4} & \multicolumn{1}{c}{128.0} & \multicolumn{1}{c}{\cellcolor[HTML]{FE0000}237.8} & \multicolumn{1}{c}{154.4} & \multicolumn{1}{c}{11.9} & \multicolumn{1}{c}{\cellcolor[HTML]{34FF34}8.5} & \multicolumn{1}{c}{11.8} & \multicolumn{1}{c}{\cellcolor[HTML]{F8FF00}8.9} \\ \cline{1-1}
\multicolumn{1}{|l|}{ \textbf{g\_05\_60.6} } & \multicolumn{1}{c}{1398.2} & \multicolumn{1}{c}{496.8} & \multicolumn{1}{c}{938.3} & \multicolumn{1}{c}{1137.5} & \multicolumn{1}{c}{\cellcolor[HTML]{FFCB2F}1766.2} & \multicolumn{1}{c}{482.2} & \multicolumn{1}{c}{1011.9} & \multicolumn{1}{c}{\cellcolor[HTML]{FE0000}1786.9} & \multicolumn{1}{c}{\cellcolor[HTML]{34FF34}48.9} & \multicolumn{1}{c}{115.6} & \multicolumn{1}{c}{92.0} & \multicolumn{1}{c}{\cellcolor[HTML]{F8FF00}55.2} \\ \cline{1-1}
\multicolumn{1}{|l|}{ \textbf{g\_05\_60.7} } & \multicolumn{1}{c}{\cellcolor[HTML]{FFCB2F}254.9} & \multicolumn{1}{c}{61.1} & \multicolumn{1}{c}{97.2} & \multicolumn{1}{c}{118.7} & \multicolumn{1}{c}{\cellcolor[HTML]{FE0000}310.1} & \multicolumn{1}{c}{62.4} & \multicolumn{1}{c}{96.6} & \multicolumn{1}{c}{105.6} & \multicolumn{1}{c}{12.2} & \multicolumn{1}{c}{\cellcolor[HTML]{F8FF00}3.4} & \multicolumn{1}{c}{\cellcolor[HTML]{34FF34}3.0} & \multicolumn{1}{c}{8.7} \\ \cline{1-1}
\multicolumn{1}{|l|}{ \textbf{g\_05\_60.8} } & \multicolumn{1}{c}{\cellcolor[HTML]{FE0000}1525.8} & \multicolumn{1}{c}{239.0} & \multicolumn{1}{c}{1001.2} & \multicolumn{1}{c}{332.7} & \multicolumn{1}{c}{\cellcolor[HTML]{FFCB2F}1365.6} & \multicolumn{1}{c}{227.6} & \multicolumn{1}{c}{1290.8} & \multicolumn{1}{c}{415.1} & \multicolumn{1}{c}{21.9} & \multicolumn{1}{c}{\cellcolor[HTML]{F8FF00}10.8} & \multicolumn{1}{c}{30.1} & \multicolumn{1}{c}{\cellcolor[HTML]{34FF34}10.6} \\ \cline{1-1}
\multicolumn{1}{|l|}{ \textbf{g\_05\_60.9} } & \multicolumn{1}{c}{\cellcolor[HTML]{FE0000}291.9} & \multicolumn{1}{c}{81.8} & \multicolumn{1}{c}{267.9} & \multicolumn{1}{c}{173.2} & \multicolumn{1}{c}{\cellcolor[HTML]{FFCB2F}283.1} & \multicolumn{1}{c}{101.6} & \multicolumn{1}{c}{258.3} & \multicolumn{1}{c}{161.7} & \multicolumn{1}{c}{17.1} & \multicolumn{1}{c}{\cellcolor[HTML]{F8FF00}11.1} & \multicolumn{1}{c}{11.6} & \multicolumn{1}{c}{\cellcolor[HTML]{34FF34}8.4} \\ \cline{1-1}
\multicolumn{1}{|l|}{ \textbf{g\_05\_80.0} } & \multicolumn{1}{c}{\cellcolor[HTML]{FE0000}496.0} & \multicolumn{1}{c}{254.5} & \multicolumn{1}{c}{197.6} & \multicolumn{1}{c}{246.7} & \multicolumn{1}{c}{\cellcolor[HTML]{FFCB2F}441.3} & \multicolumn{1}{c}{212.2} & \multicolumn{1}{c}{187.5} & \multicolumn{1}{c}{265.7} & \multicolumn{1}{c}{23.4} & \multicolumn{1}{c}{\cellcolor[HTML]{F8FF00}15.0} & \multicolumn{1}{c}{\cellcolor[HTML]{34FF34}12.0} & \multicolumn{1}{c}{28.2} \\ \cline{1-1}
\multicolumn{1}{|l|}{ \textbf{g\_05\_80.1} } & \multicolumn{1}{c}{5.3} & \multicolumn{1}{c}{12.9} & \multicolumn{1}{c}{\cellcolor[HTML]{FFCB2F}23.5} & \multicolumn{1}{c}{19.2} & \multicolumn{1}{c}{5.0} & \multicolumn{1}{c}{13.6} & \multicolumn{1}{c}{\cellcolor[HTML]{FE0000}24.8} & \multicolumn{1}{c}{16.0} & \multicolumn{1}{c}{0.5} & \multicolumn{1}{c}{\cellcolor[HTML]{F8FF00}0.3} & \multicolumn{1}{c}{0.4} & \multicolumn{1}{c}{\cellcolor[HTML]{34FF34}0.3} \\ \cline{1-1}
\multicolumn{1}{|l|}{ \textbf{g\_05\_80.2} } & \multicolumn{1}{c}{\cellcolor[HTML]{FFCB2F}2042.1} & \multicolumn{1}{c}{158.7} & \multicolumn{1}{c}{211.9} & \multicolumn{1}{c}{414.7} & \multicolumn{1}{c}{\cellcolor[HTML]{FE0000}9990.0} & \multicolumn{1}{c}{151.1} & \multicolumn{1}{c}{217.8} & \multicolumn{1}{c}{358.9} & \multicolumn{1}{c}{94.5} & \multicolumn{1}{c}{\cellcolor[HTML]{34FF34}18.3} & \multicolumn{1}{c}{\cellcolor[HTML]{F8FF00}20.6} & \multicolumn{1}{c}{67.6} \\ \cline{1-1}
\multicolumn{1}{|l|}{ \textbf{g\_05\_80.3} } & \multicolumn{1}{c}{553.7} & \multicolumn{1}{c}{667.7} & \multicolumn{1}{c}{\cellcolor[HTML]{FE0000}1821.4} & \multicolumn{1}{c}{687.4} & \multicolumn{1}{c}{440.1} & \multicolumn{1}{c}{533.9} & \multicolumn{1}{c}{\cellcolor[HTML]{FFCB2F}1497.5} & \multicolumn{1}{c}{515.8} & \multicolumn{1}{c}{79.7} & \multicolumn{1}{c}{\cellcolor[HTML]{F8FF00}67.9} & \multicolumn{1}{c}{85.6} & \multicolumn{1}{c}{\cellcolor[HTML]{34FF34}46.9} \\ \cline{1-1}
\multicolumn{1}{|l|}{ \textbf{g\_05\_80.4} } & \multicolumn{1}{c}{\cellcolor[HTML]{FFCB2F}320.5} & \multicolumn{1}{c}{119.2} & \multicolumn{1}{c}{255.3} & \multicolumn{1}{c}{221.8} & \multicolumn{1}{c}{\cellcolor[HTML]{FE0000}375.8} & \multicolumn{1}{c}{125.3} & \multicolumn{1}{c}{296.2} & \multicolumn{1}{c}{215.4} & \multicolumn{1}{c}{26.7} & \multicolumn{1}{c}{\cellcolor[HTML]{34FF34}10.4} & \multicolumn{1}{c}{\cellcolor[HTML]{F8FF00}13.0} & \multicolumn{1}{c}{55.0} \\ \cline{1-1}
\multicolumn{1}{|l|}{ \textbf{g\_05\_80.5} } & \multicolumn{1}{c}{\cellcolor[HTML]{FE0000}3871.9} & \multicolumn{1}{c}{893.2} & \multicolumn{1}{c}{1011.9} & \multicolumn{1}{c}{1735.2} & \multicolumn{1}{c}{2850.6} & \multicolumn{1}{c}{917.4} & \multicolumn{1}{c}{\cellcolor[HTML]{FFCB2F}2859.4} & \multicolumn{1}{c}{1155.8} & \multicolumn{1}{c}{91.9} & \multicolumn{1}{c}{104.8} & \multicolumn{1}{c}{\cellcolor[HTML]{F8FF00}82.7} & \multicolumn{1}{c}{\cellcolor[HTML]{34FF34}75.6} \\ \cline{1-1}
\multicolumn{1}{|l|}{ \textbf{g\_05\_80.6} } & \multicolumn{1}{c}{\cellcolor[HTML]{FE0000}426.5} & \multicolumn{1}{c}{124.6} & \multicolumn{1}{c}{269.5} & \multicolumn{1}{c}{165.6} & \multicolumn{1}{c}{\cellcolor[HTML]{FFCB2F}390.6} & \multicolumn{1}{c}{117.0} & \multicolumn{1}{c}{244.9} & \multicolumn{1}{c}{175.1} & \multicolumn{1}{c}{102.7} & \multicolumn{1}{c}{\cellcolor[HTML]{F8FF00}29.9} & \multicolumn{1}{c}{\cellcolor[HTML]{34FF34}21.3} & \multicolumn{1}{c}{122.6} \\ \cline{1-1}
\multicolumn{1}{|l|}{ \textbf{g\_05\_80.7} } & \multicolumn{1}{c}{\cellcolor[HTML]{FE0000}2159.0} & \multicolumn{1}{c}{208.2} & \multicolumn{1}{c}{166.2} & \multicolumn{1}{c}{558.6} & \multicolumn{1}{c}{\cellcolor[HTML]{FFCB2F}1803.0} & \multicolumn{1}{c}{211.6} & \multicolumn{1}{c}{149.7} & \multicolumn{1}{c}{498.6} & \multicolumn{1}{c}{64.4} & \multicolumn{1}{c}{\cellcolor[HTML]{F8FF00}19.7} & \multicolumn{1}{c}{\cellcolor[HTML]{34FF34}5.8} & \multicolumn{1}{c}{239.2} \\ \cline{1-1}
\multicolumn{1}{|l|}{ \textbf{g\_05\_80.8} } & \multicolumn{1}{c}{\cellcolor[HTML]{FFCB2F}2269.5} & \multicolumn{1}{c}{1308.5} & \multicolumn{1}{c}{1287.8} & \multicolumn{1}{c}{1321.9} & \multicolumn{1}{c}{\cellcolor[HTML]{FE0000}2655.1} & \multicolumn{1}{c}{1139.1} & \multicolumn{1}{c}{1582.2} & \multicolumn{1}{c}{1316.2} & \multicolumn{1}{c}{189.3} & \multicolumn{1}{c}{\cellcolor[HTML]{34FF34}48.3} & \multicolumn{1}{c}{\cellcolor[HTML]{F8FF00}98.9} & \multicolumn{1}{c}{239.2} \\ \cline{1-1}
\multicolumn{1}{|l|}{ \textbf{g\_05\_80.9} } & \multicolumn{1}{c}{292.8} & \multicolumn{1}{c}{262.8} & \multicolumn{1}{c}{\cellcolor[HTML]{FE0000}695.6} & \multicolumn{1}{c}{355.5} & \multicolumn{1}{c}{313.2} & \multicolumn{1}{c}{274.0} & \multicolumn{1}{c}{\cellcolor[HTML]{FFCB2F}688.5} & \multicolumn{1}{c}{487.6} & \multicolumn{1}{c}{37.8} & \multicolumn{1}{c}{\cellcolor[HTML]{34FF34}20.7} & \multicolumn{1}{c}{\cellcolor[HTML]{F8FF00}25.2} & \multicolumn{1}{c}{36.7} \\ \cline{1-1}
\multicolumn{1}{|l|}{ \textbf{g\_05\_100.0} } & \multicolumn{1}{c}{168.5} & \multicolumn{1}{c}{317.0} & \multicolumn{1}{c}{273.2} & \multicolumn{1}{c}{\cellcolor[HTML]{FE0000}586.5} & \multicolumn{1}{c}{172.3} & \multicolumn{1}{c}{318.8} & \multicolumn{1}{c}{236.8} & \multicolumn{1}{c}{\cellcolor[HTML]{FFCB2F}568.9} & \multicolumn{1}{c}{14.3} & \multicolumn{1}{c}{\cellcolor[HTML]{34FF34}14.0} & \multicolumn{1}{c}{\cellcolor[HTML]{F8FF00}14.0} & \multicolumn{1}{c}{128.8} \\ \cline{1-1}
\multicolumn{1}{|l|}{ \textbf{g\_05\_100.1} } & \multicolumn{1}{c}{230.4} & \multicolumn{1}{c}{298.3} & \multicolumn{1}{c}{\cellcolor[HTML]{FE0000}1225.8} & \multicolumn{1}{c}{344.3} & \multicolumn{1}{c}{208.8} & \multicolumn{1}{c}{308.0} & \multicolumn{1}{c}{\cellcolor[HTML]{FFCB2F}1185.6} & \multicolumn{1}{c}{372.4} & \multicolumn{1}{c}{\cellcolor[HTML]{34FF34}14.6} & \multicolumn{1}{c}{18.3} & \multicolumn{1}{c}{\cellcolor[HTML]{F8FF00}17.1} & \multicolumn{1}{c}{49.1} \\ \cline{1-1}
\multicolumn{1}{|l|}{ \textbf{g\_05\_100.2} } & \multicolumn{1}{c}{93.5} & \multicolumn{1}{c}{215.4} & \multicolumn{1}{c}{\cellcolor[HTML]{FFCB2F}550.8} & \multicolumn{1}{c}{369.5} & \multicolumn{1}{c}{94.0} & \multicolumn{1}{c}{211.8} & \multicolumn{1}{c}{\cellcolor[HTML]{FE0000}581.1} & \multicolumn{1}{c}{411.3} & \multicolumn{1}{c}{\cellcolor[HTML]{34FF34}3.3} & \multicolumn{1}{c}{\cellcolor[HTML]{F8FF00}5.1} & \multicolumn{1}{c}{5.2} & \multicolumn{1}{c}{202.4} \\ \cline{1-1}
\multicolumn{1}{|l|}{ \textbf{g\_05\_100.3} } & \multicolumn{1}{c}{1088.5} & \multicolumn{1}{c}{1219.2} & \multicolumn{1}{c}{\cellcolor[HTML]{FE0000}3938.7} & \multicolumn{1}{c}{2195.8} & \multicolumn{1}{c}{644.9} & \multicolumn{1}{c}{1499.9} & \multicolumn{1}{c}{\cellcolor[HTML]{FFCB2F}3341.4} & \multicolumn{1}{c}{2028.9} & \multicolumn{1}{c}{402.2} & \multicolumn{1}{c}{\cellcolor[HTML]{F8FF00}293.3} & \multicolumn{1}{c}{\cellcolor[HTML]{34FF34}224.3} & \multicolumn{1}{c}{312.8} \\ \cline{1-1}
\multicolumn{1}{|l|}{ \textbf{g\_05\_100.4} } & \multicolumn{1}{c}{\cellcolor[HTML]{FFCB2F}7612.2} & \multicolumn{1}{c}{227.7} & \multicolumn{1}{c}{162.2} & \multicolumn{1}{c}{422.0} & \multicolumn{1}{c}{\cellcolor[HTML]{FE0000}7877.0} & \multicolumn{1}{c}{243.8} & \multicolumn{1}{c}{170.9} & \multicolumn{1}{c}{442.8} & \multicolumn{1}{c}{55.2} & \multicolumn{1}{c}{\cellcolor[HTML]{F8FF00}22.4} & \multicolumn{1}{c}{\cellcolor[HTML]{34FF34}17.3} & \multicolumn{1}{c}{73.6} \\ \cline{1-1}
\multicolumn{1}{|l|}{ \textbf{g\_05\_100.5} } & \multicolumn{1}{c}{1427.9} & \multicolumn{1}{c}{792.1} & \multicolumn{1}{c}{213.2} & \multicolumn{1}{c}{\cellcolor[HTML]{FFCB2F}1630.7} & \multicolumn{1}{c}{1591.0} & \multicolumn{1}{c}{716.2} & \multicolumn{1}{c}{224.0} & \multicolumn{1}{c}{\cellcolor[HTML]{FE0000}2388.7} & \multicolumn{1}{c}{\cellcolor[HTML]{F8FF00}67.3} & \multicolumn{1}{c}{80.4} & \multicolumn{1}{c}{\cellcolor[HTML]{34FF34}49.3} & \multicolumn{1}{c}{81.6} \\ \cline{1-1}
\multicolumn{1}{|l|}{ \textbf{g\_05\_100.6} } & \multicolumn{1}{c}{443.0} & \multicolumn{1}{c}{285.7} & \multicolumn{1}{c}{\cellcolor[HTML]{FE0000}530.6} & \multicolumn{1}{c}{382.4} & \multicolumn{1}{c}{488.5} & \multicolumn{1}{c}{288.3} & \multicolumn{1}{c}{\cellcolor[HTML]{FFCB2F}520.5} & \multicolumn{1}{c}{248.0} & \multicolumn{1}{c}{80.6} & \multicolumn{1}{c}{\cellcolor[HTML]{34FF34}15.6} & \multicolumn{1}{c}{\cellcolor[HTML]{F8FF00}18.7} & \multicolumn{1}{c}{46.4} \\ \cline{1-1}
\multicolumn{1}{|l|}{ \textbf{g\_05\_100.7} } & \multicolumn{1}{c}{395.0} & \multicolumn{1}{c}{397.6} & \multicolumn{1}{c}{292.8} & \multicolumn{1}{c}{\cellcolor[HTML]{FE0000}774.4} & \multicolumn{1}{c}{390.9} & \multicolumn{1}{c}{338.7} & \multicolumn{1}{c}{347.5} & \multicolumn{1}{c}{\cellcolor[HTML]{FFCB2F}662.9} & \multicolumn{1}{c}{112.1} & \multicolumn{1}{c}{\cellcolor[HTML]{F8FF00}30.6} & \multicolumn{1}{c}{\cellcolor[HTML]{34FF34}14.0} & \multicolumn{1}{c}{124.0} \\ \cline{1-1}
\multicolumn{1}{|l|}{ \textbf{g\_05\_100.8} } & \multicolumn{1}{c}{1750.2} & \multicolumn{1}{c}{1091.2} & \multicolumn{1}{c}{\cellcolor[HTML]{FFCB2F}2745.3} & \multicolumn{1}{c}{1258.1} & \multicolumn{1}{c}{1798.5} & \multicolumn{1}{c}{1110.7} & \multicolumn{1}{c}{\cellcolor[HTML]{FE0000}2751.1} & \multicolumn{1}{c}{1586.6} & \multicolumn{1}{c}{368.0} & \multicolumn{1}{c}{\cellcolor[HTML]{34FF34}258.5} & \multicolumn{1}{c}{373.0} & \multicolumn{1}{c}{\cellcolor[HTML]{F8FF00}331.2} \\ \cline{1-1}
\multicolumn{1}{|l|}{ \textbf{g\_05\_100.9} } & \multicolumn{1}{c}{189.7} & \multicolumn{1}{c}{249.1} & \multicolumn{1}{c}{\cellcolor[HTML]{FFCB2F}1018.9} & \multicolumn{1}{c}{356.6} & \multicolumn{1}{c}{160.0} & \multicolumn{1}{c}{238.3} & \multicolumn{1}{c}{\cellcolor[HTML]{FE0000}1148.9} & \multicolumn{1}{c}{435.3} & \multicolumn{1}{c}{\cellcolor[HTML]{34FF34}23.9} & \multicolumn{1}{c}{25.7} & \multicolumn{1}{c}{\cellcolor[HTML]{F8FF00}24.2} & \multicolumn{1}{c}{252.7} \\ \cline{1-1}
\end{tabular}
\label{TTS_tab}
\end{table}

\section{Bayesian-optimized nonlinearity comparison for GD-IM and MOM-IM}
\label{subsec:supp_bo_gd_mom_nonlins}

Complementing the coarse grid-scan comparisons reported above, we here provide the full Bayesian-optimized comparison of the four nonlinearities for Gradient Descent (GD-IM) and Momentum (MOM-IM) on the g05 benchmark set. As in the main text, the purpose of this analysis is to move beyond the restricted \((\alpha,\beta)\) scan and allow each nonlinearity to be optimized jointly with the full set of free optimizer parameters. This makes it possible to assess the optimizer--nonlinearity interaction in the same way as for the Adam-based variants.

\begin{figure}[!t]
    \centering
    \begin{minipage}[t]{0.49\linewidth}
        \centering
        \includegraphics[width=\linewidth]{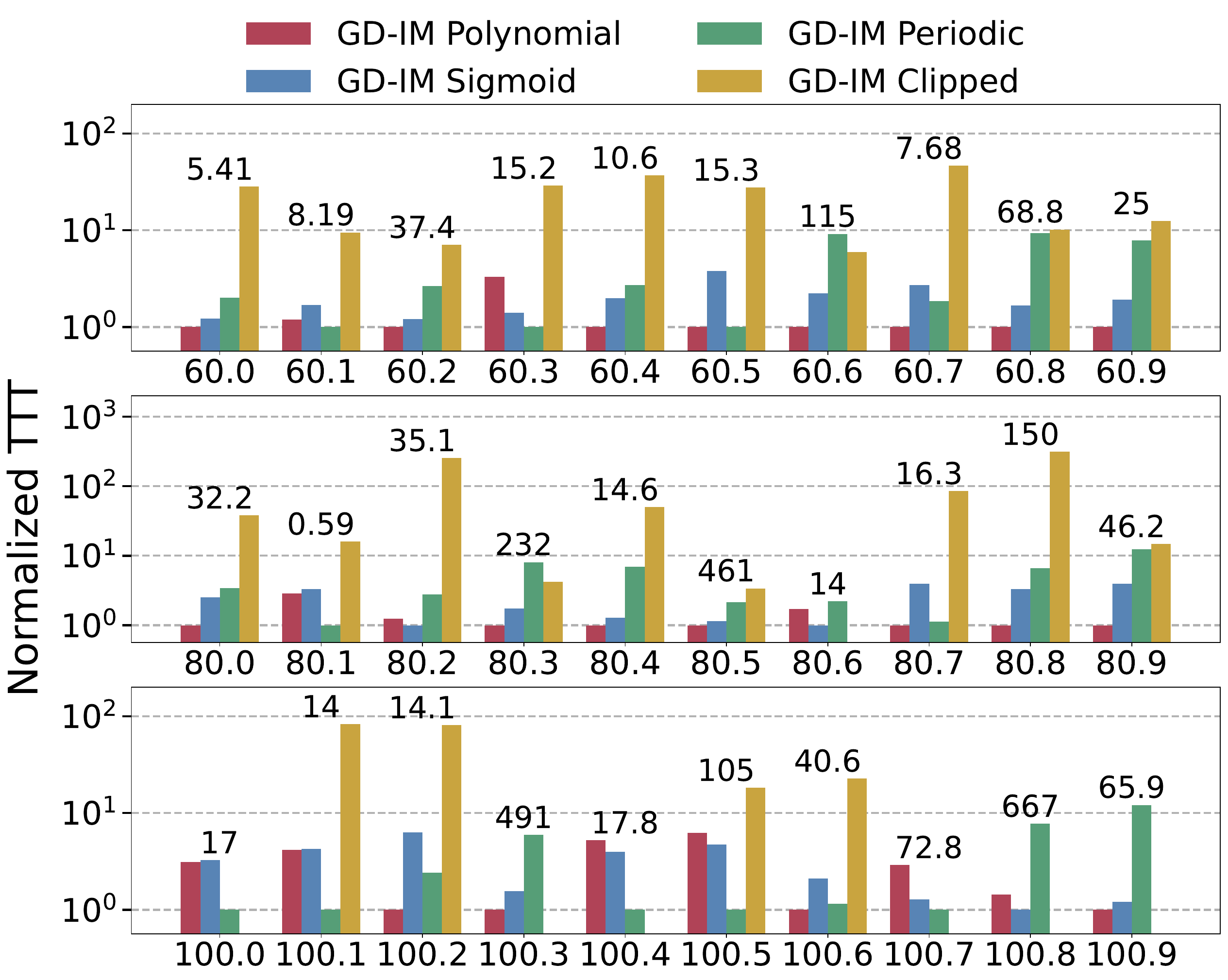}
    
        \vspace{1mm}
        \small (a) GD-IM
    \end{minipage}\hfill
    \begin{minipage}[t]{0.49\linewidth}
        \centering
        \includegraphics[width=\linewidth]{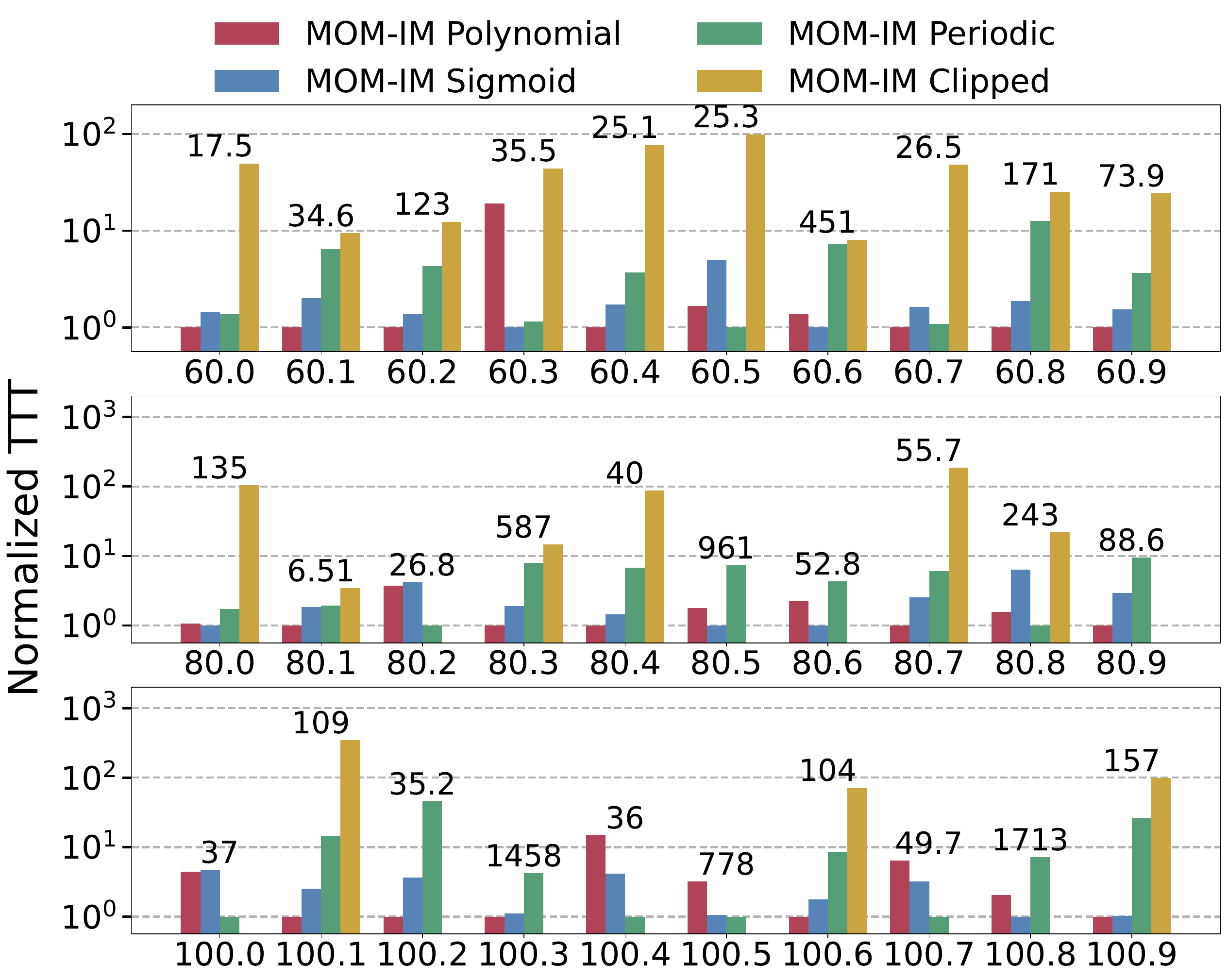}
    
        \vspace{1mm}
        \small (b) MOM-IM
    \end{minipage}
    \caption{\textbf{g05 benchmark set: Bayesian-optimized normalized time-to-solution (TTS) across all four nonlinearities for GD-IM and MOM-IM.}
    Results are shown for (a) GD-IM and (b) MOM-IM. For each benchmark instance and each nonlinearity, we first take the lowest \(\mathrm{TTS}\) found during Bayesian optimization. These per-instance minima are then normalized by the lowest \(\mathrm{TTS}\) obtained on that same instance across the four nonlinearities for the optimizer under consideration, so that the fastest nonlinearity is assigned the value \(1\). The vertical axis is shown on a logarithmic scale. The numerical labels above the bars indicate the corresponding unnormalized minimum \(\mathrm{TTS}\) values used for this normalization.}
    \label{fig:supp_g05_gd_mom_nonlins_combined}
\end{figure}

Figure~\ref{fig:supp_g05_gd_mom_nonlins_combined} shows the resulting Bayesian-optimized time-to-solution \(\mathrm{TTS}\) for (a) GD-IM and (b) MOM-IM. For each benchmark instance and each nonlinearity, we first take the lowest \(\mathrm{TTS}\) found during Bayesian optimization. These per-instance minima are then normalized by the lowest \(\mathrm{TTS}\) obtained on that same instance across the four nonlinearities for the optimizer under consideration, so that the fastest nonlinearity is assigned the value \(1\). The numerical labels above the bars indicate the corresponding unnormalized minimum \(\mathrm{TTS}\) values used for this normalization.

The Bayesian-optimized results show that, for both GD-IM and MOM-IM, the \emph{sigmoid} and \emph{polynomial} nonlinearities become much closer than in the coarse grid scan. In many instances they are nearly indistinguishable on the logarithmic \(\mathrm{TTS}\) scale, with the polynomial nonlinearity occasionally giving the lowest per-instance value. Even so, the sigmoid nonlinearity retains a slight overall advantage because its performance is more uniform across the full set of instances and less sensitive to fine tuning. By contrast, the \emph{periodic} nonlinearity, which could still appear competitive in the coarse scan, falls behind once all hyperparameters are optimized jointly, and the \emph{clipped} nonlinearity remains the least reliable of the four.

\section{Absolute cut values on the \textsc{Gset} benchmark set}
\label{subsec:supp_gset_cutvalues}

To complement the solution-quality comparison in the main text, we report here the best MaxCut values obtained on the unweighted and weighted \textsc{Gset} benchmark instances. These data provide the instance-by-instance values underlying the percentage plots shown in the main paper and allow direct comparison with the best-known cut values from the literature.

\begin{table}[H]
\centering
\begin{tabular}{l|@{\hspace{6pt}}r@{\hspace{6pt}}r@{\hspace{6pt}}r@{\hspace{6pt}}r@{\hspace{6pt}}r}
\toprule
Graph & Best known & 1-ADAM-IM & ADAM-IM & MOM-IM & GD-IM \\
\midrule
\bottomrule
G1 & \textbf{11624} & \textbf{11624} & \textbf{11624} & \textbf{11624} & \textbf{11624} \\
G2 & \textbf{11620} & \textbf{11620} & \textbf{11620} & \textbf{11620} & 11619 \\
G3 & \textbf{11622} & \textbf{11622} & \textbf{11622} & \textbf{11622} & \textbf{11622} \\
G4 & \textbf{11646} & \textbf{11646} & \textbf{11646} & \textbf{11646} & \textbf{11646} \\
G5 & \textbf{11631} & \textbf{11631} & \textbf{11631} & \textbf{11631} & \textbf{11631} \\
G6 & \textbf{2178} & \textbf{2178} & \textbf{2178} & \textbf{2178} & \textbf{2178} \\
G7 & \textbf{2006} & \textbf{2006} & \textbf{2006} & 2004 & \textbf{2006} \\
G8 & \textbf{2005} & \textbf{2005} & \textbf{2005} & \textbf{2005} & 2004 \\
G9 & \textbf{2054} & \textbf{2054} & \textbf{2054} & \textbf{2054} & 2052 \\
G10 & \textbf{2000} & \textbf{2000} & \textbf{2000} & \textbf{2000} & \textbf{2000} \\
G11 & \textbf{564} & \textbf{564} & \textbf{564} & 560 & 560 \\
G12 & \textbf{556} & \textbf{556} & \textbf{556} & 554 & 554 \\
G13 & \textbf{582} & \textbf{582} & \textbf{582} & 580 & 580 \\
G14 & \textbf{3064} & 3063 & 3061 & 3054 & 3052 \\
G15 & \textbf{3050} & \textbf{3050} & 3049 & 3038 & 3038 \\
G16 & \textbf{3052} & 3051 & 3051 & 3036 & 3039 \\
G17 & \textbf{3047} & \textbf{3047} & 3043 & 3035 & 3031 \\
G18 & \textbf{992} & \textbf{992} & \textbf{992} & 984 & 985 \\
G19 & \textbf{906} & \textbf{906} & \textbf{906} & 898 & 899 \\
G20 & \textbf{941} & \textbf{941} & \textbf{941} & 939 & 938 \\
G21 & \textbf{931} & \textbf{931} & \textbf{931} & 923 & 925 \\
G22 & \textbf{13359} & 13358 & 13358 & 13338 & 13341 \\
G23 & \textbf{13344} & 13342 & 13342 & 13326 & 13328 \\
G24 & \textbf{13337} & 13335 & 13332 & 13317 & 13319 \\
G25 & \textbf{13340} & 13339 & 13336 & 13320 & 13319 \\
G26 & \textbf{13328} & 13326 & 13325 & 13307 & 13307 \\
G27 & \textbf{3341} & \textbf{3341} & 3334 & 3327 & 3322 \\
G28 & \textbf{3298} & \textbf{3298} & \textbf{3298} & 3287 & 3289 \\
G29 & \textbf{3405} & \textbf{3405} & \textbf{3405} & 3392 & 3378 \\
G30 & \textbf{3413} & \textbf{3413} & 3412 & 3397 & 3403 \\
G31 & \textbf{3310} & 3309 & 3307 & 3293 & 3296 \\
G32 & \textbf{1410} & 1406 & 1394 & 1394 & 1396 \\
G33 & \textbf{1382} & 1376 & 1370 & 1372 & 1372 \\
G34 & \textbf{1384} & 1378 & 1374 & 1372 & 1374 \\
G35 & \textbf{7687} & 7675 & 7665 & 7622 & 7623 \\
G36 & \textbf{7680} & 7663 & 7649 & 7612 & 7615 \\
G37 & \textbf{7691} & 7681 & 7663 & 7621 & 7621 \\
G38 & \textbf{7688} & 7676 & 7663 & 7631 & 7634 \\
G43 & \textbf{6660} & \textbf{6660} & \textbf{6660} & \textbf{6660} & 6659 \\
G44 & \textbf{6650} & \textbf{6650} & \textbf{6650} & 6648 & 6648 \\
G45 & \textbf{6654} & \textbf{6654} & \textbf{6654} & 6653 & 6653 \\
G46 & \textbf{6649} & \textbf{6649} & \textbf{6649} & 6646 & 6648 \\
G47 & \textbf{6657} & \textbf{6657} & \textbf{6657} & 6656 & 6656 \\
G51 & \textbf{3848} & 3846 & 3843 & 3829 & 3829 \\
G52 & \textbf{3851} & 3848 & 3846 & 3835 & 3836 \\
G53 & \textbf{3850} & 3847 & 3844 & 3834 & 3829 \\
G54 & \textbf{3852} & 3849 & 3849 & 3828 & 3828 \\
\bottomrule
\end{tabular}
\caption{
Best MaxCut values found on the unweighted and weighted \textsc{Gset} instances for 
1-ADAM-IM, ADAM-IM, MOM-IM, and GD-IM.  
The reported values correspond to the best cut obtained across seven Bayesian 
optimization runs per optimizer, using the same hyperparameter ranges as in the 
time-to-solution analysis.  
Boldface indicates matches to the best-known values from the literature.}
\label{tab:gset_methods_selected}
\end{table}

\section{Algorithmic comparison of Euler--Maruyama and standard discrete implementations}
\label{subsec:supp_algorithmic_euler_vs_discrete}

\begin{figure*}[!t]
    \centering

    \begin{minipage}[t]{0.49\textwidth}
        \centering
        \includegraphics[width=\linewidth]{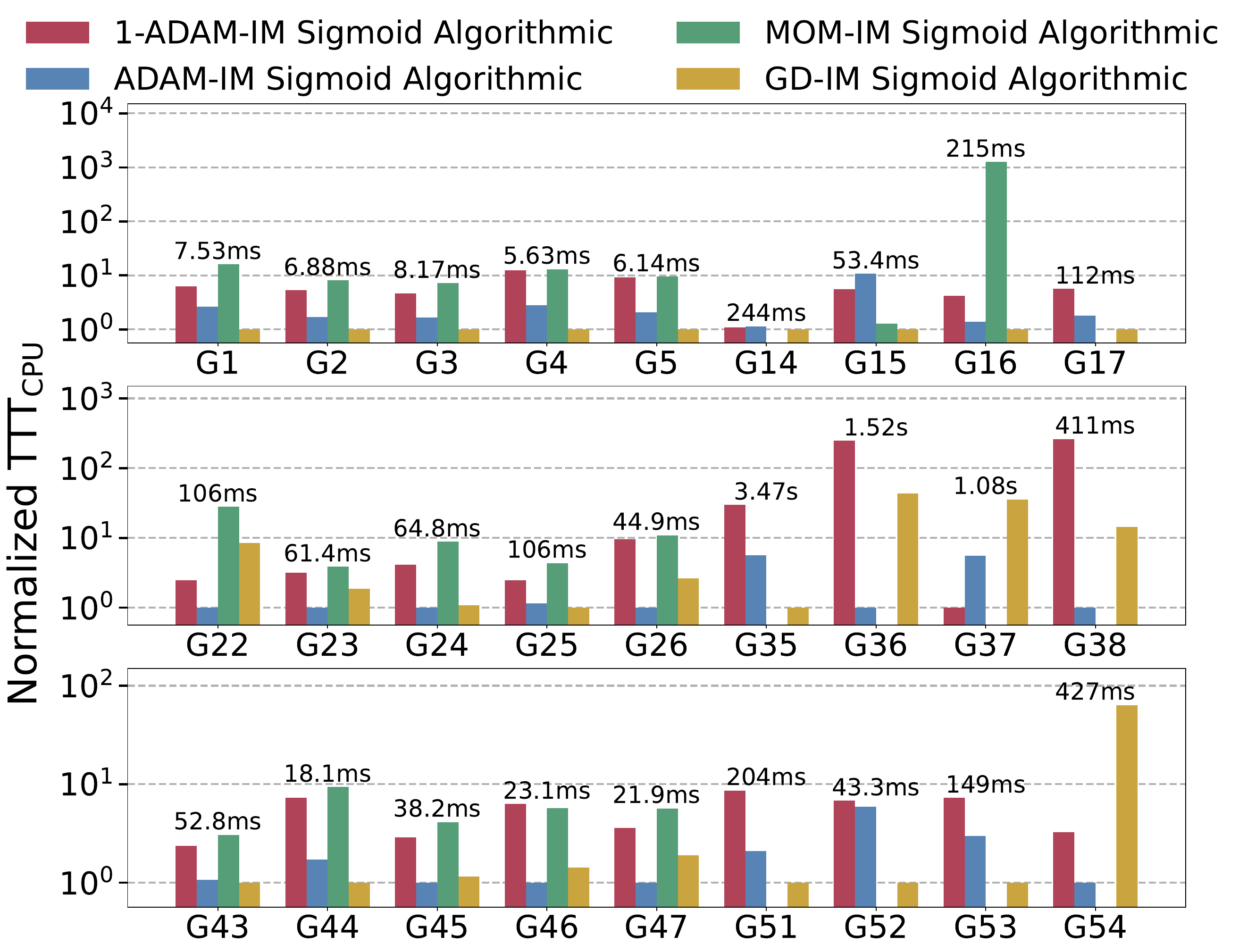}

        \vspace{1mm}
        \small (a) Euler--Maruyama, unweighted \textsc{Gset}
    \end{minipage}\hfill
    \begin{minipage}[t]{0.49\textwidth}
        \centering
        \includegraphics[width=\linewidth]{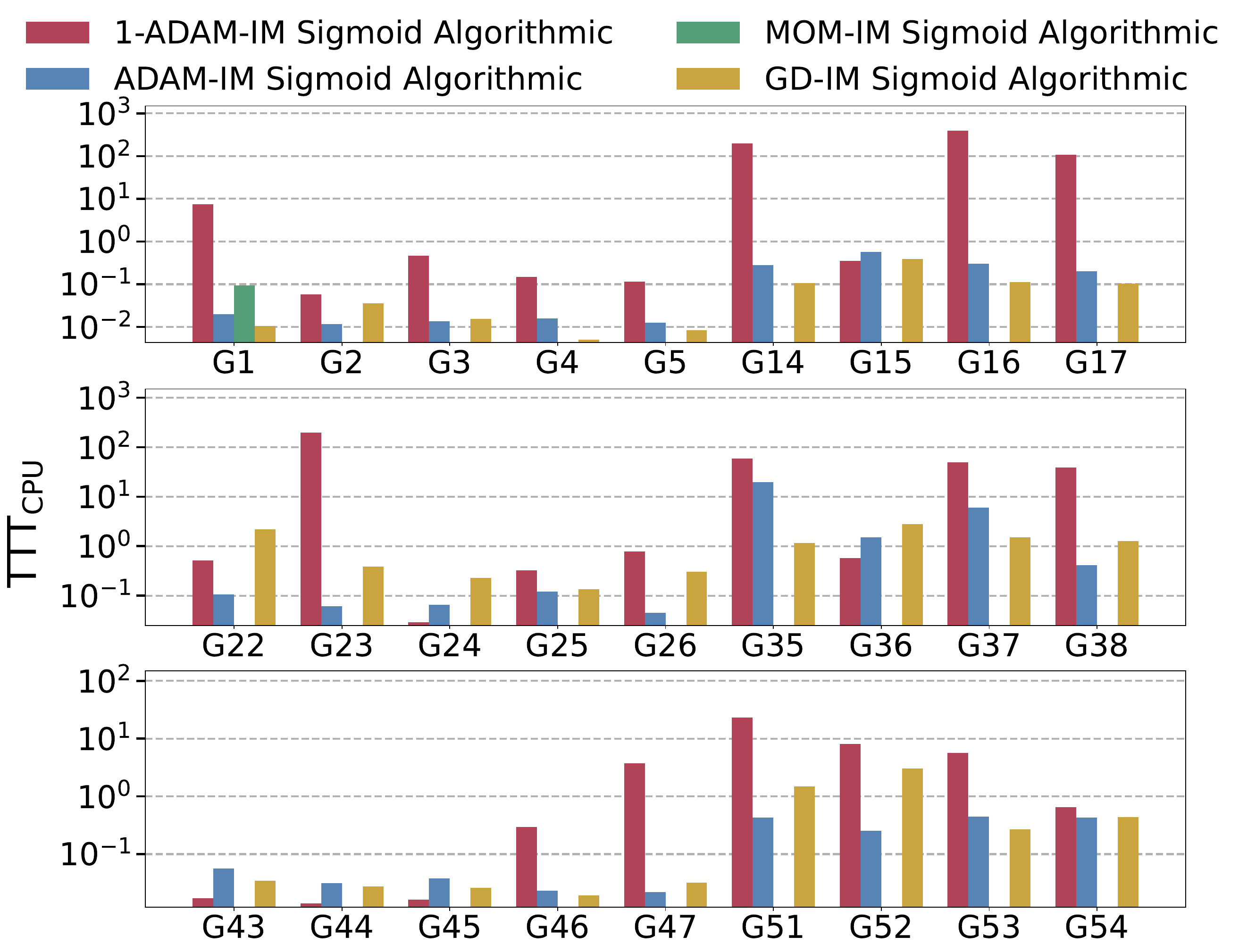}

        \vspace{1mm}
        \small (b) Standard discrete updates, unweighted \textsc{Gset}
    \end{minipage}

    \vspace{2mm}

    \begin{minipage}[t]{0.49\textwidth}
        \centering
        \includegraphics[width=\linewidth]{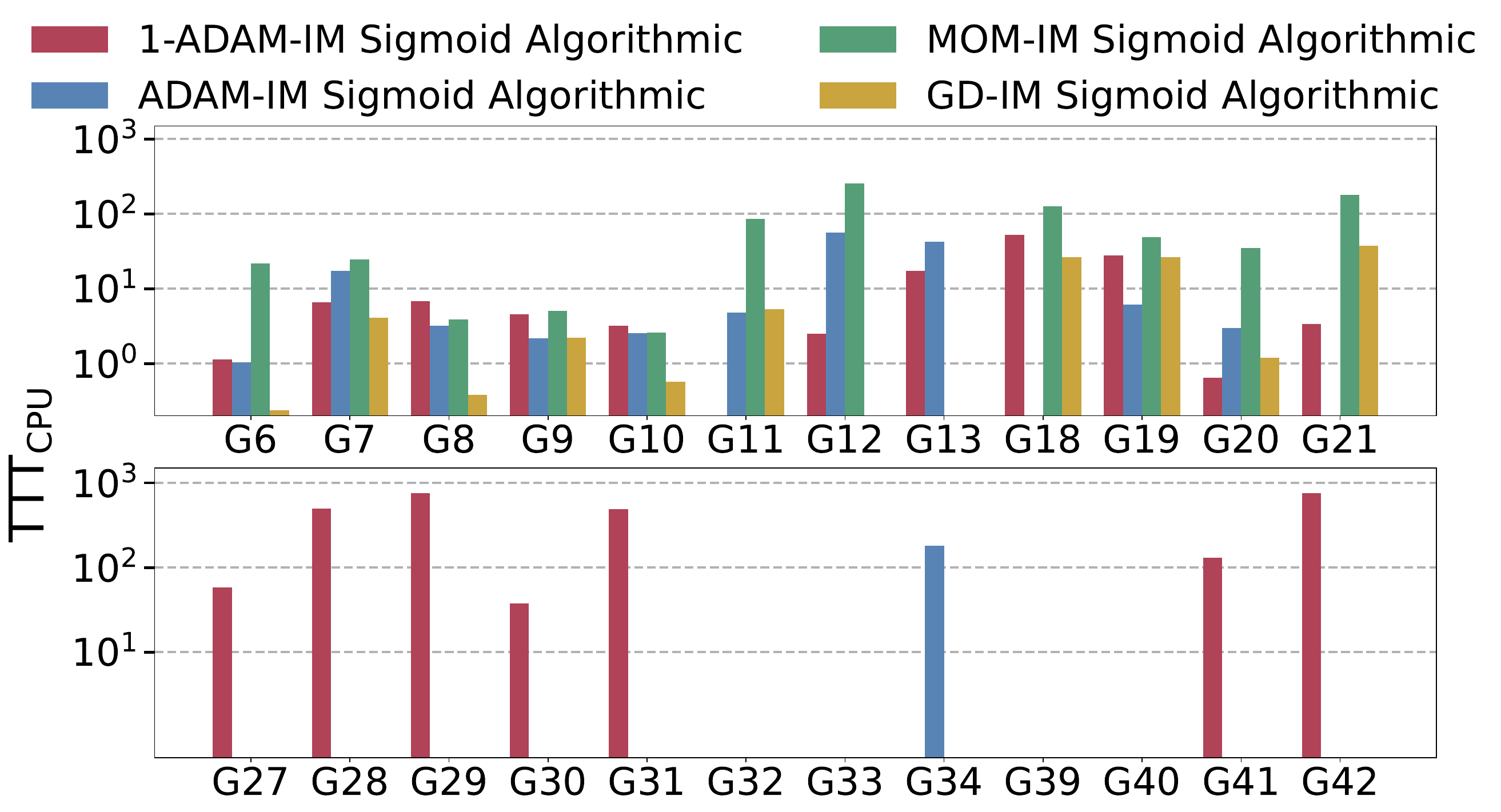}

        \vspace{1mm}
        \small (c) Euler--Maruyama, weighted \textsc{Gset}
    \end{minipage}\hfill
    \begin{minipage}[t]{0.49\textwidth}
        \centering
        \includegraphics[width=\linewidth]{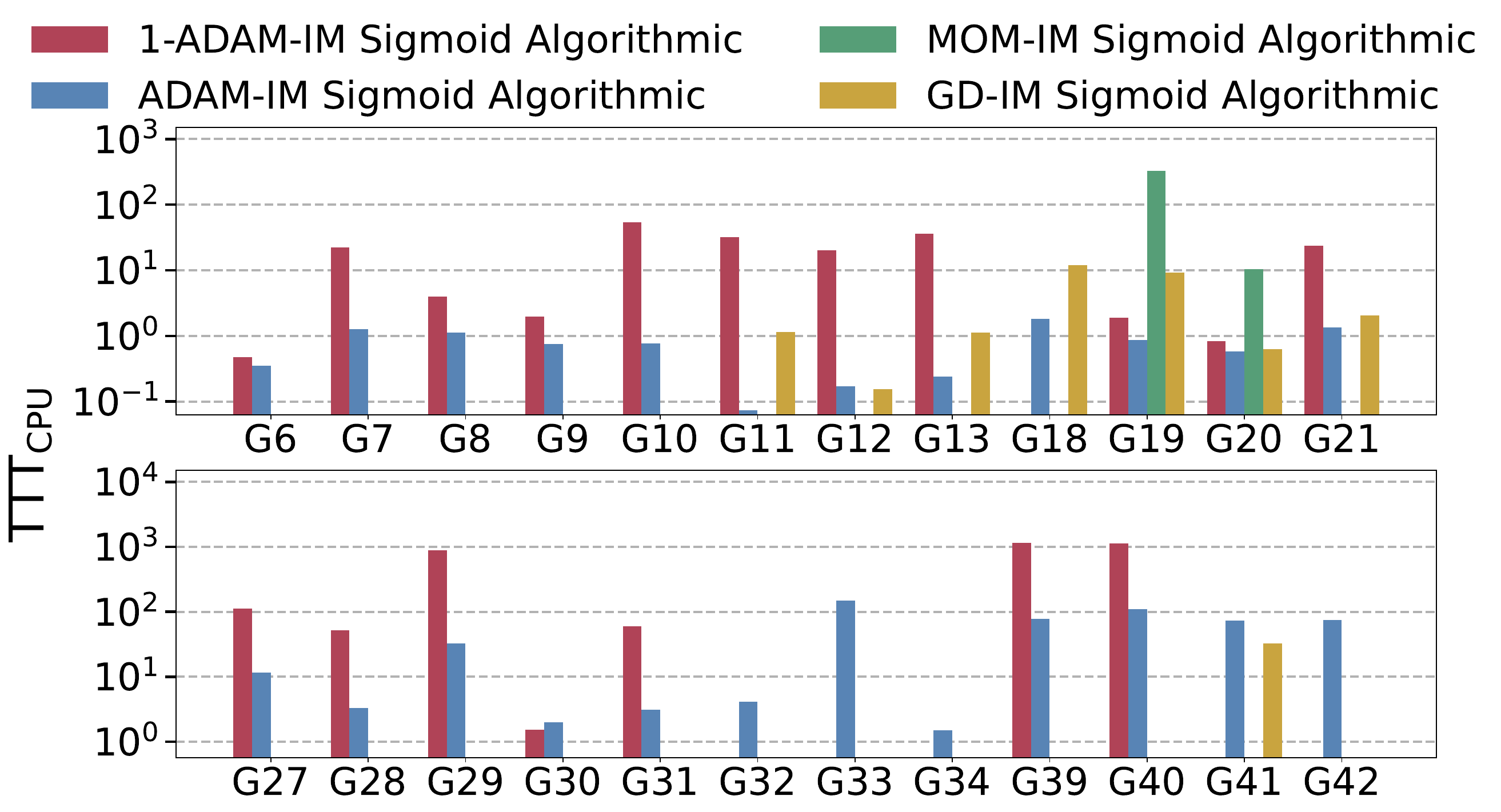}

        \vspace{1mm}
        \small (d) Standard discrete updates, weighted \textsc{Gset}
    \end{minipage}

    \caption{Comparison of Euler--Maruyama and standard discrete implementations in the algorithmic setting. Shown are the Bayesian-optimized unnormalized \(\mathrm{TTT}_{\mathrm{CPU}}\) values, in seconds and on a logarithmic scale, for (a,b) unweighted and (c,d) weighted \textsc{Gset} instances. Panels (a) and (c) show Euler--Maruyama implementations, while panels (b) and (d) show the corresponding standard discrete implementations. These results motivate the implementation choices used in the main-text algorithmic comparison.}
    \label{fig:alg_euler_vs_discrete}
\end{figure*}

For the algorithmic discrete-time setting, we compared Euler--Maruyama discretizations of the continuous-time formulations with the corresponding standard discrete update rules whenever both were available. Figure~\ref{fig:alg_euler_vs_discrete} shows the resulting unnormalized \(\mathrm{TTT}_{\mathrm{CPU}}\) values on a logarithmic scale for the unweighted and weighted \textsc{Gset} benchmarks. These comparisons motivate the implementation choices used in the main-text algorithmic results: for GD-IM, the Euler--Maruyama implementation performs at least comparably to the standard discrete update on the unweighted \textsc{Gset} instances and performs better on the weighted \textsc{Gset} instances; for MOM-IM and 1-ADAM-IM, the Euler--Maruyama implementations perform best throughout; and for ADAM-IM, the standard discrete update rule performs best throughout.

\section{\(\eta\)-rescaled time-to-target diagnostic}
\label{subsec:supp_eta_rescaled_ttt}

In the physical continuous-time comparisons, the Euler--Maruyama timestep is not treated as a free algorithmic parameter. It must remain small enough to resolve the underlying continuous-time dynamics. This matters for the interpretation of the learning-rate parameter \(\eta\). In GD-IM, \(\eta\) multiplies the deterministic update directly, so increasing \(\eta\) at fixed \(\Delta t\) is effectively equivalent to taking larger Euler steps. To keep the same physical resolution, one would instead reduce the timestep such that \(\eta\Delta t\) remains fixed. Since physical time is proportional to the timestep, this implies an approximate linear rescaling of time:
\[
\widetilde{\mathrm{TTT}}=\eta\,\mathrm{TTT}.
\]
Equivalently, this rescaling asks how the time-to-target would change if GD-IM were allowed to use the same effective learning-rate scale.

For Adam-based dynamics, this is not an exact timescale correction, because the dynamics also depend on the first- and second-moment variables and on adaptive prefactors. We therefore use the rescaling only as a diagnostic: it tests whether the observed physical time-to-target hierarchy could be removed by a simple GD-like learning-rate rescaling. For ADAM-IM, we use the optimized learning rate directly,
\[
\widetilde{\mathrm{TTT}}_{\mathrm{ADAM}}
=
\eta\,\mathrm{TTT}.
\]
For 1-ADAM-IM, the optimized learning rate is not on the same scale, because the constant prefactor from the first-order Puiseux expansion has been absorbed into \(\eta\). The leading prefactor is
\[
C(\beta_1,\beta_2)
=
\frac{\sqrt{-\ln(\beta_2)}}{-\ln(\beta_1)} .
\]
Using the reference value \(\beta_1=\beta_2=0.99\) gives \(C\simeq 9.97\approx 10\). We therefore use the effective learning rate \(\eta_{\mathrm{eff}}=\eta/10\) and define
\[
\widetilde{\mathrm{TTT}}_{\mathrm{1\text{-}ADAM}}
=
\frac{\eta}{10}\,\mathrm{TTT}.
\]
For GD-IM and MOM-IM, \(\eta=1\) in the physical simulations, so \(\widetilde{\mathrm{TTT}}=\mathrm{TTT}\).

\begin{figure*}[!t]
    \centering

    \begin{minipage}[t]{0.49\textwidth}
        \centering
        \includegraphics[width=\linewidth]{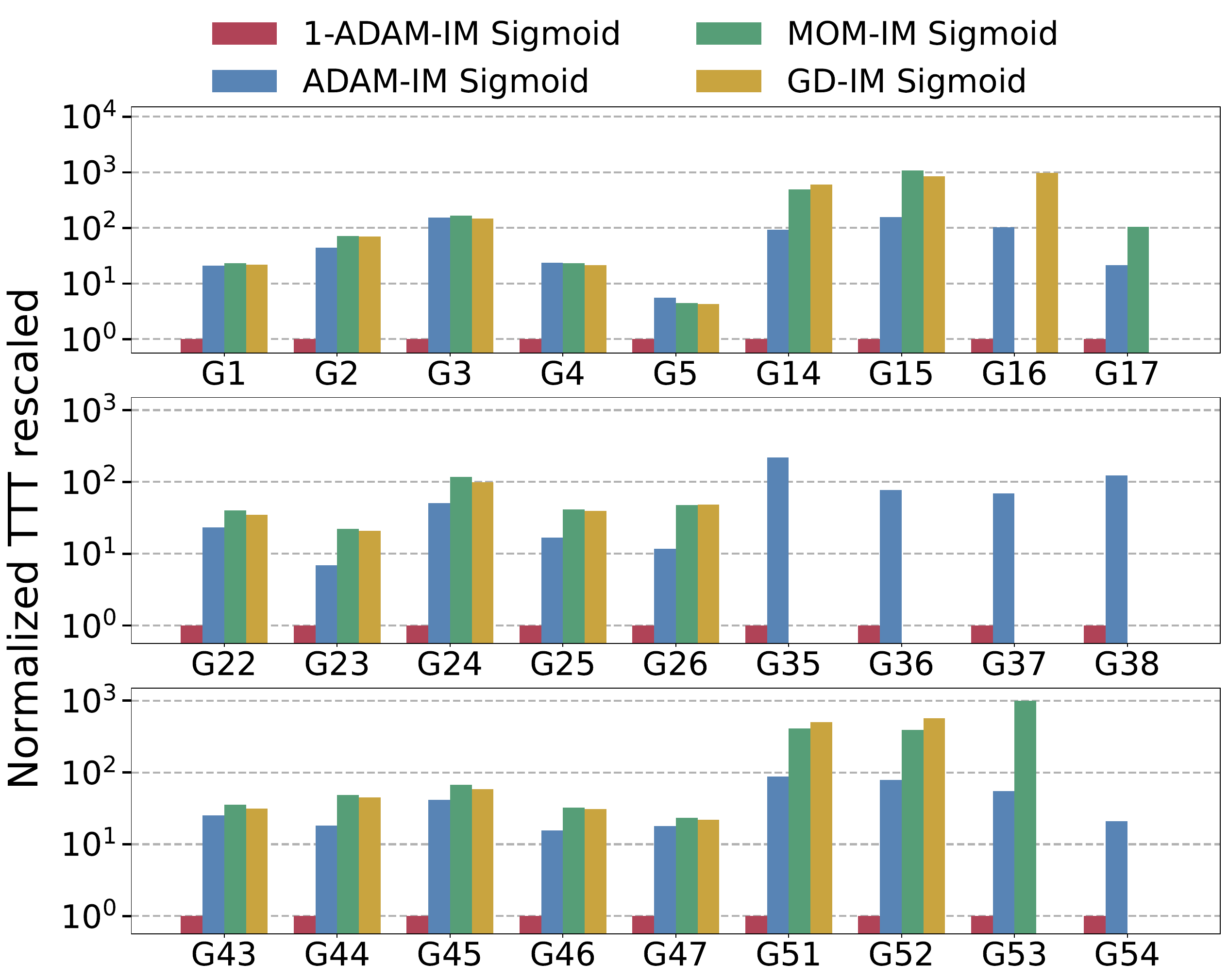}

        \vspace{1mm}
        \small (a) Unweighted \textsc{Gset}
    \end{minipage}\hfill
    \begin{minipage}[t]{0.49\textwidth}
        \centering
        \includegraphics[width=\linewidth]{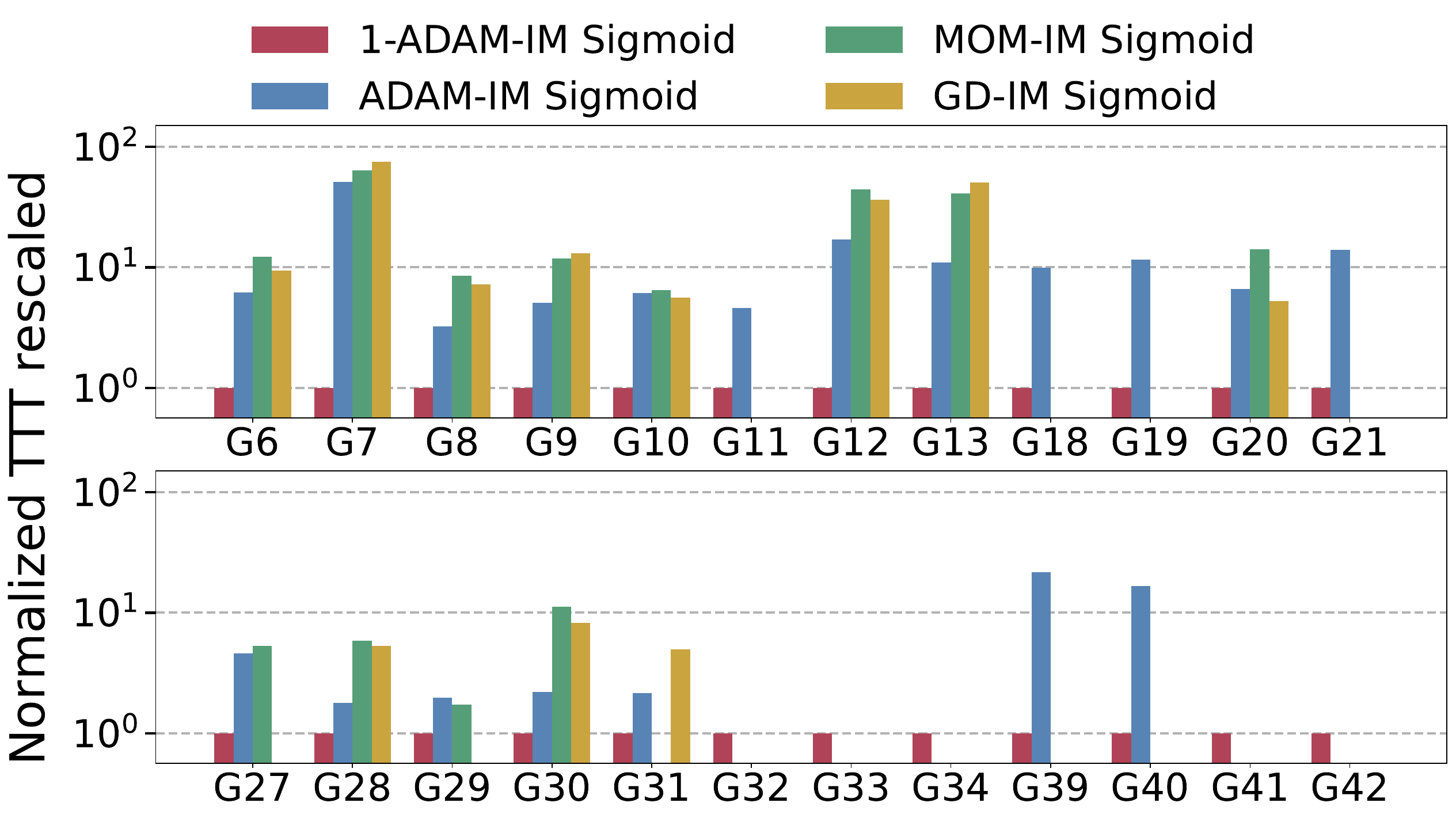}

        \vspace{1mm}
        \small (b) Weighted \textsc{Gset}
    \end{minipage}

    \caption{\(\eta\)-rescaled physical time-to-target diagnostic for the sigmoid nonlinearity on (a) unweighted and (b) weighted \textsc{Gset}. For ADAM-IM, the plotted quantity is computed sample-wise as \(\widetilde{\mathrm{TTT}}=\eta\,\mathrm{TTT}\). For 1-ADAM-IM, we use \(\widetilde{\mathrm{TTT}}=(\eta/10)\mathrm{TTT}\), because the first-order Puiseux prefactor is absorbed into the 1-ADAM-IM learning rate. For GD-IM and MOM-IM, \(\eta=1\), so \(\widetilde{\mathrm{TTT}}=\mathrm{TTT}\).}
    \label{fig:eta_rescaled_ttt}
\end{figure*}

Figure~\ref{fig:eta_rescaled_ttt} shows this \(\eta\)-rescaled diagnostic for the physical \textsc{Gset} results, separately for the unweighted and weighted benchmark instances.
The rescaled diagnostic in Fig.~\ref{fig:eta_rescaled_ttt} does not change the conclusion drawn from the physical \(\mathrm{TTT}\) comparison in the main text, Figure 5. The rescaling compresses the performance gap for some instances, as expected when part of the difference can be attributed to the optimized learning-rate scale. Nevertheless, the overall hierarchy remains the same: 1-ADAM-IM remains the best-performing physical continuous-time optimizer, followed by ADAM-IM, while GD-IM and MOM-IM remain substantially slower on the harder instances.

\begin{figure*}[!t]
    \centering

    \begin{minipage}[t]{0.49\textwidth}
        \centering
        \includegraphics[width=\linewidth]{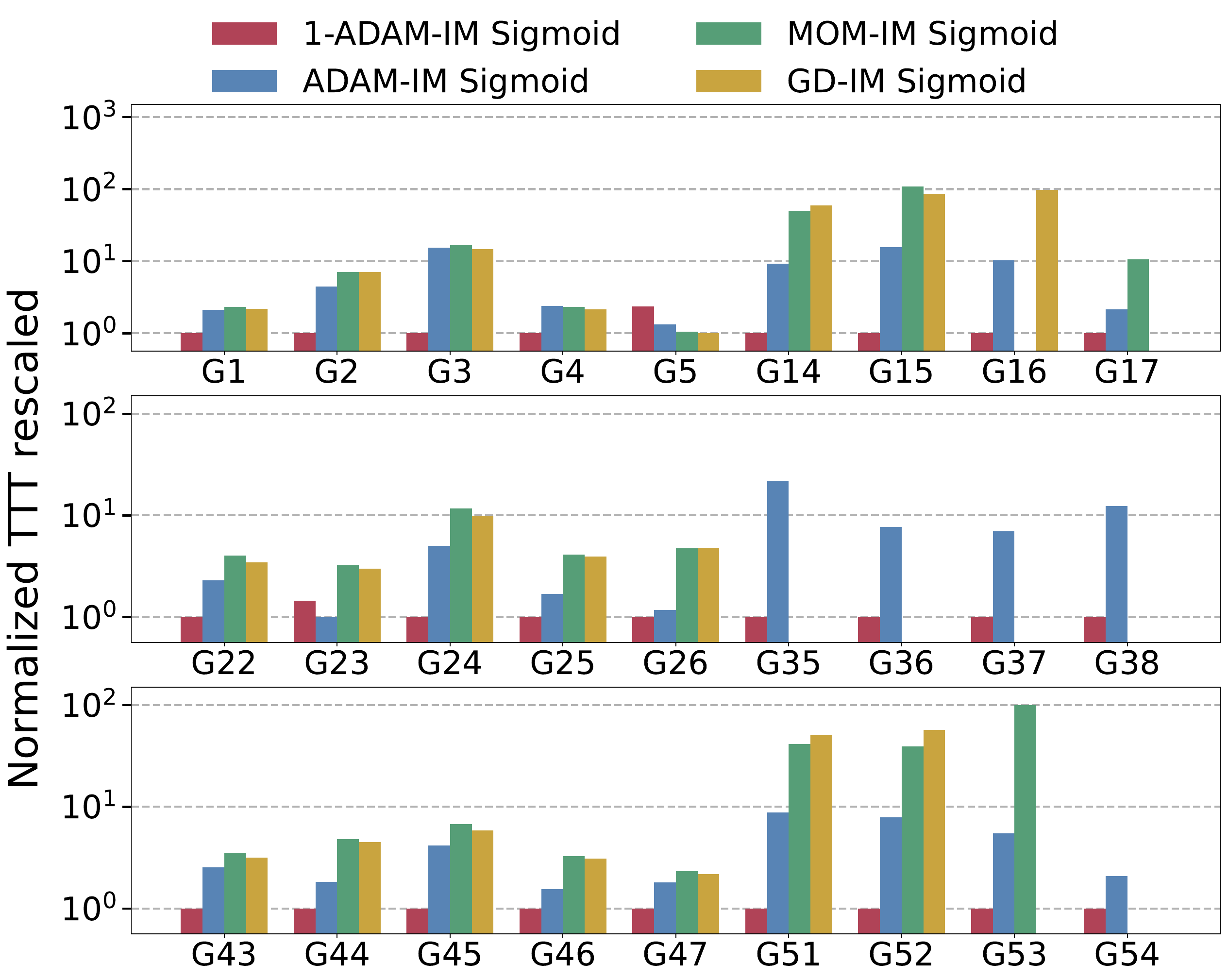}

        \vspace{1mm}
        \small (a) Unweighted \textsc{Gset}
    \end{minipage}\hfill
    \begin{minipage}[t]{0.49\textwidth}
        \centering
        \includegraphics[width=\linewidth]{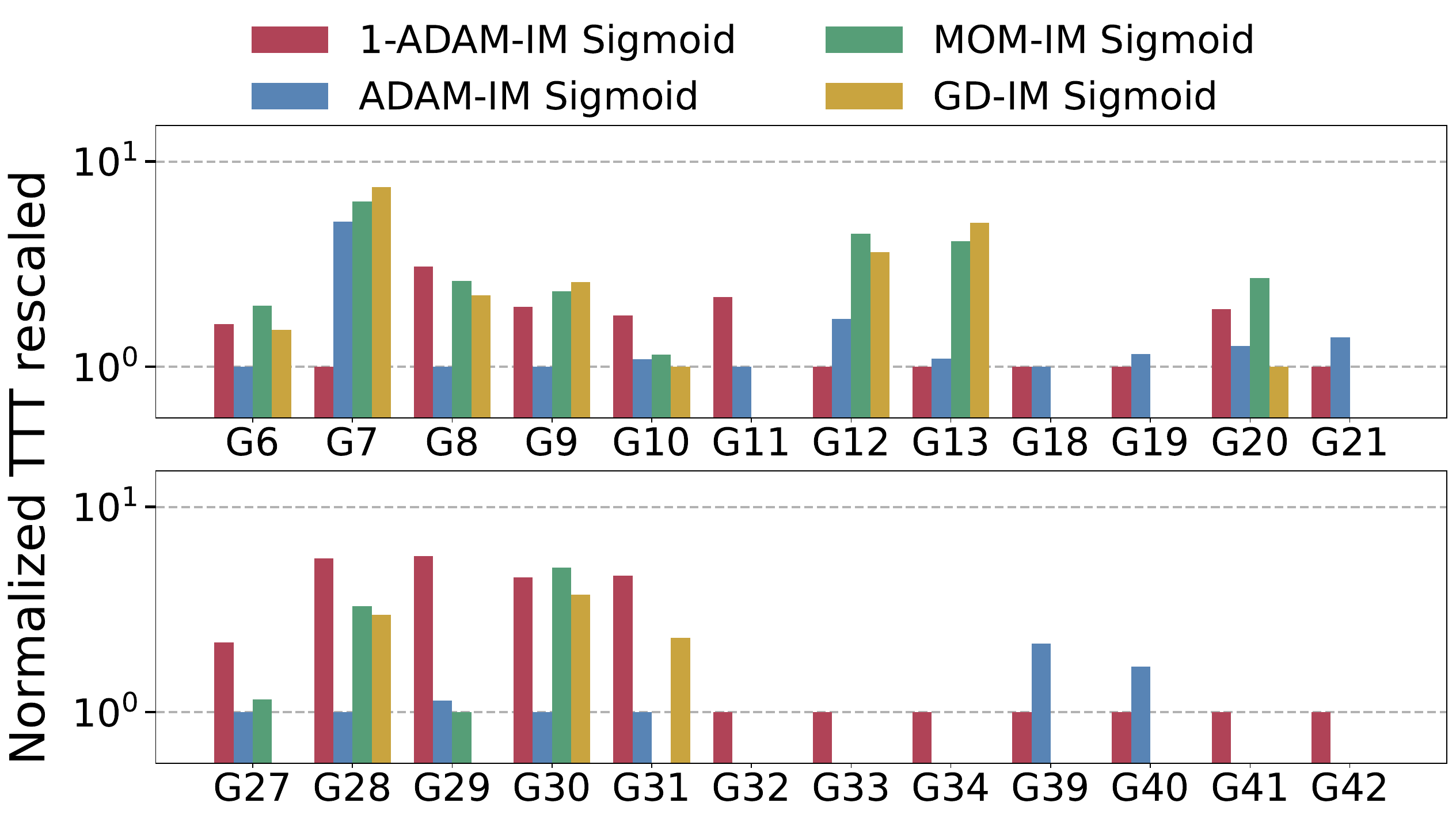}

        \vspace{1mm}
        \small (b) Weighted \textsc{Gset}
    \end{minipage}

    \caption{Alternative, more conservative, \(\eta\)-rescaled physical time-to-target diagnostic without correcting the 1-ADAM-IM learning rate for the absorbed Puiseux prefactor. In this case, both ADAM-IM and 1-ADAM-IM are rescaled sample-wise as \(\widetilde{\mathrm{TTT}}=\eta\,\mathrm{TTT}\). Results are shown for the sigmoid nonlinearity on (a) unweighted and (b) weighted \textsc{Gset}. This stronger rescaling further compresses the \(\mathrm{TTT}\) gap, especially on the unweighted instances; on the weighted instances, the rescaled \(\mathrm{TTT}\) values become similar on instances where all methods reach the target.}
    \label{fig:eta_rescaled_ttt_no_puiseux}
\end{figure*}

As a more conservative diagnostic, we also repeat the rescaling without dividing the 1-ADAM-IM learning rate by the Puiseux prefactor, i.e. using \(\widetilde{\mathrm{TTT}}=\eta\,\mathrm{TTT}\) instead of \(\widetilde{\mathrm{TTT}}=(\eta/10)\mathrm{TTT}\). This deliberately overestimates the effective learning-rate rescaling for 1-ADAM-IM and therefore gives a stringent test of whether the physical \(\mathrm{TTT}\) advantage could be attributed to learning-rate scale alone. The resulting plots are shown in Fig.~\ref{fig:eta_rescaled_ttt_no_puiseux}.
This stronger rescaling shows that part of the physical \(\mathrm{TTT}\) gap might indeed be attributed to the optimized learning-rate scale. However, it does not alter the main conclusion of the physical comparison: 1-ADAM-IM still achieves substantially better solution quality and transient success rates on the difficult weighted instances, as shown in the main text. Thus, the advantage of 1-ADAM-IM is not solely a consequence of a trivial time-rescaling effect.

\bibliography{Adam.bib}

\end{document}